\documentstyle[aps,eqsecnum,prb,multicol,epsf]{revtex}  
\newcommand{\beq}{\begin{equation}}
\newcommand{\eeq}{\end{equation}}
\newcommand{\beqa}{\begin{eqnarray}}
\newcommand{\eeqa}{\end{eqnarray}}
\newcommand{\w}{\omega}

\newcommand{\nn}{{\bf n}}
\renewcommand{\r}{{\bf r}}
\newcommand{\q}{{\bf q}}

\renewcommand{\>}{\rangle}
\newcommand{\<}{\langle}
\newcommand{\St}{\mbox{St}}
\renewcommand{\Re}{\mbox{Re} }
\renewcommand{\Im}{\mbox{Im} }
\begin{document}          

\title{Interaction corrections at intermediate temperatures: 
Longitudinal conductivity and kinetic equation}
\author{G\'abor Zala, B.N. Narozhny, and I.L. Aleiner} 
\address{Department of Physics and Astronomy, SUNY at Stony Brook, 
Stony Brook, NY, 11794} 
\date{\today}
\maketitle
\begin{abstract}                
It is well known that electron-electron interaction in two dimensional
disordered systems leads to logarithmically divergent Altshuler-Aronov
corrections to conductivity at low temperatures ($T\tau\ll 1$; $\tau$
is the elastic mean-free time). This paper is devoted to the fate of
such corrections at intermediate temperatures $T\tau \gtrsim 1$. We
show that in this (ballistic) regime the temperature dependence of
conductivity is still governed by the same physical processes as the
Altshuler-Aronov corrections - electron scattering by Friedel
oscillations. However, in this regime the correction is linear in
temperature; the value and even the {\em sign} of the slope depends on
the strength of electron-electron interaction. (This sign change may
be relevant for the ``metal-insulator'' transition observed recently.)
We show that the slope is directly related to the renormalization of
the spin susceptibility and grows as the system approaches the
ferromagnetic Stoner instability.  Also, we obtain the temperature
dependence of the conductivity in the cross-over region between the
diffusive and ballistic regimes. Finally, we derive the quantum
kinetic equation, which describes electron transport for arbitrary
value of $T\tau$.
\end{abstract}
\draft

\begin{multicols}{2} 

\section{Introduction}

Temperature dependent corrections to conductivity due to
electron-electron interactions has been a subject of theoretical
\cite{aar,fin,Lee,stern,dstern,gdl,das} 
and experimental \cite{aae,Bergman,cwh} studies for
more than two decades. Recently the interest in the matter was renewed
with appearance of new data \cite{new,nfl} showing a sign change in the
temperature dependence of conductivity in two dimensions
(2D). Theoretical discussions \cite{nfl} that followed emphasized the
question of whether that data indicated a non-Fermi liquid behavior.
However, the experiments were performed in a regime where the
temperature $T$ was of the same order of magnitude as the inverse
scattering time $\tau^{-1}$ (obtained from the Drude conductivity),
while pre-existing calculations were focused on the two limiting
cases: the diffusive regime \cite{aar,fin,Lee} $T\tau\ll 1$, and the
ballistic regime \cite{gdl,das} $T\tau\gg 1$.

In the diffusive limit one finds\cite{aar,fin,Lee} for the logarithmically
divergent correction to the diagonal conductivity $\delta \sigma$:
\beq
\delta\sigma=-
\frac{e^2}{2\pi^2\hbar}\ln\left(\frac{\hbar}{T\tau}\right)
\Big[1 +  3\left(1-\frac{\ln(1+F_0^\sigma)}{F_0^\sigma}
\right) \Big],
\label{introdiff}
\eeq
where $F_0^\sigma$ is the interaction constant in the triplet
channel which depends on the interaction strength. It is clear,
that the sign of this logarithmically divergent correction may
be positive (metallic) or negative (insulating), depending on
the value of $F_0^\sigma$\cite{footnote1}.

The result\cite{gdl,das}
 for the ballistic region frequently cited in literature
reads
\beq
\delta\sigma= -  \frac{e^2}{\pi\hbar}\left(\frac{T\tau}{\hbar}\right)
f(r_s),
\label{introbal}
\eeq where $f(r_s)$ is a positive function of the gas parameter of the
system, $r_s$. In a sharp contrast to Eq.~(\ref{introdiff}), equation
(\ref{introbal}) predicts always metallic sign of the interaction
correction.

The absence of a rigorous calculation at intermediate temperatures,
$T\tau/\hbar \simeq 1$, may have contributed to the notion that those
two limits are governed by different physical processes. In this paper
we prove that notion erroneous: the results (\ref{introdiff}) and
(\ref{introbal}) are due to the same physical process, namely elastic
scattering of electrons by the self-consistent potential created by
all the other electrons. Therefore, these two different expressions
are in fact the two limits of a single interaction correction. We
calculate the correction within assumptions of the Fermi liquid theory
(other limitations of our approach we discuss below) and present the
cross-over function between the diffusive and ballistic limits.

Moreover, we show that the existing theory for the ballistic limit
(\ref{introbal}) is incomplete. First, the results of
Ref.~\onlinecite{stern,dstern,gdl,das} account only for Hartree-like
interaction terms missing the exchange or Fock terms.  Second, this
theory essentially employs a perturbative expansion in terms of the
interaction strength, which breaks down for stronger coupling. Both
issues lead to the change in the theoretical prediction even on a
qualitative level.

The consequence of the first point is that the correction to
conductivity (\ref{introbal}) is always negative unlike the correction
in the diffusive limit that changes sign depending on the value of
$F_0^\sigma$.  This sign change is due to competition between the
universal (and positive) Fock correction and the coupling-specific
(and negative) Hartree contribution. If the Fock potential (or, to be
more precise, singlet channel) is properly taken into account, then
the sign of the correction in the ballistic limit is also not
universal (being positive for weak interaction in contrast to
Refs.~\onlinecite{stern,dstern,gdl,das}).

As follows from the second point, for the stronger interaction the
Hartree correction should be modified to include higher order
processes. For this case we show (see Section~\ref{diagrams}) that in
fact it should be replaced by the triplet channel correction, which is
characterized by the Fermi liquid constant $F_0^\sigma$. This constant
measures the strength of the spin-exchange interaction. If $F_0^\sigma
< 0$, the spin-exchange interaction tends to align electron spins and
(if it is strong enough) leads to the ferromagnetic Stoner
instability\cite{footnote1}.  Even though this constant is unknown, it
can be found experimentally by means of independent measurement of the
spin susceptibility of the system. As a function of temperature the
interaction correction to conductivity is almost always monotonous,
except for a narrow region of parameters (where it is so small that it
can hardly be observed).

The remainder of the paper is organized as follows: the following
section is devoted to qualitative discussion of the physics
involved. In the same section we summarize our results. Then we
present two alternative approaches to the microscopic calculation. In
Section~\ref{diagrams} we use the traditional perturbation
theory~\cite{aar} to derive the results presented in
Section~\ref{qualitative}, while in Section~\ref{eilenberger} the same
results are obtained using the quantum kinetic equation that we
derive.  The advantage of the kinetic equation approach is that it can
be readily used to discuss the temperature behavior of quantities
other than conductivity. These results are advertised in Conclusions
and will be published elsewhere \cite{pre}.

\section{Qualitative discussion and results}
\label{qualitative}

In this section we describe the scattering processes contributing to
the temperature dependence of conductivity.  We show that unlike the
standard Fermi liquid $T^2$ corrections, the leading correction to
conductivity is accumulated at large distances, of the order
$v_F/min(T,\sqrt{T/\tau})$.  In the ballistic limit such correction is
linear in temperature and we derive this result here using a text-book
quantum mechanical approach.  The diffusive limit is discussed in
detail in Ref.~\onlinecite{aar}. The resulting correction
$\delta\sigma\sim\ln T$ seems to be rather different from the linear
one, but we show that both corrections arise due to the same physics -
coherent scattering by Friedel oscillations. Throughout the paper we
keep the units such that $\hbar = 1$, except for the final answers.

\subsection{Scattering by Friedel oscillations}
\label{friedel}

We start with the simplest case of a weak short-range interaction
$V_0(\vec r_1 - \vec r_2)$ and show how one can obtain the correction
to conductivity in the ballistic limit, i.e. due to a single
scatterer. This discussion is similar to that of
Ref.~\onlinecite{rag}, where the correction to the one-particle
density of states (DoS) was discussed, and also of
Ref.~\onlinecite{aag}, which describes the correction to the
conductivity in the diffusive limit.

Consider a single impurity localized at some point, taken as the
origin. The impurity potential $U(\vec r)$ induces a modulation of
electron density close to the impurity. The oscillating part of the
modulation is known as the Friedel oscillation, which in 2D can be
written as

\begin{equation}
\delta\rho(\vec r) = - \frac{\nu\lambda}{2\pi r^2} \sin(2k_F r).
\label{frie}
\end{equation}

\noindent
Here $r$ denotes the distance to the impurity and its potential is
treated in the Born approximation $\lambda=\int U(\vec r)d\vec r$. In
2D the free electron DoS is given by $\nu=m/\pi\hbar^2$ and $m$ is the
electron mass, $k_F$ is the Fermi momentum.

Taking into account electron-electron interaction $V_0(\vec r_1 - \vec
r_2)$ one finds additional scattering potential due to the Friedel
oscillation Eq.~(\ref{frie}). This potential can be presented as a sum
of the direct (Hartree) and exchange (Fock) terms \cite{kit}

\begin{mathletters}
\begin{equation}
\delta V(\vec r_1, \vec r_2) = V_H(\vec r_1)
\delta(\vec r_1 - \vec r_2)-V_F(\vec r_1, \vec r_2);
\end{equation}
\begin{equation}
V_H(\vec r_1) = \int d\vec r_3 
V_0(\vec r_1 - \vec r_3)\delta\rho(\vec r_3);
\label{hp}
\end{equation}
\begin{equation}
V_F(\vec r_1, \vec r_2) = \frac{1}{2}V_0(\vec r_1 - \vec r_2)
\delta n(\vec r_1, \vec r_2),
\label{fp}
\end{equation}
\label{hf}
\end{mathletters}

\noindent
where by $\rho(\vec r)$ we denote diagonal elements of the one
electron density matrix $n$,

\begin{equation}
n(\vec r_1, \vec r_2) = \sum\limits_k \Psi^*_k(\vec r_1)
\Psi_k(\vec r_2).
\label{dm}
\end{equation}

\noindent
The factor $1/2$ indicates that only electrons with the same spin
participate in exchange interaction. As a function of distance from
the impurity the Hartree-Fock energy $\delta V$ oscillates similarly
to Eq.~(\ref{frie}).

{
\narrowtext
\begin{figure}[ht]
\vspace{0.5 cm}
\epsfxsize=6 cm
\centerline{\epsfbox{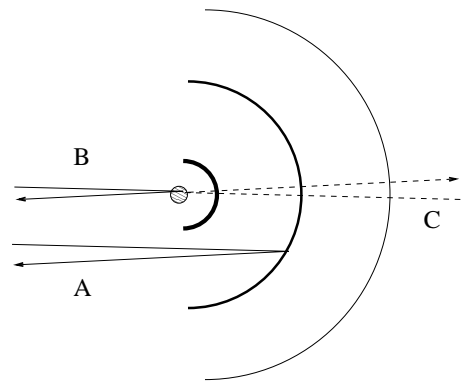}}
\vspace{0.5cm}
\caption{Scattering by the Friedel oscillation. Interference between
the two paths A and B contributes mostly to backscattering. The Friedel 
oscillation is created due to backscattering on the impurity, path C.}
\label{2}
\end{figure}
}

The leading correction to conductivity is a result of interference
between two semi-classical paths depicted on Fig.~\ref{2}. If an
electron follows path ``A'', it scatters off the Friedel oscillation
created by the impurity and path ``B'' corresponds to scattering by
the impurity itself. Interference is most important for scattering
angles close to $\pi$ (or for backscattering), since the extra phase
factor accumulated by the electron on path ``A'' ($e^{i2kR}$ with $R$
being the length of the extra path interval relative to ``B'' and $2k$
being the difference between initial and final momenta for that extra
path interval) is canceled by the phase of the Friedel oscillation
$e^{-i2k_FR}$ so that the amplitudes corresponding to the two paths
are coherent.  As a result, the probability of backscattering is
greater than the classical expectation (taken into account in the
Drude conductivity).  Therefore, taking into account interference
effects leads to a correction to conductivity.  We note that the
interference persists to large distances, limited only by temperature
$R\approx 1/|k-k_F|\le v_F/T$. Thus there is a possibility for the
correction to have a non-trivial temperature dependence.  The sign of
the correction depends on the sign of the coupling constant that
describes electron-electron interaction.

To put the above argument onto more rigorous footing and to find the
temperature dependence of the correction consider now a scattering
problem in the potential Eq.~(\ref{hf}). Following the textbook
approach \cite{llq}, we write a particle's wave function as a sum of
the incoming plane wave and the out-coming spherical wave (in 2D it is
given by a Bessel function, which we replace by its asymptotic form)

\begin{eqnarray}
\Psi = e^{i\vec k\cdot\vec r} 
+ if(\theta)\sqrt{\frac{2\pi}{kr}}e^{ikr}.
\label{wv}
\end{eqnarray}    
 
\noindent
Here $f(\theta)$ is the scattering amplitude, which we will discuss in
the Born approximation. For the impurity potential itself the
amplitude $f(\theta)$ weakly depends on the angle. At zero temperature
it determines the Drude conductivity $\sigma_D$, while the leading
temperature correction is proportional to $T^2$, as is usual for Fermi
systems. We now show that this is not the case for the potential
Eq.~(\ref{hf}). In fact, taking into account Eq.~(\ref{hf}) leads to
enhanced backscattering and thus to the conductivity correction that
is linear in temperature.

First, we discuss the Hartree potential Eq.~(\ref{hp}).  Far from the
scatterer the wave function of a particle can be found in the first
order of the perturbation theory as $\Psi = e^{i\vec k\cdot\vec r} +
\delta\Psi (\vec r)$, where the correction is given by \cite{llq}

\begin{equation}
\delta\Psi (\vec r) = i\int d\vec r_1 V_H(\vec r_1) 
e^{i\vec k\cdot\vec r_1}
\sqrt{\frac{2\pi}{k|\vec r -\vec r_1|}}e^{ik|\vec r -\vec r_1|}.
\label{dpsi}
\end{equation}

\noindent
Here $|\vec r -\vec r_1|\approx r - \vec r \cdot\vec r_1/r$, since we
are looking at large distances. Substituting the form of the potential
Eq.~(\ref{hp}) and introducing the Fourier transfer of the
electron-electron interaction $V_0$ we can rewrite Eq.~(\ref{dpsi}) as

\begin{equation}
\delta\Psi (\vec r) = -i\frac{\nu\lambda}{\sqrt{2\pi}}V_0(q)
\frac{e^{ikr}}{\sqrt{kr}}\int \frac{d\vec r_1}{r_1^2} \sin (2k_Fr_1)
e^{i\vec q\cdot\vec r_1},
\label{dp1}
\end{equation}

\noindent
where 

\[
\vec q = \vec k - k\vec r/r, \quad |q| =
2k\sin(\theta /2),
\]

\noindent
with $\theta$ being the angle of scattering. Comparing to
Eq.~(\ref{wv}) we find the scattering amplitude as a function of
$\theta$ (it also depends on the electron's energy $\epsilon =
k^2/2m$)

\begin{equation}
f(\theta) = -  \frac{\nu\lambda}{2\pi}V_0(q)
\int \frac{d\vec r}{r^2} \sin(2k_Fr)
e^{i\vec q\cdot\vec r}.
\label{ft1}
\end{equation}

\noindent
The integral can be evaluated exactly\cite{gri} and the result is
given by

\begin{equation}
f(\theta) = -  \frac{\nu\lambda}{2\pi}V_0(q)
\left\{
\matrix{
\frac{\pi}{2},  & |q|<2k_F  ;\cr
\arcsin\left(\frac{2k_F}{q}\right),  & |q|>2k_F .\cr
}
\right. 
\label{ft}
\end{equation}

\noindent
Let us examine this expression more closely. Since ${|q|\le 2k}$, the
scattering amplitude Eq.~(\ref{ft}) for small $k$ weakly depends on
the angle through the Fourier component of the interaction $V_0(q)$,
see background value of $f(\theta)$ on Fig.~\ref{1}.  However, we are
dealing with electronic excitations close to the Fermi surface, so in
fact $k$ is close to $k_F$, $|k-k_F|/k_F \ll 1$.  If $k>k_F$, then the
scattering amplitude Eq.~(\ref{ft}) has a non-trivial angular
dependence around $\theta = \pi$ shown on Fig.~\ref{1}.

According to Eq.~(\ref{ft}) such dependence is only possible in the
region $|q| > 2k_F$. This translates into the condition $|\theta -
\pi| < [2(k-k_F)/k_F]^{1/2}$, which determines the singular dependence
of the width of the feature in the scattering amplitude on the energy
of the scattered electron.  Finally, using the fact that $\arcsin(1-x)
= \pi/2 - \sqrt{2x}$, we find that the dependence of the height of the
feature in the scattering amplitude is also singular: $\delta
f(\theta) \simeq [(k-k_F)/k_F]^{1/2}$.

{
\narrowtext
\begin{figure}[ht]
\vspace{0.5 cm}
\epsfxsize=8 cm
\centerline{\epsfbox{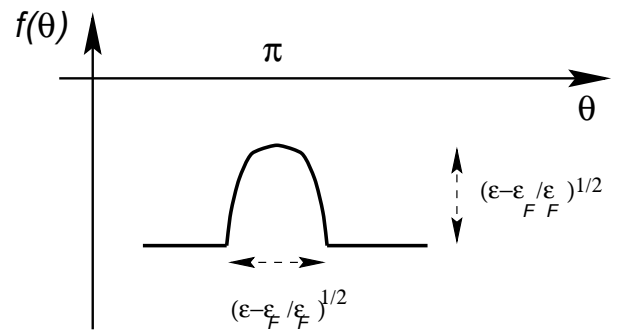}}
\vspace{0.5cm}
\caption{Scattering amplitude. The singularity for backscattering 
is due to interference of paths ``A'' and ``B'' on Fig. 1.} 
\label{1}
\end{figure}
}

The transport scattering rate $\tau^{-1}$ is determined by the
scattering cross-section and can be found with the help of the
amplitude Eq.~(\ref{ft}), as well as the constant amplitude $f_0$ of
the scattering by the impurity itself

\begin{equation}
\tau^{-1}(\epsilon) = \int \frac{d\theta}{2\pi} 
(1-\cos\theta) |f_0 + f(\theta)|^2.
\label{t1}
\end{equation}

\noindent
The leading energy dependence of $\tau^{-1}$ comes from the
interference term, which is proportional to $f(\theta)$. Then
integration around $\theta=\pi$ is dominated by the feature of
$f(\theta)$ resulting in a term of order $(\epsilon -
\epsilon_F)/\epsilon_F$. It is this term that gives rise to the linear
temperature dependence we are after. Since we are interested in this
leading correction only, in all other terms we can set $k\approx k_F$
and write the scattering rate as

\begin{equation}
\tau^{-1}(\epsilon) = \tau^{-1}_0 + \frac{\nu\lambda}{2}V_0(2k_F) 
\frac{\epsilon - \epsilon_F}{\epsilon_F} 
\eta(\epsilon - \epsilon_F)f_0.
\label{te}
\end{equation}

\noindent
Here $\eta(x)$ is the Heaviside step function and $\tau^{-1}_0$ is the
zero-temperature rate that determines the Drude conductivity (indeed,
the $\theta = \pi$ feature in $f(\theta)$ only exists for $k>k_F$ and
at $T=0$ there are no electrons with $k>k_F$).

To obtain the scattering time we have to integrate the
energy-dependent rate Eq.~(\ref{te}) with the derivative of the Fermi
distribution function $n_F(\epsilon)$

\begin{eqnarray*}
\tau = \int d\epsilon \tau(\epsilon)
\frac{\partial}{\partial\epsilon} n_F(\epsilon).
\end{eqnarray*}

\noindent
Then the second term in Eq.~(\ref{te}) leads to a linear correction to
the Drude conductivity, small as $T/\epsilon_F$.  However this is not
the only contribution to the temperature dependence. At finite
temperatures we also have to modify the Friedel oscillation
Eq.~(\ref{frie}) as follows:

\begin{eqnarray*}
\delta\rho(\vec r) = - 
\frac{\nu\lambda T^2}{2\pi v_F^2 \sinh^2\left(\frac{rT}{v_F}\right)} 
\sin(2k_F r).
\end{eqnarray*}

\noindent
Consequently, the scattering amplitude Eq.~(\ref{ft1}) becomes
temperature dependent

\begin{eqnarray}
&&f(\theta) = -  \frac{\nu\lambda}{2\pi}V_0(q)
\int \frac{d\vec r_2}{r_2^2} e^{-2r_2\frac{T}{v_F}}\sin(2k_Fr_2)
e^{i\vec q\cdot\vec r_2}
\nonumber\\
&&
\nonumber\\
&&\quad\quad\quad\quad\quad
= -  \frac{\nu\lambda}{2\pi}V_0(q)
\arcsin\left(\frac{4k_F}{p}\right);
\label{ftt}
\\
&&
\nonumber\\
&&
p=\sqrt{\left(\frac{2T}{v_F}\right)^2 + (q+2k_F)^2}+
\sqrt{\left(\frac{2T}{v_F}\right)^2 + (q-2k_F)^2}.
\nonumber
\end{eqnarray}

\noindent
Neglecting the small temperature dependent term in the denominator in
Eq.~(\ref{ftt}) brings us back to Eq.~(\ref{ft}). Keeping this term
leads to the same feature in $f(\theta)$ as the one on Fig.~\ref{1},
only now its width and magnitude are proportional to $\sqrt{T}$. The
resulting correction to the conductivity is therefore similar to the
one discussed above. Up to a numerical coefficient,

\begin{equation}
\frac{\delta\sigma}{\sigma_D} = - 2 \nu V_0(2k_F)\frac{T}{\epsilon_F}.
\label{ds1}
\end{equation}

The conductivity Eq.~(\ref{ds1}) is the same correction as the one
calculated in Ref.~\onlinecite{gdl}, see also Eq.~(\ref{introbal}), up
to a numerical factor.  It is also clear that Eq.~(\ref{ds1}) is not
the full story. We have forgotten about the Fock part of the potential
Eq.~(\ref{hf})! Substituting Eq.~(\ref{wv}) into Eq.~(\ref{dm}), we
find the perturbation of the density matrix [which appears in the Fock
potential Eq.~(\ref{fp})] $\delta n(\vec r_1, \vec r_2) \approx
\delta\rho[(\vec r_1+\vec r_2)/2]$. Then the argument can be repeated.
The only difference is that the leading temperature correction comes
from the Fourier component at $q=0$, rather than $q=2k_F$. What is
most important, the Fock potential enters with the opposite
sign. Therefore the expression for the conductivity Eq.~(\ref{ds1})
has to be corrected

\begin{equation}
\sigma = \sigma_D \left[1 - \nu \Big(2V_0(2k_F)-V_0(0)\Big)
\frac{T}{\epsilon_F}\right].
\label{ds}
\end{equation}

\noindent
The sign of the correction is thus not universal and depends on the
details of electron-electron scattering. If the weak interaction is
reasonably long-ranged, then $V_0(0)\gg V_0(2k_F)$, so that 
the correction in Eq.~(\ref{ds}) has the sign opposite to that in
Ref.~\onlinecite{gdl}.
    
{
\narrowtext
\begin{figure}[ht]
\vspace{0.5 cm}
\epsfxsize=7 cm
\centerline{\epsfbox{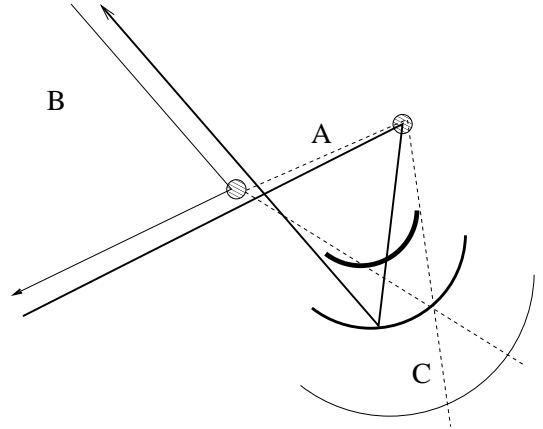}}
\vspace{0.5cm}
\caption{Scattering process with two impurities and the Friedel 
oscillation. Scattering to all angles is affected by interference.
The relevant Friedel oscillation is created by the self-intersecting 
path C.} 
\label{3}
\end{figure}
}

So far we have considered the effect of a single impurity. The
extension of the above arguments to the case of many impurities is
straightforward. In particular, one can consider a scattering process,
which involves two impurities and the Friedel oscillation shown on
Fig.~\ref{3}. It is clear that this process contributes to the
scattering amplitude at any angle, and not just for backscattering as
the single impurity process on Fig.~\ref{2} (which is typical for the
diffusive motion of electrons). Such processes were discussed in
detail (although using a slightly different language) in
Ref.~\onlinecite{aag}. Scattering by Friedel oscillations created by
multiple impurities results in a conductivity correction
(\ref{introdiff}) that is logarithmic in temperature and is typical
for 2D diffusive systems \cite{aar}.

Comparing the scattering processes on Figs.~\ref{2} and \ref{3}, one
can clearly see that conductivity corrections, which arise from these
processes are governed by the same physics: coherent scattering by the
Friedel oscillation, which means that the ballistic and diffusive
regions should be analyzed on the same footing.  In the next
subsection we present the results of such anlysis, postponing 
the actual calculations until Secs.~\ref{diagrams} and
\ref{eilenberger}.

\subsection{Results}
\label{results}

Let us first consider the case of a weak, short range interaction
potential. Then the interaction can be treated in the lowest order of
perturbation theory, so that the resulting correction is proportional
to the interaction constant:

\begin{eqnarray}
\delta\sigma_w =  \frac{e^2}{\pi\hbar}
\left[
 \gamma_1 \frac{T\tau}{\hbar} 
\left[ 1 -\frac{3}{8} w(T\tau)\right]
-\frac{\gamma_2}{4\pi}\ln\frac{E_F}{T}\right].
\label{sw}
\end{eqnarray}

\noindent
Similarly to Eq.~(\ref{ds}), it can be written as a sum of Hartree and
Fock contributions (similar expression for $\gamma_1$ in one-dimensional
systems was obtained in Ref.~\onlinecite{gla}):

\begin{eqnarray}
&&\gamma_1 = \nu \Big[V_0(0)-2V_0(2k_F)\Big], 
\nonumber\\
&&
\nonumber\\
&&
\gamma_2 = \nu \Big[V_0(0)-2\langle V_0(k)\rangle_{FS}\Big],
\label{u}
\end{eqnarray}

\noindent
where $\langle\dots\rangle_{FS}$ stands for the average over the Fermi
surface.  Here we kept the notation for the electron-electron
interaction adopted in the previous Section. Then the Hartree
correction is proportional to the Fourier component of $V_0(q)$ at
$q=2k_F$, while for the Fock correction $q=0$. The two corrections
have different sign as we discussed above.  The extra factor of $2$ in
the Hartree correction is due to electron spin degeneracy.

Note, that Eq.(\ref{sw}) is defined only up to a
temperature-independent constant which is determined by the
ultraviolet contribution.  We have chosen the argument of the
logarithm to be $E_F/T$ instead of the usual $1/T\tau$ to emphasize
that contrary to the naive expectations the logarithmic term persists
up to temperatures much larger than $1/\tau$, see also
Ref.~\onlinecite{rag}.
 
The different expressions for the Hartree terms in $\gamma_1$ and
$\gamma_2$ are related to the fact that the single impurity
scattering, see Fig.~\ref{2}, and multiple impurity case, see
Fig.~\ref{3}, allow for different possible scattering angles.

The dimensionless function $w(T\tau)$ describes the crossover between
ballistic and diffusive regimes.  In the ballistic limit $T\tau\gg 1$
it vanishes as

\begin{eqnarray*}
w(x\gg 1) \approx \frac{8}{3\pi x}\left[\ln(2x)-\frac{1}{4}
(\ln x -1)(6\ln 2 -1)\right].
\end{eqnarray*}

\noindent
In the opposite limit $T\tau\ll 1$ it approaches a constant value
(${\cal C}\approx 0.577\dots$ is the Euler's constant and
$\zeta^\prime(x)$ is a derivative of the Riemann zeta function),

\begin{eqnarray*}
w(x\ll 1) \approx && 1 + 
\frac{2\pi x}{9}
\left(\ln x - \ln 2 - {\cal C} + \frac{3}{4} 
+ 6\zeta^\prime(2)\right),
\end{eqnarray*}

\noindent
so that the linear correction does not completely vanish in the
diffusive limit, but competes with the logarithmic term and in
semiconductor structures with low Fermi energy it might be important
except for the lowest temperatures. The full function $w(x)$ is
plotted on Fig~\ref{plot:w}.
   
{
\narrowtext
\begin{figure}[ht]
\vspace{0.5 cm}
\epsfxsize=7 cm
\centerline{\epsfbox{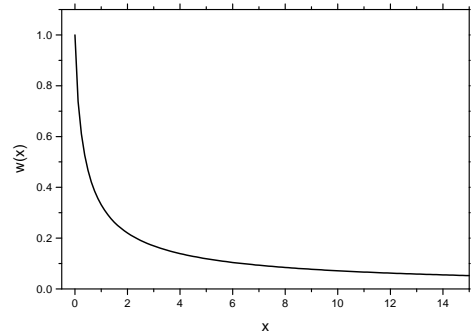}}
\vspace{0.5cm}
\caption{Dimensionless function $w(x)$, which is defined so that
$w(0)=1$.}
\label{plot:w}
\end{figure}
}

If the Coulomb interaction is considered, then the lowest order in
interaction is not sufficient since for the long range interaction
$\nu V_0(q \approx 0) \gg 1$.  Although the interaction itself is
still independent of the electron spin, summation of the perturbation
theory depends on the spin state of the two electrons involved. In the
first order correction Eq.~(\ref{sw}) all spin channels gave identical
contributions.  The total number of channels is $4$ and they can be
classified by the total spin of the two electrons: one state with the
total spin zero (``singlet'') and three states with the total spin $1$
(``triplet''; the three states differ by the value of the
$z$-component of the total spin). For long range interaction the
perturbation theory for the Hartree correction singlet and triplet
channels is different. It is known \cite{aar,fin} (see also
Section~\ref{diagrams}), that the singlet channel contribution should
be combined with the Fock correction as a renormalization of the
coupling constant. However, the final result is universal due to
dynamical screening: the singlet channel modification of the coupling
does not affect the result. What remains of the Hartree term is the
triplet channel contribution, which now depends on the corresponding
Fermi liquid constant $F_0^\sigma$. Thus, the total correction to the
conductivity can be written as a sum of the ``charge'' (which combines
Fock and singlet part of Hartree) and triplet contributions

\begin{mathletters}
\label{all}
\begin{equation}
\sigma = \sigma_D + \delta\sigma_T + \delta\sigma_C,
\end{equation}

\noindent
where the charge channel correction is given by

\begin{equation}
\delta\sigma_C =
\frac{e^2}{\pi\hbar} \frac{T\tau}{\hbar} 
\left[ 1 -\frac{3}{8}f(T\tau)\right]
 -\frac{e^2}{2\pi^2\hbar}\ln\frac{E_F}{T}
,
\label{fc}
\end{equation}

\noindent
and the triplet channel correction is

\begin{eqnarray}
\delta\sigma_T = &&
\frac{3F_0^\sigma}{(1+F_0^\sigma)}
\frac{e^2}{\pi\hbar}\frac{T\tau}{\hbar} 
\left[ 1 -\frac{3}{8}t(T\tau; F_0^\sigma)\right]
\nonumber\\
&&
\nonumber\\
&&
-3\left(1-\frac{1}{F_0^\sigma}
\ln(1+F_0^\sigma)\right)
\frac{e^2}{2\pi^2\hbar}\ln\frac{E_F}{T}.
\label{tc}
\end{eqnarray}
\label{s}
\end{mathletters}

\noindent
here the factor of three in the triplet channel correction
Eq.~(\ref{tc}) is due to the fact that all three components of the
triplet state contribute equally. We reiterate that the corrections
Eqs.(\ref{s}) are defined only up to a temperature independent 
(however not necessarily Fermi-liquid constant independent) term,
see also discussion after Eq.~(\ref{sw}).

{
\narrowtext
\begin{figure}[ht]
\epsfxsize=9 cm
\centerline{\epsfbox{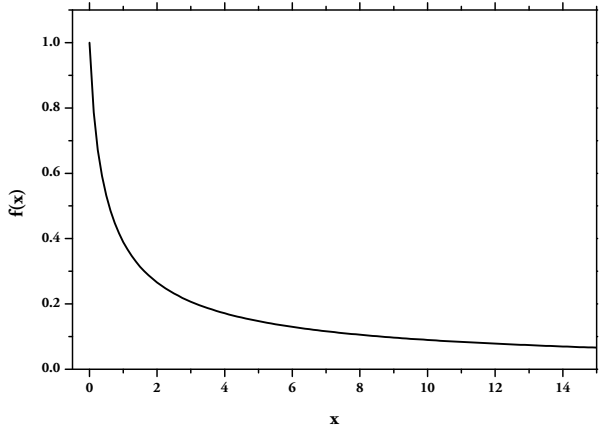}}
\vspace{0.2cm}
\caption{Dimensionless function $f(x)$, defined so that $f(0)=1$} 
\label{plot:f}
\end{figure}
}
  
We should warn the reader here, that we describe the interaction in
the triplet channel by one coupling constant $F_0^\sigma$.  For the
weak coupling limit, it corresponds to the approximation $V_0(2k_F)
\simeq \langle V_0(k) \rangle_{FS}$. This approximation overestimates
the triplet channel contribution to the ballistic case for $r_s =
\sqrt{2}e^2/(\kappa\hbar v_F) \ll 1$. However, in this limit
contribution itself is much smaller than the singlet one. For better
estimates in this regime one should use

\[
F_0^\sigma \to - \frac{1}{2}\frac{r_s}{r_s + \sqrt{2}}
\]

\noindent
in the first line of Eq.~(\ref{tc}) and

\begin{eqnarray*}
F_0^\sigma \to - \frac{1}{2\pi}
\frac{r_s}{ \sqrt{2-r_s^2} }
\ln \left(
\frac{\sqrt{2}+\sqrt{2-r_s^2} }{\sqrt{2} - \sqrt{2-r_s^2} }
\right), 
\ \ 
&& r_s^2 < 2; 
\\
F_0^\sigma \to - \frac{1}{\pi}\frac{r_s}{\sqrt{r_s^2-2}}
\arctan\sqrt{\frac{1}{2}r_s^2-1},
\ \ 
&& r_s^2 > 2
\end{eqnarray*}

\noindent
in the second line. For $r_s \gtrsim 1$ our replacament is well
justified even within weak coupling scheme.

Similar to Eq.~(\ref{sw}) the dimensionless functions $f(x)$ and
$t(x;F_0^\sigma)$ describe the cross-over between ballistic and
diffusive limits. They are plotted on Figs.~\ref{plot:f} and
\ref{plot:t} and full expressions are given by Eqs.~(\ref{f2}). The
universal function $f(x)$ has the following limits

\begin{mathletters}
\begin{equation}
f(x\gg 1) \approx -\frac{1}{3\pi x}\Big(2(\ln x - 1)\ln 2 -
\frac{7}{2}\ln (2x)\Big);
\end{equation}
\begin{eqnarray}
&&f(x\ll 1) \approx 1 - \gamma_1 x 
+ \frac{\pi}{6} x \ln x;
\\
&&
\nonumber\\
&&
\gamma_1 = -\frac{\zeta^\prime(2)}{\pi}+\frac{\pi}{6}
\left({\cal C} +\frac{1}{3}\ln 2 \right)
\approx 0.7216;
\nonumber
\end{eqnarray}
\label{fl}
\end{mathletters}

{
\narrowtext
\begin{figure}[ht]
\epsfxsize=9 cm
\centerline{\epsfbox{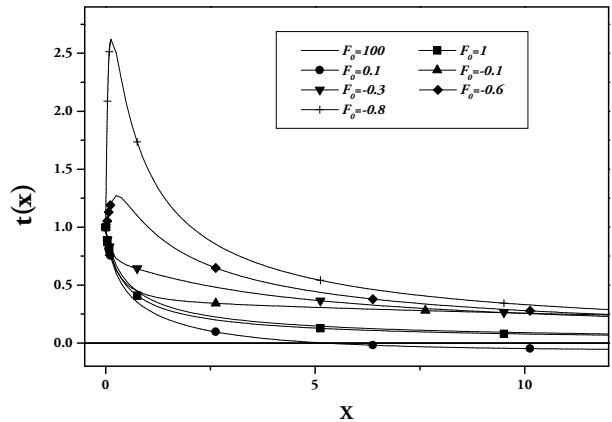}}
\vspace{0.2cm}
\caption{Dimensionless function $t(x,F_0^\sigma)$ defined so that
$t(0,F_0^\sigma)=1$.}
\label{plot:t}
\end{figure}
}

The function $t(x;F_0^\sigma)$ depends on the coupling constant and
therefore its asymptotic form also depends on $F_0^\sigma$. For very
small $x\ll 1+F_0^\sigma$ the asymptotic form is

\begin{eqnarray}
&&t(x\ll 1+F_0^\sigma) \approx 1 -\gamma_2 x 
+\frac{\pi}{18} x \ln x \left(3+\frac{1}{1+F_0^\sigma}\right);
\nonumber\\
&&
\nonumber\\
&& \quad
\gamma_2 = -\frac{\zeta^\prime(2)}{3\pi}
\left(3+\frac{1}{1+F_0^\sigma}\right)
-\frac{\pi\gamma_3}{9(1+F_0^\sigma)}
\label{tsx}\\
&&
\nonumber\\
&& \quad\quad\quad
+\frac{\pi}{18}
\left[{\cal C}\left(3+\frac{1}{1+F_0^\sigma}\right )
+\ln 2 \left(1+\frac{3}{1+F_0^\sigma}\right)\right];
\nonumber\\
&&
\nonumber\\
&& \quad
\gamma_3=1-\frac{5F_0^\sigma -3}{1+F_0^\sigma} 
- \left(\frac{5}{2} - 2F_0^\sigma 
\right)\frac{\ln (1+F_0^\sigma)}{F_0^\sigma}.
\nonumber
\end{eqnarray}
Notice that at $T\tau \to 0$, Eqs. (\ref{all}) reproduce the known
result (\ref{introdiff}).
Let us point out that for numerical reasons contributions
of scaling functions $w,\ f,\ t$  change the result only by few
percents and they can be neglected for all the practical purposes.

Notice that while the charge channel correction Eq.~(\ref{fc}) is
universal, the triplet channel correction Eq.~(\ref{tc}) is
proportional to $F_0^\sigma$, which might be negative. That leads to
the conclusion, that the overall sign of the total correction
Eqs.~(\ref{s}) depends on value of
$F_0^\sigma$: it can be either positive or negative, see 
Fig.~\ref{sigma}.

{
\narrowtext
\begin{figure}[ht]
\epsfxsize=5 cm
{\epsfbox{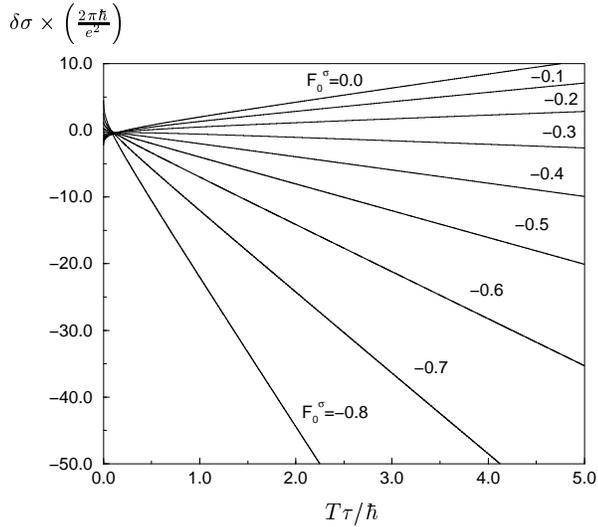}}
\vspace{0.2cm}
\caption{Total interaction correction to conductivity. The divergence 
at $T\tau/\hbar\protect\rightarrow 0$ is due to the usual logarithmic 
correction
\protect\cite{aar}. 
Curve $F_0^\sigma=0$ corresponds to the universal behavior of
completely spin polarized electron gas.
The correction is defined up to a temperature
independent part, see 
Eq.~(\protect\ref{ultra}) and discussion after Eq.~(\protect\ref{sw}).} 
\label{sigma}
\end{figure}
}

Combining together all of the above results we plot the total
correction to the conductivity on Fig.~\ref{sigma} for different
values of $F_0^\sigma$. The divergence at low temperature is due to
the usual logarithmic correction \cite{aar}. Although the exact value
of $F_0^\sigma$ can not be calculated theoretically (in particular,
its relation to the conventional measure of the interaction strength,
$r_s$, is unknown for $r_s>1$), in principle it can be found from a 
measurement of the Pauli spin susceptibility

\begin{equation}
\chi = \frac{\nu}{1+F_0^\sigma},
\label{pauli}
\end{equation}

\noindent
where the density of states $\nu$ should be obtained from a
measurement of the specific heat (at $\tau^{-1}\ll T \ll E_F$).  The
constant $F_0^\sigma$ is the only parameter in our theory which
describes all the data, including the Hall coefficient and the
magneto-resistance in the parallel field.  The theory for interaction
corrections in the magnetic field will be addressed in the forthcoming
paper \cite{pre}.

The correction in Fig.~\ref{sigma} is almost always monotonous, except
for a narrow region $-0.45 < F_0^\sigma < -0.25$. A typical curve in
this region is shown in Fig.~\ref{nonmon}. Note, however, that the
overall magnitude of the correction in the range of $T\tau$ in
Fig.~\ref{sigma} is so small that it can hardly be observed.
 
{
\narrowtext
\begin{figure}[ht]
\epsfxsize=5 cm
{\epsfbox{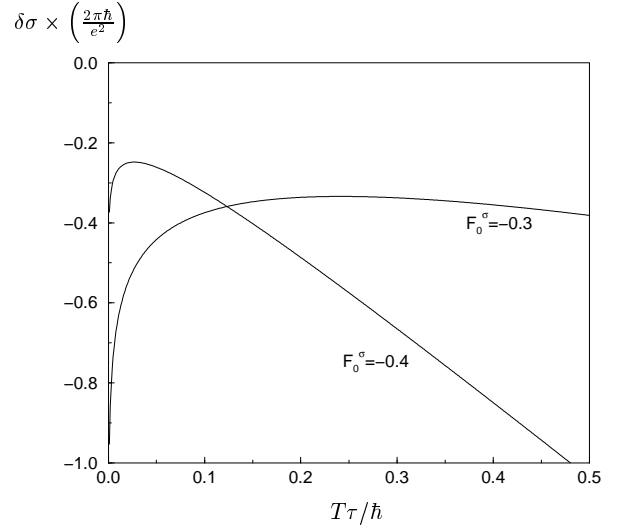}}
\vspace{0.2cm}
\caption{The non-monotonous correction to conductivity. Note the difference
in the overall scale relative to the previous figure.} 
\label{nonmon}
\end{figure}
}

When the interaction becomes so strong that the system approaches the
Stoner instability, $F_0^\sigma$ ceases to be a constant and becomes
momentum-dependent. Thus the result Eq.~(\ref{s}) is no longer
valid. Although the simple condition $\delta\sigma_T < \sigma_D$
suggests that this happens at $T\approx (1+F_0^\sigma)E_F$, the more
detailed analysis (see Section~\ref{triplet}) shows that it happens
much earlier. In fact, the approximation of the constant $F_0^\sigma$
is valid in the parameter region defined by the inequality

\begin{equation}
\frac{T}{E_F} < (1+F_0^\sigma)^2,
\label{lim}
\end{equation}

\noindent
see Section~\ref{triplet} for the origin of this inequality. We 
were not able to make a reliable calculation of 
$\delta \sigma(T)$ at higher temperatures.

\section{Perturbation theory}
\label{diagrams}

In this section we show how the announced results Eq.~(\ref{s}) can be
obtained with the help of the traditional perturbation theory.  We try
to explain the most important points of the calculation in detail.
The comprehensive review of the diagrammatic technique for disordered
systems can be found in Ref.~\onlinecite{aar}. We start by a brief
discussion of the case of a weak, short-range interaction
potential. Although this case is artificial and is unrelated to any
experiment, it is governed by the same physics as the general problem,
and it is simple enough to allow a transparent presentation.  To
generalize to stronger coupling, we need to recall the basic ideas of
the Landau Fermi liquid theory and to identify the soft modes in the
system. Then we present the calculation leading to Eq.~(\ref{s}).
Finally, to establish the relation of our results to existing
literature, we briefly discuss scattering on a single impurity (this
discussion is completely analogous to the one in
Section~\ref{qualitative} but uses the language of diagrams).

\subsection{Hartree-Fock considerations.}
\label{h-f}


The static conductivity of a system of electrons is given by the Kubo
formula

\begin{eqnarray}
&&\sigma_{\alpha\beta} = 
\label{kubo}\\
&&
\nonumber\\
&&
-
\lim_{\omega \to 0} 
{\rm Re}
\left[
\frac{1}{\Omega_n} 
\int\limits_{0}^{1/T} d\tau \langle {\bf T_\tau} \hat j_\alpha(\tau)
\hat j_\beta(0) \rangle e^{i\Omega_n\tau}
\right]_{i\Omega_n \to \omega},
\nonumber
\end{eqnarray}

\noindent
where $\hat j_\alpha(\tau)$ is the operator of the electric current at
imaginary time $\tau$ and the analytic continuation of the function
defined at Matzubara frequencies $\Omega_n = 2\pi Tn$ to function
analytic at ${\rm Im} \omega >0$ is performed.

{
\narrowtext
\begin{figure}[ht]
\vspace{0.5 cm}
\epsfxsize=8 cm
\centerline{\epsfbox{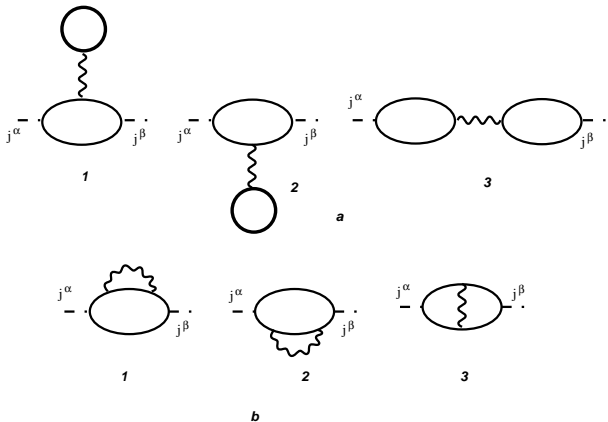}}
\vspace{0.5cm}
\caption{Interaction correction to conductivity in the lowest order of
perturbation theory. Here solid lines correspond to Matsubara Green's
functions $-G(i\epsilon_n; \vec r_1, \vec r_2)$ and the wavy line
represents the interaction potential, $-V(\vec r_1- \vec r_2)$.}
\label{sr}
\end{figure}
}   

Assuming that electrons interact by means of a weak, short-range
interaction (range shorter than $v_F min(\tau, 1/T )$, $V(r)$ it is
sufficient to consider the lowest order of the perturbation theory.
The perturbation theory can be conveniently expressed in terms of
Feinman diagrams. The lowest order diagrams for the interaction
correction to the conductivity are shown on Fig.~\ref{sr}. The Hartree
term corresponds to the diagrams ``a'', while the Fock contribution
corresponds to diagrams ``b''. Evaluation of the correction consists
of two main steps: (i) analytic continuation to real time, and (ii)
disorder averaging. While these two steps can be performed in any
order without affecting the result, it is more convenient (for
technical reasons) to start with step (i).

Although analytic continuation in Eq.~(\ref{kubo}) is now a textbook
task, we include a brief discussion of the standard procedure in the
Appendix to make the paper self-contained. After the continuation any
physical quantity is expressed in terms of exact (i.e. not averaged
over disorder) retarded and advanced Green's functions of the
electronic system, which are defined as

\begin{eqnarray}
G^{R(A)}_{12}(\epsilon)=\sum\limits_j
{{\Psi^*_j(\vec{r}_1)\Psi_j(\vec{r}_2)}
\over{\epsilon-\epsilon_j\pm\imath 0}},
\label{gra}
\end{eqnarray}

\noindent
where $j$ labels the exact eigenstates of the system and $\epsilon_j$
are the exact eigenvalues, counted from the Fermi energy:

\[
\left(\frac{-\nabla^2}{2m} + U(\vec r) \right) \Psi_j(\vec{r}) =
(\epsilon_j + \epsilon_F) \Psi_j(\vec{r}).
\]

\noindent
Here $U(\vec r)$ is the disorder potential.

The resulting expression for the correction to the symmetric part of
the conductivity (the Hall conductivity will be discussed in a
separate publication \cite{pre}) can be written as \cite{aag}

\begin{eqnarray}
\delta&&\sigma_{\alpha\beta} = \int\limits_{-\infty}^{\infty} 
\frac{d\Omega}{8\pi^2}
\left[\frac{\partial}{\partial\Omega}
\left(\Omega\coth\frac{\Omega}{2T}\right)\right] 
\int d^2 r_3 d^2 r_4
\label{cc}
\\
&&
\nonumber\\
&&
\times
{\rm Im} \Bigg\{ V(\vec r_3- \vec r_4)
\Big ( B^{\alpha\beta}_F (\Omega; \vec r_3, \vec r_4) 
-2 B^{\alpha\beta}_H (\Omega; \vec r_3, \vec r_4) 
\nonumber\\
&&
\nonumber\\
&& \hspace{5 cm}
+ \{\alpha\leftrightarrow\beta\}
\Big)\Bigg\},
\nonumber
\end{eqnarray}

\noindent
where the extra factor of $2$ in the Hartree term is due to the
summation over electron spin. Here we denoted products of four 
Green's functions as $B_{F(H)}$. For the Fock term we have

\begin{mathletters}
\begin{eqnarray}
B&&^{\alpha\beta}_F(\Omega; \vec r_3, \vec r_4) = 
\int \frac{d^2 r_1 d^2 r_5}{{\cal V}} 
\nonumber\\
&&
\nonumber\\
&&
\times\Big\{
\hat{J}_1^\alpha G^R_{15}(\epsilon) 
\hat{J}_5^\beta G^A_{53}(\epsilon)
G^R_{34}(\epsilon-\Omega) G^A_{41}(\epsilon)
\label{f11}
\\
&&
\nonumber\\
&&\quad
+ \hat{J}_1^\alpha G^A_{15}(\epsilon) 
\hat{J}_5^\beta G^R_{53}(\epsilon)
G^R_{34}(\epsilon-\Omega) G^R_{41}(\epsilon)
\label{f12}
\\
&&
\nonumber\\
&&\quad
+ 2 \hat{J}_1^\alpha G^R_{13}(\epsilon) 
G^R_{35}(\epsilon-\Omega) \hat{J}_5^\beta 
G^R_{54}(\epsilon-\Omega) G^A_{41}(\epsilon)
\label{f32}
\\
&&
\nonumber\\
&&\quad
- \hat{J}_1^\alpha G^A_{15}(\epsilon) \hat{J}_5^\beta 
G^A_{53}(\epsilon)
G^R_{34}(\epsilon-\Omega) G^A_{41}(\epsilon)
\label{f13}
\\
&&
\nonumber\\
&&\quad
- \hat{J}_1^\alpha G^A_{13}(\epsilon) G^R_{35}(\epsilon-\Omega)
\hat{J}_5^\beta G^R_{54}(\epsilon-\Omega) 
G^A_{41}(\epsilon)
\Big\},
\label{f31}
\end{eqnarray}
\label{fck}
\end{mathletters}
 
\noindent
where ${\cal V}$ is the area of the system.  Equations~(\ref{f32}) and
(\ref{f31}) come from the diagram ``b3'' on Fig.~\ref{sr} and the rest
of Eq.~(\ref{fck}) correspond to diagrams ``b1'' and ``b2''. For the
Hartree term the expression is similar,

\begin{mathletters}
\begin{eqnarray}
B&&^{\alpha\beta}_H (\Omega; \vec r_3, \vec r_4) = 
\int \frac{d^2 r_1 d^2 r_5}{{\cal V}} 
\nonumber\\
&&
\nonumber\\
&&
\times\Big\{
\hat{J}_1^\alpha G^R_{15}(\epsilon) \hat{J}_5^\beta 
G^A_{53}(\epsilon)
G^R_{44}(\epsilon-\Omega) G^A_{31}(\epsilon)
\label{h11}
\\
&&
\nonumber\\
&&\quad
+ \hat{J}_1^\alpha G^A_{15}(\epsilon) \hat{J}_5^\beta 
G^R_{53}(\epsilon)
G^R_{44}(\epsilon-\Omega) G^R_{31}(\epsilon)
\label{h12}
\\
&&
\nonumber\\
&&\quad
+ 2 \hat{J}_1^\alpha G^R_{13}(\epsilon) G^R_{45}(\epsilon-\Omega)
\hat{J}_5^\beta G^R_{54}(\epsilon-\Omega) G^A_{31}(\epsilon)
\label{h32}\\
&&
\nonumber\\
&&\quad
- \hat{J}_1^\alpha G^A_{15}(\epsilon) \hat{J}_5^\beta 
G^A_{53}(\epsilon)
G^R_{44}(\epsilon-\Omega) G^A_{31}(\epsilon)
\label{h13}
\\
&&
\nonumber\\
&&\quad
- \hat{J}_1^\alpha G^A_{13}(\epsilon) G^R_{45}(\epsilon-\Omega)
\hat{J}_5^\beta G^R_{54}(\epsilon-\Omega) G^A_{31}(\epsilon)
\Big\}.
\label{h31}
\end{eqnarray}
\label{hrt}
\end{mathletters}
 
\noindent
Again, Eqs.~(\ref{h32}) and (\ref{h31}) correspond to the diagram
``a3'' in Fig.~\ref{sr}. The current operator is defined as

\begin{eqnarray}
f_1(\vec r) \hat{\vec J} f_2(\vec r) && = \frac{ie}{2m}
\left[ \left( \vec\nabla f_1 \right) f_2 - 
\left( f_1\vec\nabla f_2 \right) \right] 
\nonumber\\
&&
\nonumber\\
&&
- \frac{e\vec A(\vec r)}{m} f_1(\vec r) f_2(\vec r).
\label{jop-def}
\end{eqnarray}

\noindent
In the above expressions terms corresponding to diagrams ``b3'' and
``a3'' on Fig.~\ref{sr} allow for at least one of the spatial
integrations to be performed with the help of the identity

\begin{equation}
\int d\vec r_5 G^R_{35}(\epsilon)\hat{J}_5^\beta G^R_{54}(\epsilon)
=-ie(\vec r_3 - \vec r_4)^\beta G^R_{34}(\epsilon).
\label{id}
\end{equation}

\noindent
Now it is clear that Hartree terms Eqs.~(\ref{h32}) and (\ref{h31})
vanish identically, since there the identity (\ref{id}) should be
applied with coordinates $\vec r_3$ and $\vec r_4$ being equal to each
other. In the Fock terms Eqs.~(\ref{f31}) and (\ref{f32}) one needs to
further multiply the result of Eq.~(\ref{id}) by the interaction
potential $V(\vec r_3 - \vec r_4)$. In the case of the short range
interaction potential this also gives vanishing contribution.  Thus we
conclude, that the diagram ``a3'' on Fig.~\ref{sr} does not contribute
for any form of the interaction, while the diagram ``b3'' vanishes for
the short-range interaction.

The same identity can also be applied to terms Eqs.~(\ref{f13}) and
(\ref{h13}), which also vanish by the same reason. Thus the task of
averaging over disorder is now simplified because we only need to
average two Fock terms Eqs.~(\ref{f11}) and (\ref{f12}) and two
Hartree terms Eqs.~(\ref{h11}) and (\ref{h12}). These expressions
contain only Green's functions of non-interacting electrons and can be
averaged using the standard diagrammatic technique of the theory of
disordered systems (see Ref.~\onlinecite{aar} for review). The
diagrams for averaged quantities can be constructed using the four
``building blocks'' (we use the momentum representation since
translational invariance is restored after averaging):

\noindent
(1) the average electronic Green's function (denoted as a solid line;
there should be no confusion with the previous use of the solid line
for exact Green's functions before averaging), which in momentum space
can be written as

\begin{eqnarray}
\langle G^{R(A)}\rangle (k,\epsilon) = 
\frac{1}{\epsilon - \xi_k \pm \frac{i}{2\tau}};
\label{green}
\end{eqnarray}

\noindent
(2) the disorder potential, which is assumed to be Gaussian with the
correlator 

\begin{eqnarray*}
\langle U(\vec r_1)U(\vec r_2)\rangle = 
\frac{1}{2\pi\nu\tau} \delta(\vec r_1 - \vec r_2).
\end{eqnarray*}

\noindent
In the diagrams this correlator is represented by the dotted line;

{
\narrowtext
\begin{figure}[ht]
\vspace{0.2 cm}
\epsfxsize=8.5 cm
\centerline{\epsfbox{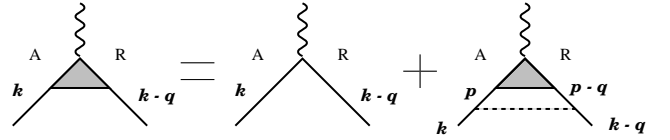}}
\vspace{0.2cm}
\caption{Dressed interaction vertex.} 
\label{8}
\end{figure}
}

\noindent
(3) the dressed interaction vertex $\Gamma$ ($q$ and $\Omega$ are
momentum and frequency of the interaction propagator), which
represents a geometric series in disorder potential shown on
Fig.~\ref{8};

\begin{mathletters}
\begin{equation}
\Gamma ( \vec q; \Omega)= 1+ \frac{1/\tau}{S- \frac{1}{\tau}}
\label{gamma}
\end{equation}

\noindent
where we denote

\begin{eqnarray}
S=\sqrt{\left(i\Omega + \frac{1}{\tau}\right)^2 + v_F^2 q^2},
\label{sqrt}
\end{eqnarray}
\label{gsd}
\end{mathletters}

\noindent
(4) the averaged product of a retarded and an advanced Green's
functions (sometimes referred to as the diffuson), where we have
summed up a geometric series shown on Fig.~\ref{9}
 
{
\narrowtext
\begin{figure}[ht]
\vspace{0.2 cm}
\epsfxsize=8 cm
\centerline{\epsfbox{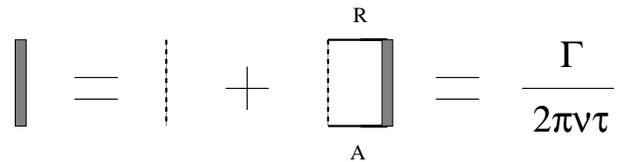}}
\vspace{0.2cm}
\caption{Diffuson - geometric series of impurity lines} 
\label{9}
\end{figure}
}

Using these building blocks we can average the products of Green's
functions as shown on Fig.~\ref{hfav}. It is convenient to write the
averaged $B_{F(H)}$ in the momentum representation. The product $B_F$,
which appears in the Fock term, can be viewed as a function of
coordinates of the two interaction vertices and can be transformed to
the momentum space as

\begin{equation}
\langle B_F (\Omega; \vec r_1,\vec r_2) \rangle 
 = \int \frac{d^2 q}{(2\pi)^2}
e^{i\vec q (\vec r_3 - \vec r_4)} 
\langle B_F (q, \Omega) \rangle.
\label{noname1}
\end{equation}

\noindent
Using the explicit expressions Eqs.~(\ref{green}) and (\ref{gsd}) we
can write the analytic form of the averaged $B_F$

\begin{eqnarray}
\frac{1}{\sigma_D}B_F (q, \Omega) = 
\frac{(\Gamma^2-1)\tau}{S}
\label{bf}
+\frac{\Gamma(\Gamma+1)}{v_F^2 q^2}
\left(\frac{i\Omega + \frac{1}{\tau}}{S} - 1 \right)^2.
\nonumber
\end{eqnarray}

In the absence of magnetic field,
$B^{\alpha\beta}_{F(H)}=\delta^{\alpha\beta} B_{F(H)}$, which is why
we did not include the Greek indices in Eq.~(\ref{bf}).

{
\narrowtext
\begin{figure}[ht]
\vspace{0.2 cm}
\epsfxsize=8.5 cm
\centerline{\epsfbox{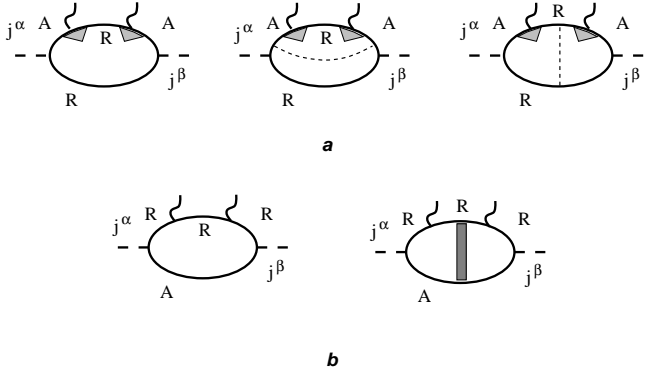}}
\vspace{0.2cm}
\caption{Averaged product of four Green's functions. 
The wavy lines indicate $\delta(\vec p_{in}- \vec p_{out}+ \vec q)$
for the Fock contribution $B_F(q)$
and $2\pi\delta(\vec p_{in}- \vec p_{out}+ \vec q)
\delta[\widehat{\vec{n}\vec p_{in}}]$
for the Hartree contribution $B_H(q, n_1,n_2)$.} 
\label{hfav}
\end{figure}
}

The Hartree contribution is considered analogously.
One can write

\beqa
\langle B_H(\Omega; \vec r_1, \vec r_2) \rangle =
\int \frac{d \theta_1}{(2\pi)}&&
\frac{d \theta_2}{(2\pi)}
\frac{d^2 q}{(2\pi)^2} 
e^{ik_F(\vec n_1-\vec n_2)(\vec r_1-\vec r_2)}
\nonumber\\
&&
\nonumber\\
&&
\times
B_H(\Omega; \vec n_1, \vec n_2, \vec q),
\label{Ht1}
\eeqa

\noindent
where $\vec n_i = (\cos \theta_i, \sin \theta_i)$ indictates
the direction of the momentum.
Then, disorder averaging of $B_H(\Omega; \vec n_1, \vec n_2, \vec q)$
is performed with the help of the same diagrams (see Fig.~\ref{hfav})
but the expression for the vertices changed as indicated in the figure
caption.

Accordingly, the expression for the dressed vertex 
(\ref{gamma}), see also Fig.~\ref{8}, is changed to

\beqa
\Gamma_H&&(\vec n,\vec n_k; \vec q; \Omega) = 
2\pi \delta(\widehat{\vec n \vec n_k})
+  \frac{1}{S_H}
\frac{S}{ S{\tau}-1},
\label{gammah}
\\
&&
\nonumber\\
&&
S_H(\vec n, \vec q; \Omega)={i\Omega -iv_F \vec q \vec n + 1/\tau},
\nonumber
\eeqa

\noindent
where $\vec n_k$ corresponds to the direction of the momentum $k$ on
Fig.~\ref{8}. The final expression for $B_H$ is similar to
Eq.~(\ref{bf})

\begin{eqnarray}
&& \frac{1}{\sigma_D} B_H ( \Omega; 
{\vec n}_1, {\vec n}_2,
{\vec q}) = -
\frac{-
2\pi\delta\left(\widehat{\vec n_1\vec n_2}\right)\tau}{S}
\nonumber\\
&&
\nonumber\\
&&
\quad\quad  +
\frac{\tau}{S}
\int \frac{d \theta_3}{2\pi}
\Gamma_H(\vec n_1, \vec n_3)\Gamma_H(\vec n_2, n_3)
\label{bh}
\\
&&
\nonumber\\
&&\quad\quad  +
\int \frac{d \theta_3}{2\pi}
\int \frac{d \theta_4}{2\pi}
\left(\vec n_3\vec n_4\right)
\frac{\Gamma_H(\vec n_1,n_3)\Gamma_H(\vec n_2,n_4)}
{S_H(\vec n_3)S_H(\vec n_4)}
\nonumber\\
&&
\nonumber\\
&&\quad\quad 
+\left(\vec n_1\vec n_2\right) 
\frac{\Gamma}{S_H(\vec n_1)S_H(\vec n_2)},
\nonumber
\end{eqnarray}

\noindent
and we suppressed the arguments $q,\Omega$ in the right-hand-side
of the equation.

We note in passing, that by construction of Eqs.~(\ref{f11}),
(\ref{f12}), (\ref{h11}) and (\ref{h12}) that

\[
B_F(\Omega; \vec r, \vec r)=B_H(\Omega; \vec r, \vec r),
\]

\noindent
and,  therefore, according to Eqs.~(\ref{noname1}) and (\ref{Ht1})
the relation

\[
B_F(\Omega;  \vec q) =
\int \frac{d \theta_1}{2\pi}\frac{d \theta_2}{2\pi}
B_H(\Omega; \vec n_1, \vec n_2, \vec q)
\]

\noindent
must hold [this can be easily verified using explicit expresions
(\ref{bf}) and (\ref{bh})].

We are now prepared to calculate the temperature dependence
of the conductivity from Eq.~(\ref{cc}). We substitute Eqs.~(\ref{bf})
and (\ref{bh}) into Eq.~(\ref{cc}). As we will see, the
main contribution to the temperature dependence is provided by
wave-vectors $q_{T} \simeq max(T,(T/\tau)^{1/2})/v_F$. On the other hand the
potential $V(\vec r)$ has a range much shorter than $1/q_T$.
This enables us to use the following approximations

\begin{eqnarray*}
\int \frac{d^2 r_3 d^2 r_4}{\cal V} 
V(\vec r_3-\vec r_4)e^{i\vec q(\vec r_3-\vec r_4)}
\approx V(0);
\end{eqnarray*}
\begin{eqnarray*}
&&
\int \frac{d^2 r_3 d^2 r_4}{\cal V} 
V(\vec r_3-\vec r_4)e^{i k_F (\vec n_1 - \vec n_2) (\vec r_3-\vec r_4)}
\\
&&
\\
&&
\hspace*{4cm}\approx 
V\left(2k_F\sin \frac{\widehat{\vec n_1\vec n_2}}{2}
\right),
\end{eqnarray*}
where $V(k)$ in the right-hand-side of the above equations denotes
the Fourier transform of the interaction potential.

Altogether, we now write the conductivity correction as

\begin{eqnarray}
&&\delta\sigma = \int\limits_{-\infty}^{\infty} 
\frac{d\Omega}{8\pi^2} 
\frac{\partial}{\partial\Omega}
\left(\Omega\coth\frac{\Omega}{2T}\right)
\int\frac{d^2 q}{(2\pi)^2}
\label{cc5}\\
&&
\nonumber\\
&&
\times {\rm Im} \Bigg\{
V_0(0)  B_F (q,\Omega) 
\nonumber\\
&&
\nonumber\\
&&
- 2 \int\frac{d\theta_1}{2\pi} \int\frac{d\theta_2}{2\pi}
V_0\left(2k_F\sin\frac{\widehat{\vec n_1 \vec n_2}}{2}\right) 
B_H (\Omega; \vec n_1, \vec n_2, \vec q)\Bigg\}.
\nonumber
\end{eqnarray}

\noindent
Evaluating this integral (where we only keep the temperature 
dependent part, see Section~\ref{single} for details) 
one arrives to the same result
Eq.~(\ref{sw}), but with the coefficient in the form Eq.~(\ref{u}), in
agreement with the discussion of Section~\ref{qualitative}.

Let us now turn to the case of the Coulomb potential, where the scheme
of the calculation (as described so far) breaks down. In the Fock term
we have $V(0)$, which diverges for the Coulomb interaction ($V(q)\sim
1/q$). To obtain meaningful results one needs to take into account the
effect of dynamical screening. The Hartree term seems to work better
since using just the static screening makes the result
finite. However, this is wrong also, since in this case diagrams with
extra interaction lines do not contain any smallness (see e.g.
Fig.~\ref{cor}; there the correction is $\sim V(2k_F) V(0)$). Thus one
can not justify the perturbation theory in the interaction
potential. The way out of this problem is the standard theory of 
Landau Fermi liquid, which we briefly discuss in the following subsection.

{
\narrowtext
\begin{figure}[ht]
\vspace{0.5 cm}
\epsfxsize=7 cm
\centerline{\epsfbox{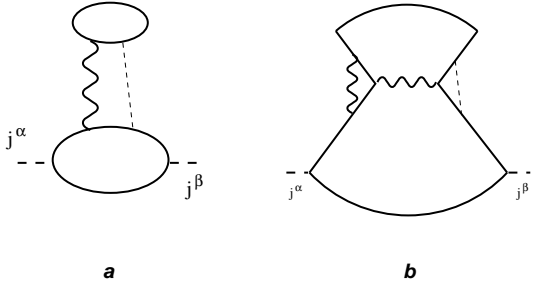}}
\vspace{0.5cm}
\caption{(a) Single impurity contribution to the Hartree term, see
Section~\ref{single} for a detailed discussion 
(b) Second order correction to the Hartree term (a).} 
\label{cor}
\end{figure}
}

\subsection{Soft modes}
\label{soft}

As we already discussed, the main contribution to the temperature
dependence of physical quantities comes from the processes
characterized by spatial scales much larger than the Fermi wave-length
$\lambda_F$.  Therefore, there is a scale separation in the problem;
all the Fermi liquid parameters\cite{Fermiliquid} $F_i$ are established
at small distances of the order of $\lambda_F$, and are not affected
by  disorder if the relation
 $\epsilon_F\tau \gtrsim 1$ holds. On the other hand, all the
temperature and disorder dependence is determined by infrared behavior
of the system where $F_i$ can be considered as fixed.

Therefore, our first step is to identify the terms in the interaction
Hamiltonian, which may produce the biggest contributions at
temperatures much smaller than the Fermi energy. This procedure
contains nothing new in comparison with the standard identification of
singlet, triplet and Cooper channels, see Ref.~\onlinecite{AGD}, and
we present here the main steps to make the paper self-contained.

The original interaction Hamiltonian has the form

\begin{eqnarray*}
\hat{H}_{int} = 
\sum_{\vec q, \vec p_i}
\frac{V(q)}{2}\psi^\dagger_{\sigma_1}(\vec p_1)
\psi^\dagger_{\sigma_2}(\vec p_2)
\psi_{\sigma_2}(\vec p_2+\vec q)\psi_{\sigma_1}(\vec p_1-\vec q),
\end{eqnarray*}

\noindent
and we imply summation over repeated spin indices.  Soft modes of the
system correspond to the situation when two of the fermionic operators
have momenta close to each other.  The difference of the momenta $q*$
defines the scale $1/q* \gg \lambda_F$, which is the smallest length
scale allowed in the theory. Therefore, we explicitly separate the
Hamiltonian 
into a part that contains all the soft modes (first three terms) and a
correction $\delta\hat{H}$, which does not contain such pairs of
fermionic operators:

\beq
\label{101}
\hat{H}_{int} = 
\hat{H}_\rho + \hat{H}_\sigma + \hat{H}_{pp} + \delta\hat{H}. 
\label{hamiltonian}
\eeq

\noindent

The explicit expressions for the entries of the Hamiltonian 
(\ref{hamiltonian}) are the
following. The interaction in the singlet channel (charge dynamics) is
described by

\beqa
&&\hat{H}_\rho =
\frac{1}{2}\sum_{|\vec q| < q*, \vec p_i}
\left[V(q)
+\frac{ F^\rho\left(\widehat{\vec n_1 \vec n_2}\right)}{ \nu }
\right] \label{Hrho}\label{102} 
\\
&&
\nonumber\\
&&
\times 
\left[\psi^\dagger_{\sigma_1}\left(\vec p_1\right)
\psi_{\sigma_1}\left(\vec p_1-{\vec q}\right)
\right]
\left[\psi^\dagger_{\sigma_2}\left(\vec p_2\right)
\psi_{\sigma_2}\left(\vec p_2+{\vec q}\right)
\right],
\nonumber
\eeqa

\noindent
where $\vec n_i = \vec p_i/|p_i|$, the dimensionless parameter
$F^\rho\left(\widehat{\vec n_1 \vec n_2}\right)$ is related to the
original interaction potential $V(q)$ by

\beq
\label{103}
F^\rho\left(\theta\right) = - \frac{\nu}{2}
V\left(2k_F\sin\frac{\theta}{2} \right),
\eeq

\noindent
and $\nu$ is the thermodynamic density of states of non-interacting
electrons (introduced here to make $F^\rho$ dimensionless).

Interaction in the triplet channel (spin density dynamics) is governed
by

\beqa
\label{104}
&&\hat{H}_\sigma =
\frac{1}{2}\sum_{\matrix{ { }_{\vec p_i} \cr { }_{|\vec q| < q*} }}
\sum_{ j=x,y,z}
\frac{F^\sigma\left(\widehat{\vec n_1 \vec n_2}\right)}{\nu}
\times
\label{hsigma}
\\
&&
\nonumber\\
&&
\left[\psi^\dagger_{\sigma_1}\left(\vec p_1\right)
\hat\sigma^j_{\sigma_1\sigma_2}
\psi_{\sigma_2}\left(\vec p_1-{\vec q}\right)
\right]
\left[\psi^\dagger_{\sigma_3}\left(\vec p_2\right)
\hat\sigma^j_{\sigma_3\sigma_4}
\psi_{\sigma_4}\left(\vec p_2+{\vec q}\right)
\right]
\nonumber
\eeqa

\noindent
where parameters 
$F^\sigma\left(\widehat{\vec n_1 \vec n_2}\right)$ are

\beq
\label{105}
F^\sigma\left(\theta\right) = - \frac{\nu}{2}
V\left(2k_F\sin\frac{\theta}{2} \right).
\eeq

Finally, the Hamiltonian

\beqa
\label{106}
&&\hat{H}_{pp} = 
\sum_{|\vec q| < q*, \vec p_i} \Bigg\{
\frac{F^e\left(\widehat{\vec n_1 \vec n_2}\right)}{\nu}
\left[\psi^\dagger_{\sigma_1}\left(\vec p_1\right)
\hat\sigma^y_{\sigma_1\sigma_2}
\psi^\dagger_{\sigma_2}\left({\vec q}-\vec p_1\right)
\right]
\nonumber\\
&&
\nonumber\\
&&
\quad\quad\quad\quad
\times\left[\psi_{\sigma_3}\left(\vec p_2\right)
\hat\sigma^y_{\sigma_3\sigma_4}
\psi_{\sigma_4}\left({\vec q}-\vec p_2\right)
\right]  
\\
&& 
\nonumber\\
&&
+
\sum_{ j=x,y,z}
\frac{F^o\left(\widehat{\vec n_1 \vec n_2}\right)}{\nu}
\left[\psi^\dagger_{\sigma_1}\left(\vec p_1\right)
\tilde\sigma^j_{\sigma_1\sigma_2}
\psi^\dagger_{\sigma_2}\left({\vec q}-\vec p_1\right)
\right]
\nonumber\\
&&
\nonumber\\
&&
\quad\quad\quad\quad
\times
\left[\psi_{\sigma_3}\left(\vec p_2\right)
\left(\tilde\sigma^j\right)_{\sigma_4\sigma_3}^\dagger
\psi_{\sigma_4}\left({\vec q}-\vec p_2\right)
\right]\Bigg\}
\nonumber
\eeqa

\noindent
describes singlet, $F^e$, and triplet, $F^o$, pairing
fluctuations. The parameters in this Hamiltonian are

\beq
\label{107}
F^{e,o}\left(\theta\right) = \frac{\nu}{4}
\left[
V\left(2k_F\sin\frac{\theta}{2} \right) \pm
V\left(2k_F\cos\frac{\theta}{2} \right)
\right],
\label{fefo}
\eeq

\noindent
where plus and minus signs correspond to even ($e$) and odd ($o$)
pairing respectively. Here $\hat\sigma_{\sigma_1\sigma_2}^j$ are the
elements of the Pauli matrices in spin space

\[
\hat\sigma^x =\pmatrix{0 & 1\cr 1&0}, \quad
\hat\sigma^y =\pmatrix{0 & -i\cr i&0}, \quad
\hat\sigma^z =\pmatrix{1 & 0\cr 0&-1},
\]

\noindent
and $\tilde\sigma^j= \hat{\sigma}^y\hat\sigma^j$. 

Deriving Eqs.~(\ref{Hrho}) -- (\ref{fefo}), we used the condition, $q*
\ll k_F$. This condition allowed us to make the following
approximation

\begin{eqnarray*}
(\vec p_1 - \vec p_2)^2 \approx 4 k_F^2 
\sin^2\left(\frac{\widehat{\vec n_1 \vec n_1}}{2}\right).
\end{eqnarray*}

\noindent
We also used the identity

\begin{eqnarray*}
2\delta_{\sigma_1\sigma_2}\delta_{\sigma_3\sigma_4}
&=&\delta_{\sigma_1\sigma_3}\delta_{\sigma_2\sigma_4} 
+ \hat\sigma_{\sigma_1\sigma_3}^j\hat\sigma_{\sigma_4\sigma_2}^j
\\
&&
\\
&=&\hat\sigma_{\sigma_1\sigma_3}^y\hat\sigma_{\sigma_2\sigma_4}^y
+ \tilde{\sigma}_{\sigma_1\sigma_3}^j
\left(\tilde{\sigma}^j\right)^\dagger_{\sigma_4\sigma_2}.
\end{eqnarray*}
   
So far, the representation (\ref{hamiltonian}) of original interaction
is exact. The only advantage of this representation is that it
explicitly separates the term $\delta H$ which does not contain
coupling to the low energy excitations of the fermionic
system. Therefore, the contribution of $\delta H$ to physical
quantities is regular and not infrared divergent [like
$(T/v_Fq*)^2$]. Therefore, for the electron system with weak short
range interaction, $\delta H$ can be disregarded at all.

Moreover, even if the interaction is not weak or long range, $\delta
H$ can be treated in all the orders of perturbation theory without
generating a soft mode. If this term does not break the translational
symmetry at short distances, its only effect is to renormalize the
interaction parameters $F$'s in Eqs.~(\ref{102}), (\ref{104}) and
(\ref{106}) and the Fermi velocity in the non-interacting part of the
Hamiltonian. For instance, one obtains for the two-dimensional
electron gas with the Coulomb interaction $V(q)= 2\pi e^2/(\kappa
|q|)$

\beqa
F^\rho\left(\theta\right) =F^\sigma\left(\theta\right)
= -\frac{1}{2}
\frac{r_s}{\sqrt{2}|\sin \frac{\theta}{2}| + r_s},
\label{108}
\eeqa

\noindent
where 

\beq
\label{109}
r_s \equiv\frac{\sqrt{2}e^2}{\kappa\hbar v_F}
\eeq

\noindent
is the conventional parameter characterizing interaction strength and
$\kappa$ is the low frequency dielectric constant of the host
material. Expression (\ref{108}) is applicable only for $r_s \ll 1$,
however, keeping it in denominator is legitimate for small angle
scattering.

For stronger interaction $r_s \gtrsim 1$, but still far from the
Wigner crystal instability\cite{Wignercrystal}, $r_s \lesssim 37$,
exact calculation of the parameters $F$ from the first principles (as
well as their explicit expressions in terms of $r_s$) is not possible.
Nevertheless, to study the behavior of the system at distances much
larger than $\lambda_F$, one can still disregard the term $\delta H$
in Eq.~(\ref{101}). Then parameters $F$ are no longer bound by
Eqs.~(\ref{103}), (\ref{105}), and (\ref{107}) [or by Eq.~(\ref{108})
for the Coulomb interaction] but rather should be treated as starting
parameters for the low energy theory.  The form of Eqs.~(\ref{102}),
(\ref{104}) and (\ref{106}) is guarded by symmetries of the system:
Eq.~(\ref{102}) is guarded by translational symmetry and charge
conservation; Eq.~(\ref{104}) is guarded by translational symmetry and
symmetry with respect to spin rotations; and Eq.~(\ref{106}) is
guarded by all above symmetries and the 
electron-hole symmetry, which holds approximately at low energies.

All the consideration above essentially repeats the basics of the
Landau Fermi-liquid theory\cite{Fermiliquid}. We reiterate, that this
theory does not imply that the interaction is weak; the only
assumption here is that no symmetry is broken at small distances.

\subsection{Disorder averaging}
\label{prelim}
\label{averaging}

To study the interaction correction to conductivity due to charge and
triplet channel interactions introduced in the previous subsection, we
follow the same route as in the case of the short-range
interaction. In particular, the charge channel correction is a direct
generalization of the Fock term. We start however with the discussion
of disorder averaging.

The correction to conductivity Eq.~(\ref{cc}) represents the first
order perturbation theory in the original potential $V(q)$, valid when
the potential is weak.  For stronger coupling we make use of the
effective Hamiltonian Eq.~(\ref{101}). Although the diagrams for
conductivity look similar to the Fock term ``b'' on Fig.~\ref{sr},
their content is now quite different. First, the wavy line now
represents the propagator for one of the soft modes in
Eq.~(\ref{101}). Therefore the expression for the conductivity
Eq.~(\ref{cc}) should be rewritten as

\begin{eqnarray}
&&\delta\sigma_{\alpha\beta} = - \int\limits_{-\infty}^{\infty} 
\frac{d\Omega}{8\pi^2}
\left[\frac{\partial}{\partial\Omega}
\left(\Omega\coth\frac{\Omega}{2T}\right)\right] 
\label{cci}
\\
&&
\nonumber\\
&&
\times
{\rm Im} 
\int \frac{d^2r_3d^2r_4}{\cal V}
\left\{ \Bigg [{\cal D}^A(\Omega, \vec r_3, \vec r_4) +
{\bf Tr}\widehat{\cal D}_T^A(\Omega, \vec r_3, \vec r_4) 
\Bigg ]\right.
\nonumber\\
&&
\nonumber\\
&&
\quad\quad\quad\times
\Big ( B^{\alpha\beta}_F (\Omega, \vec r_3, \vec r_4 ) 
+ \{\alpha\leftrightarrow\beta\}\Big)\Bigg\},
\nonumber
\end{eqnarray}

\noindent
where ${\cal D}^A$ and $\widehat{\cal D}_T^A$ are advanced propagators
for charge and triplet channels [$\widehat{\cal D}_T$ is a $3\times3$
matrix as follows from Eq.~(\ref{hsigma}), see also
Sec.~\ref{triplet}] and $B_F$ is the product of electronic Green's
functions given by Eq.~(\ref{fck}), the same as in the Fock term.
Deriving Eq.~(\ref{cci}) we assumed that the spin rotational symmetry
is preserved, i.e. no Zeeman splitting or the spin-orbit interaction
is present. We also neglected the dependence of the interaction
propagators on the direction of the electron momenta.
Lifting of those two assumptions is straightforward but it will not
be done in the present paper.
To the leading order in $1/k_Fl$ we can average the propagators
independently of $B_F$ (see e.g. Ref.~\onlinecite{aar}). Here we
proceed with averaging $B_F$ and the discussion of the propagators
follows.

We have already averaged the product $B_F$ of four Green's functions
for the case of the short-range potential. There the three terms
Eqs.~(\ref{f31}), (\ref{f32}), and (\ref{f13}) vanished due to the
particular form of the potential. Now we have to take these terms into
account and consider the full set of diagrams shown on Figs.~\ref{10}
and \ref{11}. These diagrams can be evaluated in exactly the same way
as those in Section~\ref{h-f} (where we considered a subset of these
diagrams).

As a result, the averaged $B_F$ has a form similar to Eq.~(\ref{bf})
and can again be expressed in terms of the dressed vertex $\Gamma$
[see Eq.~(\ref{gsd})].  We are still interested in the longitudinal
conductivity and thus disregard the Hall contribution. Thus, after
averaging the correction Eq.~(\ref{cci}) takes the form

\begin{eqnarray}
&&\delta\sigma = -e^2 v_F^2 \pi\nu\int\limits_{-\infty}^{\infty} 
\frac{d\Omega}{4\pi^2} \frac{\partial}{\partial\Omega}
\left(\Omega\coth\frac{\Omega}{2T}\right) \int \frac{d^2 q}{(2\pi)^2}
\nonumber\\
&&
\nonumber\\
&&\quad\quad\quad
\times{\rm Im} \Bigg\{ 
\Big[{\cal D}^A(\Omega, q)+{\bf Tr}\widehat{\cal D}_T^A(\Omega, q)\Big]
\tilde B_F(\Omega,q)
\Bigg\};
\nonumber\\
&&
\nonumber\\
&&\tilde B_F(\Omega,q)=-
\frac{2\tau (i\Omega + \frac{1}{\tau})\Gamma}
{S^3}
+\frac{(\Gamma^2-1)\tau^2}{S}
\nonumber\\
&&
\nonumber\\
&&
+\frac{v_F^2 q^2-2\left(i\Omega + \frac{1}{\tau}\right)^2  }
{S^5}\frac{\Gamma^2}{2}
+\frac{\tau\Gamma(\Gamma+1)}{v_F^2 q^2}
\left(\frac{i\Omega + \frac{1}{\tau}}
{S}
- 1 \right)^2
\nonumber\\
&&
\nonumber\\
&&
-\frac{2\Gamma^2}
{ S^3}
\left(\frac{i\Omega + \frac{1}{\tau}}
{S}
- 1 \right)
+\frac{\Gamma^3v_F^2 q^2}{
\tau S^6}.
\label{fock-int}
\end{eqnarray}

\noindent
where quantities  $\Gamma$ and $S$
are defined in Eq.~(\ref{gsd}).

{
\narrowtext
\begin{figure}[ht]
\vspace{0.2 cm}
\epsfxsize=7 cm
\centerline{\epsfbox{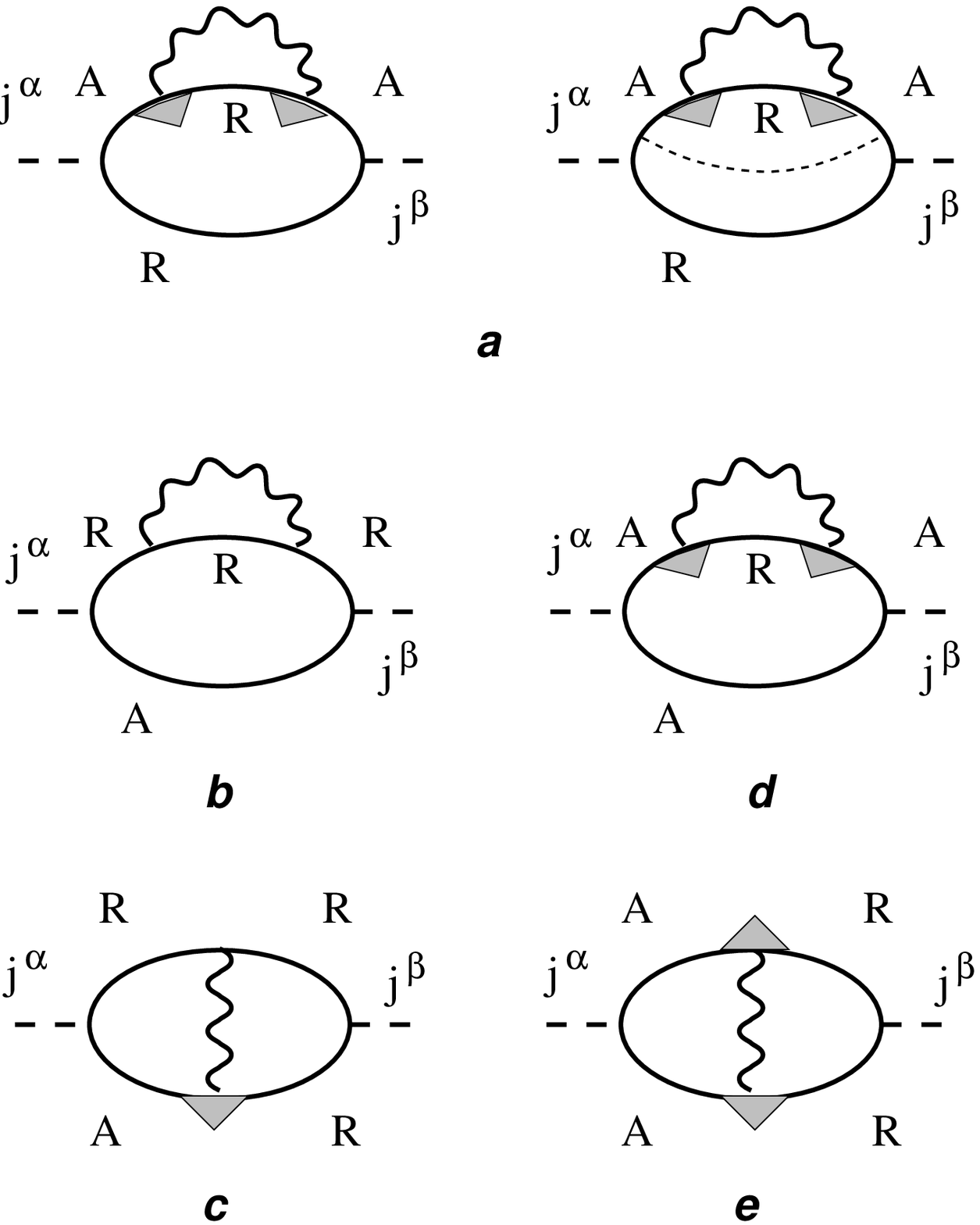}}
\vspace{0.2cm}
\caption{Conductivity diagrams, group I. Diagrams (a), (b), and (d)
were evaluated for the short range interaction in Sec.~\protect\ref{h-f}.
In the diffusive regime\protect\cite{aar} 
only diagrams (a), (d), (e), were considered at $\omega, qv_F \ll 1/\tau$.
} 
\label{10}
\end{figure}
}

{
\narrowtext
\begin{figure}[ht]
\vspace{0.2 cm}
\epsfxsize=9 cm
\centerline{\epsfbox{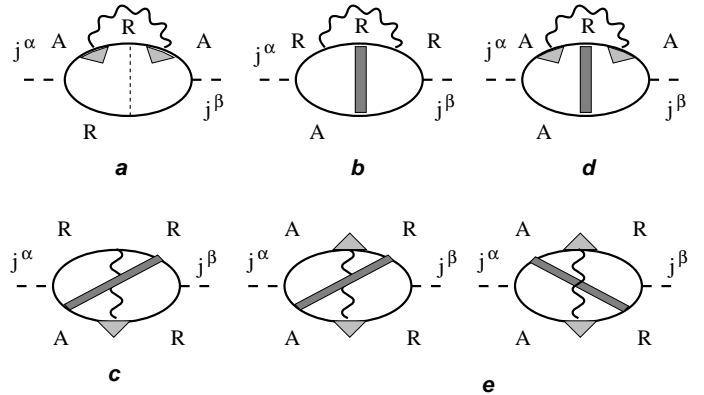}}
\vspace{0.2cm}
\caption{Conductivity diagrams, group II.
Diagrams (a) and (b)
were evaluated for the short range interaction in Sec.~\protect\ref{h-f}.
In the diffusive regime\protect\cite{aar} 
only diagrams (a), (d), (e), were considered at $\omega, qv_F \ll 1/\tau$.
}
\label{11}
\end{figure}
}

It is important to emphasize that 

\beq
\tilde B_F(\Omega,q=0) =0,
\label{ginvariance}
\eeq  

\noindent
[to see this one should use explicit
expressions (\ref{gsd}) in Eq.~(\ref{fock-int})].
This property is not accidental -- it is guarded by the gauge invariance
of the system: no interaction with zero momentum transfer can affect
the value of the closed loop.

To proceed further with the actual
calculation of the correction (\ref{fock-int})
we need to specify the interaction propagator. It will be done
in the following two subsections.

\subsection{Charge channel}
\label{charge}

In this section we discuss the charge channel correction,
described by the Hamiltonian (\ref{102}). Because the effective
interaction is characterized by the momentum transfer much smaller
than the Fermi wave vector, the Random Phase Approximation (RPA), see 
Fig.~\ref{15}, is applicable.

To simplify further considerations, we approximate the
Fermi liquid parameter $F^\rho$ by its zero angular harmonic

\beq
F^\rho(\theta) \approx F^\rho_0,
\label{rhoapprox}
\eeq 

\noindent
this approximation does not affect the final result because of
the long range nature of the Coulomb potential $V(q \to 0) \to
\infty$.

Consequently, we write the charge channel propagator in the form
 
\begin{mathletters}
\label{int-full}
\begin{equation}
{\cal D}^A(\Omega, q) = 
- \frac{\nu V(q)+F_0^\rho}{\nu+ (\nu V(q)+F_0^\rho) \Pi^A},
\label{int-dyson}
\end{equation}

\noindent
where the polarization operator is given by

\begin{equation}
\Pi^A(\Omega, q) = 
\nu \left[ 1 - \frac{i\Omega}{
S-\frac{1}{\tau}}\right],
\label{pol}
\end{equation}

\noindent
using the notation (\ref{sqrt}).

The polarization operator (\ref{pol}) differs from the more standard one
(used for instance in Ref.~\onlinecite{aag}) since the diffusion
approximation has not been made yet. Indeed, expanding the
polarization operator in small $\Omega$ and $q$ we can recover the
usual diffusive form. In terms of the scattering time it corresponds
to the limit $T\tau \ll 1$.  We do not do that here since we want
to calculate the conductivity for all values of $T\tau$.
\end{mathletters}

The form of the propagator~(\ref{int-full})
and expression for the conductivity correction (\ref{fock-int})
suggests that there could be two contributions.
First, the propagator Eq.~(\ref{int-full}) has a
pole which corresponds to the 2D plasmon. 
However, the plasmon
dispersion relation is
 
\[
(v_Fq_{pl})^2 \nu V(q_{pl}) 
=2\Omega \left(\Omega + \frac{i}{\tau}\right),
\]

\noindent
i.e. $(v_Fq_{pl})^2  \ll |\Omega \left(\Omega +
\frac{i}{\tau}\right)|$ at all distances larger than the screening
radius. According to the gauge invariance condition
(\ref{ginvariance}) this contribution is strongly suppressed 
(by a factor of the order of $max[T,(T/\tau)^{1/2}]d_{sc}/v_F$,
with $d_{sc}$ being the screening radius, $\nu V(1/d_{sc})=1$)
and we will not take it into account.
 
Second, at frequencies smaller than the
plasmon frequency we can neglect the unity in the denominator in
Eq.~(\ref{int-full}), which corresponds to the unitary limit, i.e.

\begin{eqnarray}
{\cal D}^A = -  
\frac{1}{\Pi^A} = - \frac{1}{\nu}
\frac{S - \frac{1}{\tau}}{S- \frac{1}{\tau}-i\Omega}.
\label{f-prop}
\end{eqnarray}

\noindent
Thus the original coupling $V(q)$ as well as the renormalization of the
coupling by the Fermi liquid parameter Eq.~(\ref{102}) does not affect
the resulting propagator. In other words the propagator becomes
universal.

It is important to emphasize that Eq.~(\ref{f-prop}) gives the upper
bound for the strength of the repulsive interaction. This is
guaranteed by stability of the electron system with respect to
the Wigner crystallization, i.e. by the condition $\nu V(q)+F_0^\rho >
0$ at $q<q*$. Therefore, we always have

\begin{eqnarray*}
\frac{\nu V+F_0^\rho}{\nu+ (\nu V+F_0^\rho) \Pi} < \frac{1}{\Pi},
\end{eqnarray*}

\noindent
so that Eq.~(\ref{f-prop}) is indeed the upper bound for the
propagator Eq.~(\ref{int-dyson}). Note, that the above condition is
satisfied regardless of the sign of $F_0^\rho$. 
In particular, it is possible to have
$F_0^\rho<-1$ so that the so-called compressibility of the system 
$\nu/(1+F_0^\rho)$ is negative. This fact, however,
has nothing to do with stability of the Fermi liquid and does not
affect transport phenomena \cite{vsi}.
 
Using the propagator Eq.~(\ref{f-prop}) in the expression for the
correction Eq.~(\ref{fock-int}) we obtain after momentum integration

\begin{eqnarray}
\delta\sigma_C =&& -e^2\tau\int\limits_{0}^{\infty} 
\frac{d\Omega}{2\pi} \frac{\partial}{\partial\Omega}
\left(\Omega\coth\frac{\Omega}{2T}\right) 
\nonumber\\
&&
\nonumber\\
&&
\left\{ \frac{2}{\pi}\arctan\Omega\tau + \frac{1}{\pi\Omega\tau}
+ \frac{\Omega\tau}{2\pi} H(\Omega\tau) \ln 2
\right.
\nonumber\\
&&
\nonumber\\
&&
+\frac{1}{\pi}\Big[1+H(\Omega\tau)\Big]\arctan\frac{1}{\Omega\tau} 
\nonumber\\
&&
\nonumber\\
&&\left.
+\frac{\Omega\tau}{4\pi}\left[ \frac{1}{2} +H(\Omega\tau)\right]
\ln\left(1+\frac{1}{\Omega^2\tau^2}\right)\right\},
\label{fock-int-3}
\end{eqnarray}

\noindent
where the dimensionless function $H(x)$ is defined as

\begin{eqnarray*}
H(x) = \frac{1}{4+x^2}.
\end{eqnarray*}

In the frequency integral Eq.~(\ref{fock-int-3}) we single out the
first two terms as being dominant in the ballistic and diffusive
limits respectively with the rest being the crossover function. The
diffusive limit is given by

\begin{eqnarray}
\delta\sigma_C(T\tau\ll 1) =&& -e^2\tau\int\limits_{0}^{\infty} 
\frac{d\Omega}{2\pi} \frac{\partial}{\partial\Omega}
\left(\Omega\coth\frac{\Omega}{2T}\right) \frac{1}{\pi\Omega\tau}
=
\nonumber\\
&&
\nonumber\\
&&
= -\frac{e^2}{2\pi^2}\ln\left(\frac{E_F}{T}\right).
\label{cd}
\end{eqnarray}

\noindent
In the opposite limit we can replace $\arctan\Omega\tau$ by $\pi/2$.
Then the integral is divergent in the ultra-violet, but that large
constant can be incorporated in the definition of $\tau$. This is done
as follows:

\begin{eqnarray}
\int\limits_{0}^{\infty} d\Omega \frac{\partial}{\partial\Omega}
\left(\Omega\coth\frac{\Omega}{2T}\right)
\rightarrow -2T + E_F\coth\frac{E_F}{2T},
\label{ultra}
\end{eqnarray}

\noindent
where $E_F$ is put for the upper limit of the integral. This is
consistent with the approximations in momentum integration, where one
typically relies on fast convergence in order to set the integration
limit (otherwise determined by the Fermi energy) to infinity and to
set all momenta in the numerator to the Fermi momentum in magnitude.
Since we are interested in temperatures $T\ll E_F$, the second term is
essentially a temperature independent (although infinite)
constant. The temperature dependent correction to the conductivity is
determined by the first term. As a result

\begin{eqnarray}
\delta\sigma_C(T\tau\gg 1) =&& -e^2\tau\int\limits_{0}^{\infty} 
\frac{d\Omega}{2\pi} \frac{\partial}{\partial\Omega}
\left(\Omega\coth\frac{\Omega}{2T}\right) 
=
\nonumber\\
&&
\nonumber\\
&&
= e^2\frac{T\tau}{\pi}.
\label{cb}
\end{eqnarray}

Integrating the full expression Eq.~(\ref{fock-int-3}) we find the
correction valid at all values of $T\tau$, 

\begin{equation}
\delta\sigma_C = -\frac{e^2}{2\pi^2}\ln\left(\frac{E_F}{T}\right)
+ e^2\frac{T\tau}{\pi}
\left[ 1 -\frac{3}{8} f(T\tau)\right],
\label{fc1}
\end{equation}

\noindent
where the dimensionless function $f(x)$ is defined as a dimensionless
integral:

\begin{eqnarray}
f(x) = \frac{8}{3}
\int\limits_{0}^{\infty}&& dz \left[ \frac{\partial}{\partial z}
\left(z\coth z\right) - 1\right]
\left\{\frac{xz}{\pi} H(2xz) \ln 2 \right.
\nonumber\\
&&
\nonumber\\
&&
+\frac{1}{\pi}\Big[1+H(2xz)\Big]\arctan\frac{1}{2xz}
\nonumber\\
&&
\nonumber\\
&&
+\frac{xz}{2\pi}\left[\frac{1}{2}+H(2xz)\right]
\ln\left(1+\frac{1}{(2xz)^2}\right)
\nonumber\\
&&
\nonumber\\
&&\left.
+
\frac{2}{\pi}\arctan 2xz - 1\right\}.
\label{f2}
\end{eqnarray}

\noindent
The factor $3/8$ is introduced for convenience, so that $f(0)=1$. The
integral can be evaluated analytical in the two limiting cases and the
result is given by Eq.~(\ref{fl}). In the intermediate regime the
integral can be evaluated numerically and the result is plotted on
Fig.~\ref{plot:f}.

\subsection{Triplet channel}
\label{triplet}


In this section we discuss the correction in the triplet channel.
Similar to the case of the charge channel, we need to derive the
interaction propagator in the triplet channel and then use
Eq.~(\ref{fock-int}). As follows from the Hamiltonian
Eq.~(\ref{hsigma}), the triplet channel propagator is now a $3 \times
3$ matrix. Apart from this minor complication, the propagator can be
found using the same RPA approximation as the one used in
Section~\ref{charge}, see Fig.~\ref{15}.

{
\narrowtext
\begin{figure}[ht]
\vspace{0.1 cm}
\epsfxsize=8.5 cm
\centerline{\epsfbox{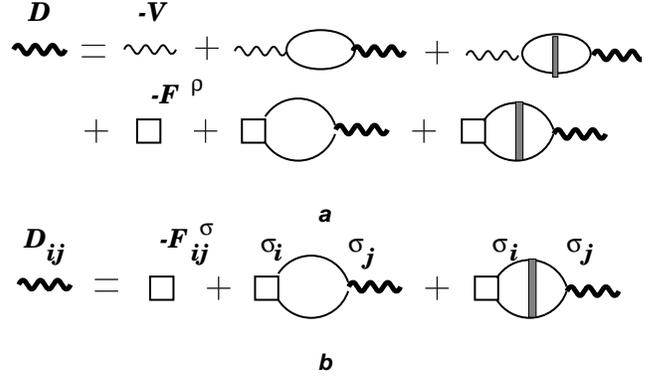}}
\vspace{0.5cm}
\caption{Interaction propagator in the (a) singlet and (b)
triplet channel} 
\label{15}
\end{figure}
}

Similarly to the charge channel, we take the Fermi liquid coupling
$\hat F^\sigma$ to be independent of electron momenta

\begin{eqnarray}
F^\sigma\left(\theta\right) \approx F^\sigma_0.
\label{appr}
\end{eqnarray}

\noindent
Unlike the case of the charge channel, this approximation slightly
affects final results (see discussion after Eqs.~(\ref{s}) for the
drawbacks of this approximation as well as for its remedies).  Then
the matrix equation for the triplet propagator has the form

\begin{equation}
\left[{\cal D}_T\right]_{ij} = 
-\delta_{ij}\frac{F^\sigma_0}{\nu}
- \frac{F^\sigma_0}{\nu}\sum_{k=x,y,z}\hat 
\Pi_{ik}
\left[{\cal D}_T\right]_{kj}
\label{prop-eq}
\end{equation}

\noindent
where $i,j=x,y,z$.

In the absence of the magnetic field and spin-orbit scattering each
electronic Green's function is a diagonal matrix in the spin space,
and therefore

\beq
\Pi_{ik}^A(q,\Omega)=\delta_{ik}\Pi^A(q,\Omega),
\label{psigma}
\eeq 

\noindent
where $\Pi^A(q,\Omega)$ is the polarization operator given by
Eq.~(\ref{pol}).

Altogether, using Eq.~(\ref{psigma}) in the equation (\ref{prop-eq}),
we find the triplet channel propagator as

\begin{eqnarray}
\left[{\cal D}_T^A(\Omega, q)\right]_{ij} = - \delta_{ij}
\frac{F^\sigma_0}{\nu + F^\sigma_0 \Pi^A(\Omega, q)}.
\label{trip-prop}
\end{eqnarray}

Before we continue, let us discuss the validity of the approximation
Eq.~(\ref{appr}). Consider the situation close to the Stoner
instability $F^\sigma_0 \rightarrow -1$. In this case the pole of the
propagator Eq.~(\ref{trip-prop}) describes a magnetic excitation in
the system.  In the ballistic case ($q>1/l$) it is a slow, over-damped
spin wave

\begin{eqnarray*}
-i\omega \approx (1+F^\sigma_0)v_F|q|.
\end{eqnarray*}
 
\noindent
The main contribution to the temperature dependent conductivity
correction comes from this pole at $\omega\sim T$. The corresponding
typical momenta are $k^*\sim T / [(1+F^\sigma_0)v_F]$.  Although we
are using the momentum independent $F^\sigma_0$, it is known
\cite{bkm} that fluctuations in the triplet channel produce a
non-analytic correction to the spin susceptibility, so up to a
numerical coefficient $F^\sigma\approx F^\sigma_0 (1 - |q|/k_F)$. Such
momentum dependence can only be neglected if $k^*\le
k_F(1+F^\sigma_0)$. This translates into a limitation for the
temperature range where the results listed in Section~\ref{results}
are valid\cite{lar}:

\begin{eqnarray}
T\ll T^* \approx (1+F^\sigma_0)^2 E_F.
\end{eqnarray}

\noindent
At higher temperatures $T>T^*$ our theory is not applicable.

Having discussed the validity of the approach, we proceed with the
straightforward calculation: one has to substitute the propagator
Eq.~(\ref{trip-prop}) into the expression for the correction
Eq.~(\ref{fock-int}) and evaluate the integral. The result of the
momentum integration is given by

\begin{mathletters}
\begin{eqnarray}
&&\delta\sigma_T = -3e^2\tau
\int\limits_{-\infty}^{\infty} 
\frac{d\Omega}{4\pi} \frac{\partial}{\partial\Omega}
\left(\Omega\coth\frac{\Omega}{2T}\right) 
\nonumber\\
&&
\nonumber\\
&& 
\left\{ \left(1-\frac{1}{F_0^\sigma}\ln(1+F_0^\sigma)\right)
\frac{1}{\pi\Omega\tau}
 + \frac{F_0^\sigma}{(1+F_0^\sigma)}\left[
\frac{2}{\pi}\arctan\Omega\tau
\right.\right.
\nonumber\\
&&
\nonumber\\
&& 
+ \frac{\Omega\tau}{2\pi} \left(\Big[H(\Omega\tau)+h_1(\Omega\tau)\Big] \ln 2
+ h_4(\Omega\tau)\right)
\label{fock-int-t}
\\
&&
\nonumber\\
&&
+\frac{1}{\pi}\Big[1+H(\Omega\tau)
+ (\Omega\tau)^2h_3(\Omega\tau)\Big]\arctan\frac{1}{\Omega\tau} 
\nonumber\\
&&
\nonumber\\
&&\left.\left.
+\frac{\Omega\tau}{4\pi}\left[ \frac{1}{2} +H(\Omega\tau)
+ h_2(\Omega\tau)\right]
\ln\left(1+\frac{1}{\Omega^2\tau^2}\right)\right]\right\},
\nonumber
\end{eqnarray}

\noindent
where we introduce notations: 

\begin{eqnarray}
h_1(x)=\tilde H&&(x; 1+2F_0^\sigma)
\nonumber\\
&&
\nonumber\\
&&\times
\Big[5+6F_0^\sigma-4(2+3F_0^\sigma)H(x)\Big],
\end{eqnarray}
\begin{eqnarray}
h_2(x)=&&h_1(x)+\tilde H(x; F_0^\sigma)
\left[-\frac{1}{2}(1+F_0^\sigma)\right.
\nonumber\\
&&
\nonumber\\
&&\left.
+F_0^\sigma x^2
\left(\frac{1}{2}-(1+F_0^\sigma)\tilde H(x; F_0^\sigma)\right)\right],
\end{eqnarray}
\begin{eqnarray}
h_3(x)=&&\tilde H(x; 1+2F_0^\sigma)
\Big[-1-2F_0^\sigma+(2+3F_0^\sigma)H(x)\Big]
\nonumber\\
&&
\nonumber\\
&&
+\frac{F_0^\sigma}{2}\tilde H(x; F_0^\sigma)
\Big[1+F_0^\sigma x^2\tilde H(x; F_0^\sigma)\Big],
\end{eqnarray}
\begin{eqnarray}
h_4(x)=\tilde H(x; F_0^\sigma)&&\left[\frac{5F_0^\sigma-3}{2}+
\frac{1-(F_0^\sigma)^2}{F_0^\sigma}\ln(1+F_0^\sigma)\right]
\nonumber\\
&&
\nonumber\\
&&
+h_5(x)\frac{1+F_0^\sigma}{F_0^\sigma}\ln(1+F_0^\sigma),
\end{eqnarray}
\begin{eqnarray}
&&h_5(x)=(2F_0^\sigma-1)\tilde H(x; 1+2F_0^\sigma) + 
\tilde H^2(x; F_0^\sigma)
\\
&&
\nonumber\\
&&\quad\quad
\times
\left[\left(\frac{1}{2}-2F_0^\sigma\right)(1+F_0^\sigma)^2 - 
(F_0^\sigma)^2x^2 \left(\frac{1}{2}+2F_0^\sigma\right)\right].
\nonumber
\end{eqnarray}
\end{mathletters}

\noindent
Here we introduce a dimensionless function $\tilde H(x; y)$

\begin{eqnarray*}
\tilde H(x; y) = \frac{1}{(1+y)^2+(xy)^2},
\end{eqnarray*}

\noindent
which is related to the function $H(x)$ introduced in Section~\ref{charge}
simply by $H(x)=\tilde H(x; 1)$.

The expression in brackets turns into its counterpart in the charge
channel in the unitary limit ($F^\sigma_0\rightarrow\infty$). Its
first term describes the diffusive limit described in
Ref.~\onlinecite{aar} (the formal difference in the coefficient stems
from the difference in the definition of the coupling constant).  The
frequency integral is evaluated in the same way as in Eq.~(\ref{cd}).
Similar to our discussion of the charge channel correction [see e.g.
Eq.~(\ref{cb})], we identify the second term in Eq.~(\ref{fock-int-t})
with the ballistic limit (which we discuss in more detail in the next
Section). The intermediate temperature regime is described by the
expression [which appeared previously in Section~\ref{results},
Eq.~(\ref{tc})]:

\begin{eqnarray}
\delta\sigma_T = &&-3\left(1-\frac{1}{F_0^\sigma}
\ln(1+F_0^\sigma)\right)
\frac{e^2}{2\pi^2}\ln\left(\frac{E_F}{T}\right)
\nonumber\\
&&
\nonumber\\
&&
+\frac{3F_0^\sigma}{(1+F_0^\sigma)}e^2\frac{T\tau}{\pi} 
\left[ 1 -\frac{3}{8}t(T\tau; F_0^\sigma)\right].
\label{tc2}
\end{eqnarray}

\noindent
where the dimensionless function $t(x; F_0^\sigma)$ is defined as

\begin{eqnarray}
t(x; &&F_0^\sigma) = \frac{8}{3}
\int\limits_{0}^{\infty} dz \left[ \frac{\partial}{\partial z}
\left(z\coth z\right) - 1\right]
\nonumber\\
&&
\nonumber\\
&&\times
\Bigg\{\frac{xz}{\pi} 
\left(\Big[H(2xz)+h_1(2xz)\Big] \ln 2 + h_4(2xz)\right)
\nonumber\\
&&
\nonumber\\
&&\quad
+\frac{1}{\pi}\Big[1+H(2xz)+4x^2z^2h_3(2xz)\Big]\arctan\frac{1}{2xz}
\nonumber\\
&&
\nonumber\\
&&\quad
+\frac{xz}{2\pi}\left[\frac{1}{2}+H(2xz)+h_2(2xz)\right]
\ln\left(1+\frac{1}{(2xz)^2}\right)
\nonumber\\
&&
\nonumber\\
&&\quad
+
\left[\frac{2}{\pi}\arctan(2xz) - 1\right]\Bigg\}.
\label{t2}
\end{eqnarray}

\noindent
Except for the limiting cases [see Eq.~(\ref{tsx})] the integral in
Eq.~(\ref{t2}) has to be evaluated numerically. We plot the result
for several values of $F_0^\sigma$ in Fig.~\ref{plot:t}.

\subsection{Single impurity limit}
\label{single}

In the previous Sections we obtained the expression for the correction
to conductivity averaged over disorder. To complete the calculation we
needed to separately average the interaction propagator and use the
result to evaluate the integral in Eq.~(\ref{fock-int}). In doing this
we assumed that the dimensionless conductance of the system is large
or in terms of the scattering time $\tau E_F\gg 1$. We have not,
however, assumed anything about the relative value of the scattering
rate and temperature. In other words, the correction
Eq.~(\ref{fock-int}) is valid in both the diffusive $T\tau\ll 1$ and
ballistic $T\tau\gg 1$ limits. It also describes the cross-over
behavior at intermediate temperatures.

The temperature behavior of the interaction correction in the limiting
cases can of course be obtained from the general result
Eq.~(\ref{s}). As we pointed out in Section~\ref{results}, in the
diffusive limit our results coincide with the standard theory,
Ref.~\onlinecite{aar}. On the other hand the correction in the
ballistic limit is subject to conflicting claims in literature
\cite{gdl,rei}.  Unfortunately, neither result is completely
correct. Therefore we discuss the ballistic limit in some detail,
starting with diagrams before averaging (i.e. diagrams on
Figs.~\ref{sr}). This way we are able to point out exactly which
diagram produces the dominant result and which diagrams were missed in
existing theories.

We begin by discussing the Hartree term. This contribution was
considered in Ref.~\onlinecite{gdl} in the framework of the
temperature dependent dielectric function. The physical idea was that
electrons tend to screen the charged impurities and thus modify the
scattering rate. In what follows we show which diagrams describe this
process and how to calculate the resulting correction, which appears
to be the same (up to a numerical factor miscalculated in
Ref.~\onlinecite{gdl}; see below for detailed explanation).  The
important difference between the two approaches is that the impurity
screening picture described only the direct (Hartree) interaction,
while missing on the exchange part. The latter was later considered in
Ref.~\onlinecite{rei}. We think that this consideration is erroneous,
and we discuss the Fock term in Section~\ref{sif}.

\subsubsection{Single impurity limit for Hartree term}
\label{sih}

The goal of this discussion is to show which diagrams correspond to
the ballistic limit of the Hartree term (as discussed in
Section~\ref{qualitative}) and how it relates to other interaction
corrections we discuss in this paper.

The Hartree term corresponds to averaging the two diagrams on
Fig.~\ref{sr}, where the wavy line represents a weak interaction
potential. In this case the diagram ``a3'' of Fig.~\ref{sr} is equal
to zero even before the averaging (as a total derivative) and we only
need to average the diagrams ``a1'' and ``a2''. The rigorous procedure
would involve dressing the interaction vertices according to
Fig.~\ref{8} and adding diffusons Fig.~\ref{9} as it was done in
Section~\ref{h-f} (see Fig.~\ref{hfav}), evaluating the resulting
expression, and finally taking the limit $T\tau\rightarrow\infty$.
However, the same result can be obtained by making the expansion by
noticing that impurity line brings smallness $1/T\tau$.  Therefore,
high temperature limit may be studied by considering diagrams on
Fig.~\ref{13} directly. Such approach is completely equivalent to that
of Ref.~\onlinecite{gdl}.  The result [which can also be obtained from
the general expression Eq.~(\ref{tc})] is similar to the one obtained
in Ref.~\onlinecite{gdl} (the difference is the extra factor of $\ln
2$ found in Ref.~\onlinecite{gdl} due to an error in this reference,
which consists in putting the energy of the scattered electron on the
Fermi shell rather than integrating over it):

\begin{eqnarray}
\delta\sigma_H = - 4\sigma_D\left(\frac{T}{E_F}\right)
\left[-\nu{\cal D}(2k_F)\right]
\label{hart-si}
\end{eqnarray}

\noindent
(for weak coupling ${\cal D}(2k_F) \equiv - V(2k_F)$ The factor of $4$
in Eq.~(\ref{hart-si}) can be interpreted as a result of a summation
over four spin configurations. Although correct for weak coupling,
this factor should be modified when stronger interaction is
considered, see discussion above.

{
\narrowtext
\begin{figure}[ht]
\vspace{0.5 cm}
\epsfxsize=7 cm
\centerline{\epsfbox{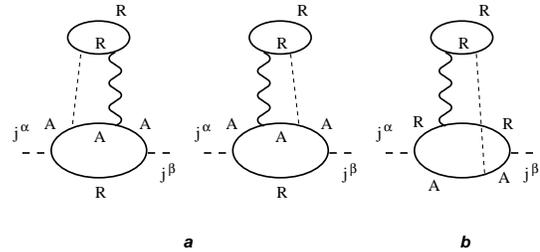}}
\vspace{0.5cm}
\caption{Single impurity diagrams for Hartree channel.} 
\label{13}
\end{figure}
}

\subsubsection{Fock contribution}
\label{sif}

In the similar manner one can discuss the single impurity contribution
to the Fock term. Again, for weak interaction we could simply expand
the result of disorder averaging for the Fock term Eq.~(\ref{bf})
to the leading order in $1/T\tau$. For Coulomb interaction we would
expand Eq.~(\ref{fock-int}), since in Eq.~(\ref{bf}) the special form
of the delta-function potential was utilized to eliminate the diagram
``b3'' on Fig.~\ref{sr}. Diagrammatically, such expansions equivalent
to direct evaluation of diagrams without impurity lines (but with
averaged electron Green's functions) shown in Fig.~\ref{3} and
diagrams with only one impurity line shown on Figs.~\ref{4}, \ref{5},
and \ref{6}.

The evaluation of the single impurity diagrams for the Fock term is 
straight-forward and is completely analogous to the Hartree term
discussed in the previous subsection. The result can be written as

\begin{equation}
\sigma_F = \frac{e^2\tau T}{\pi}.
\label{fock-si}
\end{equation}

\noindent
This result contradicts (even in sign) that of
Ref.~\onlinecite{rei}. Here we briefly discuss the reason for this
contradiction. We notice that one has to be careful to keep track of
gauge invariance while evaluating diagrams for the Fock term. Gauge
invariance manifests itself in the fact that any interaction at zero
momentum gives no contribution to physical quantities, which are
expressed diagrammatically as closed loops, see
Eq.~(\ref{ginvariance}).  This is indeed the case for
Eq.~(\ref{fock-int}), where we summed up all the diagrams. On the
other hand, any individual diagram is not gauge invariant. In
particular, each subset of diagrams in Figs.~\ref{33}-\ref{6} is not
gauge invariant.  Therefore to obtain the result Eq.~(\ref{fock-si})
from these diagrams one has to disregard terms which contain higher
than second powers of the scattering rate $1/\tau$.

{
\narrowtext
\begin{figure}[ht]
\vspace{0.2 cm}
\epsfxsize=7.2 cm
\centerline{\epsfbox{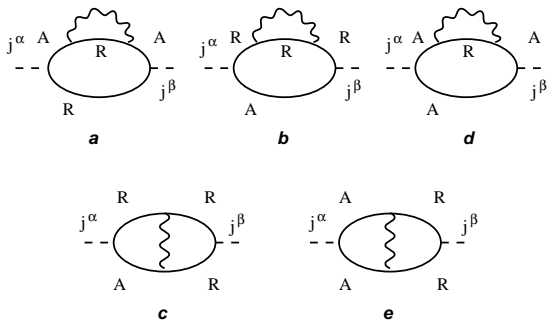}}
\vspace{0.2cm}
\caption{Fock channel diagrams without impurity lines.} 
\label{33}
\end{figure}
}

As we already mentioned, the contribution from the plasmon pole is
small due to the condition (\ref{ginvariance}).  However, in
Ref.~\onlinecite{rei} it was claimed otherwise.  Namely, diagrams in
Fig.~\ref{33} were claimed to be important for the plasmon correction
and to give a large result, while diagrams in Figs.~\ref{4}-\ref{6}
were alleged to be not important for the plasmon correction. This
claim explicitly violates gauge invariance and leads to incorrect
conclusions. In particular, the plasmon contribution to the
conductivity was overestimated by a factor of order of $(v_F/d_s T)
\simeq (E_F/ T)$.

{
\narrowtext
\begin{figure}[ht]
\vspace{0.1 cm}
\epsfxsize=6.5 cm
\centerline{\epsfbox{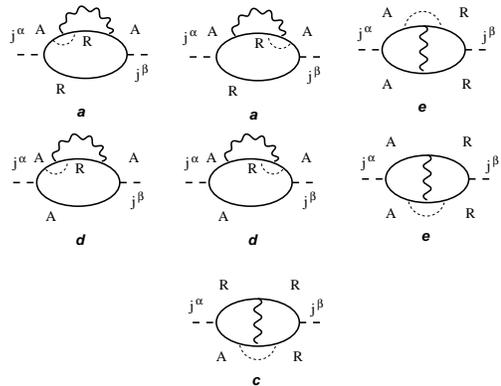}}
\vspace{0.2cm}
\caption{Single impurity diagrams for Fock channel with the impurity
line dressing one interaction vertex.}
\label{4}
\end{figure}
}
{
\narrowtext
\begin{figure}[ht]
\vspace{0.1 cm}
\epsfxsize=6.5 cm
\centerline{\epsfbox{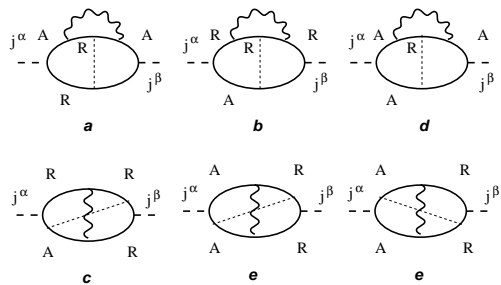}}
\vspace{0.2cm}
\caption{Single impurity diagrams for Fock channel with the impurity
line connecting a retarded and an advanced Green's functions across the 
diagram.} 
\label{5}
\end{figure}
}
{
\narrowtext
\begin{figure}[ht]
\vspace{0.2 cm}
\epsfxsize=2.7 cm
\centerline{\epsfbox{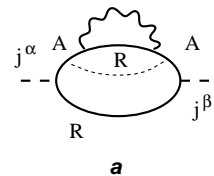}}
\vspace{0.1cm}
\caption{The single impurity diagram for Fock channel with the
impurity line connecting two advanced Green's functions.}
\label{6}
\end{figure}
}

\section{Kinetic equation approach}
\label{eilenberger}

Our purpose in this Section is to put the treatment of the interaction
effects in disordered systems into the framework of the kinetic
equation.  Even though at this point this will not produce any further
physical results, this proves to be more convenient for practical
calculations of more sophisticated quantities, such as the Hall
coefficient, the thermal conductivity, energy relaxation, etc. The
kinetic equation approach is also applicable for the description of
non-linear effects. The main technical advantage of the kinetic
equation is that it operates with gauge invariant quantities from the
very beginning, unlike the perturbation theory, where each diagram
taken separately is not gauge invariant (and may produce non-physical
divergences).

We will present the final form of the kinetic equation in subsection
\ref{kinetic}, and show how to operate with this equation for the
conductivity calculation in subsection \ref{cond}. The derivation of
this equation based on the Keldysh technique for non-equilibrium
systems \cite{Keldysh} is presented in subsection \ref{derivation}.

\subsection{Final form of the kinetic equation}
\label{kinetic}

As usual in the kinetic equation approach, averages of observable
quantities are expressed as certain integrals of the distribution
function $f(t; \epsilon,\vec r,\vec n)$.  For instance the averaged
density is

\begin{mathletters}
\label{eq:4.1}
\beq
\rho (t,\vec r) = \nu\int\limits_{-\infty}^{\infty} d\epsilon 
\langle f(t; \epsilon , \vec r, \vec n)\rangle_n
\eeq

\noindent
and the average current is

\beq
{\vec J}(t,\vec r) =e\nu v_F \int\limits_{-\infty}^{\infty} d\epsilon
\langle \vec n f(t; \epsilon , \vec r ,\vec n)\rangle_n
\label{current}
\eeq 

\noindent
and so on. Here $\nu$ is the density of states (entering into linear
specific heat of the clean system) at the Fermi surface and $v_F$ is
the Fermi velocity, $\vec n = (\cos\theta, \sin\theta)$ is the unit
vector in the direction of the electron momentum and angular averaging
is introduced as

\[
\langle \dots \rangle_n =
\int 
\frac{d\theta}{2\pi} \dots
\]

\end{mathletters}

The Boltzmann-like equation for the distribution function has the form

\beq
\label{eq:4.2}
\left[ \partial_t + v_F \vec n \vec\nabla + 
e v_F (\vec n {\vec E})\frac{\partial}{\partial \epsilon}
+\vec\omega_c  
\left(\vec n \times \frac{\partial}{\partial \vec n} \right) 
\right] f =
\St\left\{f\right\},
\label{Boltzmann}
\eeq

\noindent
where ${\vec E}$ denotes the external electric field and 
$\vec\omega_c$ is a vector with the magnitude equal to the cyclotron
frequency corresponding to an external magnetic field perpendicular to
the plane and the direction along the field.

Equations (\ref{eq:4.1}) and (\ref{Boltzmann}) neglect energy
dependence of the velocity of electrons, which makes it inapplicable
for quantities associated with electron-hole asymmetry, such as the
thermopower. On the other hand, any component of the thermal and
electrical conductivities is still within our description.

All of the interaction effects are taken into account in the collision
integral

\beq
\St\left\{f\right\} = \St_{el}\left\{f\right\} 
+ \St_{in}\left\{f\right\}.
\label{eq:4.3}
\eeq

The elastic part of the collision integral describes scattering of
electrons by static impurities (we assume point-like scattering;
generalization to the finite range is straightforward) as well as by
the self-consistent field generated by all the other electrons:

\beqa 
\St_{el}\left\{f\right\}=&& -\frac{f(t; \epsilon,\vec r,\vec n)
-\langle f(t; \epsilon, \vec r,\vec n)\rangle_n}{\tau} 
\nonumber\\
&&
\nonumber\\
&&
+ \frac{8}{\tau}I_0(t; \epsilon,\vec n, \vec r) 
\langle f(t; \epsilon, \vec r,\vec n)\rangle_n
\label{eq:4.4}
\\
&&
\nonumber\\
&&
+ \frac{8}{\tau}
n_\alpha I_1^{\alpha\beta} (t; \epsilon, \vec r) \langle n_\beta 
f(t;\epsilon, \vec r,\vec n)\rangle_n.  
\nonumber 
\eeqa 

\noindent
The effect of the self-consistent field is described by the last two
terms, where we introduce notations:

\begin{mathletters}
\label{eq:4.5}
\beqa
&&I_0(t; \epsilon,\vec n, \vec r)= -   
\int \frac{d\omega}{2\pi} 
\Big\{n_\alpha K_0^{\alpha\beta}(\omega)  
\langle n_\beta f(t; \epsilon-\omega, \vec r,\vec n)\rangle_n
\nonumber \\
&&
\nonumber \\
&&
+\frac{n_\alpha L_0^{\alpha\beta}(\w)}{2} 
\left[\nabla_\beta + e E_\beta\frac{\partial}{\partial \epsilon} \right] 
\langle f(t; \epsilon-\omega, \vec r,\vec n)\rangle_n
\Big\} 
\label{eq:4.5a}
\eeqa
\beqa
I_1^{\alpha\beta}(t; \epsilon, \vec r)= - 
\int \frac{d\w}{2\pi} K_1^{\alpha\beta}(\omega) 
\langle f(t; \epsilon-\omega, \vec r,\vec n)\rangle_n. 
\label{eq:4.5b}
\eeqa
\end{mathletters}

\noindent
The collision integral Eq.~(\ref{eq:4.4}) preserves the number of
particles on a given energy shell: integrating $\St_{el}\{f(t;
\epsilon,\vec r,\vec n)\}$ over directions of $\vec n$ gives zero for
any value of $\epsilon$ [see also Eq.(\ref{elastic2})].

The term $I_1$ expresses enhanced momentum relaxation due to static
disorder. The physics of this term was discussed in detail in Section
~\ref{qualitative}. The term $I_0$ describes electron scattering by
non-equilibrium non-local Fock like potential created by all other
electrons. This process is responsible for generation of the finite
drift velocity of electrons. One can easily see that $I_0$
vanishes in the equilibrium situation $f(\epsilon, \vec n, \vec r)=
f[\epsilon + e\varphi(\vec r)],\ \nabla_\alpha \varphi = - E_\alpha$.

The kernels $K_0$, $K_1$, and $L_0$ entering into Eqs.~(\ref{eq:4.5})
can be expressed in terms of interaction propagators and the
propagators describing 
semi-classical dynamics of non-interacting electrons. Explicitly:

\begin{mathletters}
\label{eqs4.6}
\beqa
&&K_1^{\alpha\beta}(\omega) =
\Im \int \frac{d^2q}{(2\pi)^2}  {\cal{D}}^R(\omega,\vec q) 
\label{K1}
\\
&&
\nonumber\\
&&\quad \times\left\{ \< n_\alpha D \> \< D n_\beta \> -  
\frac{\delta_{\alpha \beta}}{2} \left( \< D \> \< D \> + 
i \frac{\partial}{\partial \omega} \< D \> \right) \right\}
\nonumber
\eeqa
\beqa
&&
K_0^{\alpha\beta}(\omega)  = 
\Im \int \frac{d^2q}{(2\pi)^2} {\cal{D}}^R(\omega,\vec q)
\label{K0}\\
&&
\nonumber\\
&&
\quad \times 
\left\{ \< n_\alpha D n_\beta \>\<  D \>
 -\frac{i}{v_F}\frac{\partial}{\partial q_\alpha}  \< D n_\beta \>
- \< D n_\alpha \>\< D n_\beta \>
\right\} 
\nonumber
\eeqa
\beqa
&&
L_0^{\alpha\beta}(\omega) =
- \Re \int \frac{d^2q}{(2\pi)^2} {\cal{D}}^R(\omega,\vec q) 
\label{L0} 
\label{eq:4.6c}\\
&&
\nonumber\\
&&
\quad\times 
\left\{\<  D \>  \frac{\partial}{\partial q_\beta} \< n_\alpha D \>
-\< D n_\alpha \>\frac{\partial}{\partial q_\beta}\<  D \>
- \< D n_\alpha \frac{\partial}{\partial q_\beta} D \>\right\}
\nonumber
\eeqa
\label{kernels}
\end{mathletters}

Here, ${\cal D}^R(\omega)={\cal D}^A(-\omega)$ 
denotes the retarded interaction propagator [see
i.e. Eq.~(\ref{f-prop})] and we introduce the
short hand notation for the angular averaging

\begin{eqnarray*}
\<a  D b\> &\equiv& \int\frac{d \theta d\theta'}{(2\pi)^2}
a(\vec n)
D(\vec n, \vec n'; \omega, q)b(\vec n'), 
\\ 
&&
\nonumber\\
\<a  D b D c\> &\equiv& \int\frac{d \theta d\theta'd\theta''}{(2\pi)^3}
a(\vec n)
D(\vec n, \vec n')b(\vec n')D(\vec n', \vec n'')c(\vec n'')
\end{eqnarray*}

\noindent
for arbitrary functions $a,b$.  The function $D(\vec n, \vec n';
\omega, \vec q)$ describes the classical motion of a particle on the
energy shell $\epsilon_F$ in a magnetic field:

\beqa
&&\Big[-i\omega + i v_F \vec n \vec q + 
\vec \omega_c
\left(\vec n \times 
\frac{\partial}{\partial \vec n} \right)\Big] 
D(\vec n, \vec n'; \omega, \vec q)
\label{Diffuson}
\\ 
&&
\nonumber\\
&&
+\frac{1}{\tau} \Big( D(\vec n, \vec n'; \omega, \vec q)
 - \langle  D(\vec n, \vec n'; \omega, \vec q) \rangle_n \Big) = 
2\pi \delta(\widehat{\vec n \vec n'}).
\nonumber
\eeqa

As we have already mentioned, the elastic part of the collision
integral is nulled by a distribution function of the form 
$f[\epsilon +e\varphi(\vec r)]$ for an arbitrary $f$. 
It is the inelastic term
that is responsible for establishing the local thermal equilibrium
and it has the standard form

\begin{mathletters}
\label{eq:4.8}
\beqa
\St_{in}&&\left\{f\right\} =
\int d\omega \int d \epsilon_1 A(\omega) 
f( \epsilon_1) 
\left[1 -  f(\epsilon_1 - \omega)\right]
\label{inelastic}
\\
&&
\nonumber\\
&&
\times 
\Big\{ - f(\epsilon) 
\left[1 -  f( \epsilon + \omega)\right]
+ \left[1 - f( \epsilon)\right]f( \epsilon - \omega)
\Big\} 
\nonumber
\eeqa
\begin{eqnarray*}
f(\epsilon) = \langle f(t; \epsilon, \vec r,\vec n)\rangle_n.
\end{eqnarray*}

\noindent
The kernel $A(\omega)$ describes matrix elements for inelastic
processes in both ballistic and diffusive limits. The explicit
expression for this kernel is

\beq A(\omega) = \frac{2\nu}{\pi} \int\frac{d^2q}{(2\pi)^2} 
\left[ \Re \langle D \rangle \right]^2 
\left|{\cal{D}}^R(\omega, \vec q) \right|^2,
\label{eq:A}
\eeq
\end{mathletters}

\noindent
where $ \langle D \rangle$ is given by the solution to 
Eq.~(\ref{Diffuson}) averaged over  angles.

The above equations are written for the interaction in the singlet
channel only. In a situation where both triplet and singlet channels
are present, but the distribution function does not have a spin
structure (no Zeeman splitting or non-equilibrium spin occupation
present), one has to replace

\beq
{\cal D}^R \to {\cal D}^R +  {\bf Tr} \widehat{\cal D}^R_T
\label{r1}
\eeq

\noindent
in Eqs.~(\ref{eqs4.6}) and

\beq
|{\cal D}^R|^2 \to |{\cal D}^R|^2 + {\bf Tr}
\left\{ \widehat{\cal D}^R_T 
\left[\widehat{\cal D}^R_T\right]^\dagger\right\}
\label{r2}
\eeq

\noindent
in Eq.~(\ref{eq:A}).

Equations (\ref{eq:4.2}) - (\ref{eq:4.8}) constitute the complete
system of transport equations with the leading interaction corrections
taken into account. They may be used to study both linear and
non-linear response. We reiterate that they do not include effects of
electron-hole asymmetry and in this form can not produce finite
thermopower. The Hall effect, the thermal conductivity, and energy
relaxation, however, are  included and will be studied in a
subsequent publication \cite{pre}. In the following Subsection we
apply the kinetic equation approach to study the interaction
correction to the conductivity at intermediate and low temperatures
and reproduce the results obtained in Section~\ref{diagrams} by means
of diagrammatic technique. The reason for doing so is to show how the
kinetic equation works and to demonstrate explicitly that both
approaches are equivalent.

Closing our description of the structure of the kinetic equation, we
discuss the range of its applicability.  Any kinetic equation implies
that the distribution function changes slowly
on the spatial scale of the
Fermi wave length $\lambda_F$ and on the time scale $1/\epsilon_F$. In
our case, the conditions are more restrictive.  First, in the
interaction correction to the elastic collision integral we take
into account only the effect of the interaction on the zeroth and
first angular harmonics of the distribution function.  This implies
that the equation gives the correct description for the interaction
effects on the conductivity and diffusion, whereas it is not correct
for description of the quantities involving higher angular
harmonics. Second, we made a gradient expansion in Eq.~(\ref{eq:4.5a})
and only took into account terms linear in the electric field. This
implies that the distribution function changes slowly on the spatial
scale $L_T = {\rm min}(\hbar v_F/T,\ v_F (\hbar\tau/T)^{1/2}),$ and
on the time scale of the order of $\hbar/T$. The electric field
expansion is justified by the condition $eE L_T \ll T$.  One can check
that both these conditions are satisfied, if the energy relaxation
time is much longer than the time for the elastic collisions.
We also did not include quantum effects of the magnetic field.
This is justified at $\omega_c \ll max(T/\hbar, \tau^{-1})$. 

Finally, the interaction part of Eq.~(\ref{eq:4.4}) is calculated in
the first loop approximation. It means, that it has to be considered
as the first order correction to $1/\tau$. If one is interested in the
next order interaction correction to the elastic part, one should
take into account the second loop correction, which is not considered
in the present paper. On the contrary, the inelastic part (\ref{eq:4.8})
can be considered in all orders to find the zero angular momentum
part of the distribution function; the only assumption here is the
validity of the Fermi liquid description at energies smaller than
$\epsilon_F$.

\subsection{Conductivity calculation}
\label{cond}

In order to calculate the conductivity at zero magnetic field $\omega_c
=0$ we look for the solution of
Eqs.~(\ref{eq:4.2}) - (\ref{eq:4.4})  in a form

\beq
f(\vec n,\epsilon) = f_F(\epsilon) + \vec n 
\vec \Gamma
(\epsilon),
\label{eq:4.9}
\eeq 

\noindent
where $f_F(\epsilon)=1/(e^{\epsilon/T}+1)$ is the Fermi distribution
function (all the energies are counted from the Fermi level), and 
$\Gamma$ is the quantity to be found 
and it is proportional to the electric field.

We substitute Eq.~(\ref{eq:4.9}) into Eqs.~(\ref{eq:4.2}),
(\ref{eq:4.4}), (\ref{eq:4.5}) and (\ref{eq:4.8}). The inelastic part
of the collision integral [see Eq.~(\ref{eq:4.8})] obviously vanishes,
as effects of the heating are proportional to at least the second
power of the electric field. As a result, we obtain an equation for
$\Gamma$:

\beqa
e v_F E_\alpha \frac{\partial f_F(\epsilon) }{\partial \epsilon}
&&= - \frac{ \Gamma_\alpha (\epsilon)}{\tau} 
\label{eq:4.10}\\
&&
\nonumber\\
&&
- \frac{4}{\tau}\int\frac{d\omega}{2\pi}
\Big[ K_1^{\alpha\beta}(\omega) 
f_F(\epsilon-\omega)\Gamma_\beta (\epsilon) 
\nonumber\\
&&
\nonumber\\
&&
\quad\quad\quad\quad+
K_0^{\alpha\beta}(\omega) f_F(\epsilon)\Gamma_\beta (\epsilon-\omega)
\Big] 
\nonumber\\
&&
\nonumber\\
&&
-\frac{4f_F(\epsilon)}{\tau}\int\frac{d\omega}{2\pi}
L_0^{\alpha\beta}(\omega)
e E_\beta\frac{\partial}{\partial \epsilon} 
f_F(\epsilon-\omega)
\nonumber
\eeqa

\noindent
We solve Eq.~(\ref{eq:4.10}) by iterations. As usual for kinetic
equations, the solution is expressed in terms of the unperturbed
distribution function $f_F(\epsilon)$ and the kernels, which in this
case are given by Eq.~(\ref{kernels}):

\beqa
\Gamma_\alpha (\epsilon)
&&
= -  e v_F\tau E_\alpha \frac{\partial f_F(\epsilon) }{\partial \epsilon}
\label{eq:4.11}
\\
&&
\nonumber\\
&&
+ {4 e v_F\tau}\int\frac{d\w}{2\pi}
\left[ K_1^{\alpha\beta}(\w) 
f_F(\epsilon-\w)\frac{\partial f_F(\epsilon) }{\partial \epsilon} 
\right. 
\nonumber\\
&&
\nonumber\\
&&
\left.
\quad\quad\quad\quad\quad\quad+
K_0^{\alpha\beta}(\w) 
f_F(\epsilon)\frac{\partial f_F(\epsilon-\w) }{\partial \epsilon}
\right] E_\beta 
\nonumber\\
&&
\nonumber\\
&&
- {4f_F(\epsilon)}\int\frac{d\w}{2\pi}
L_0^{\alpha\beta}(\w)
e E_\beta\frac{\partial}{\partial \epsilon} 
f_F(\epsilon-\w)
\nonumber
\eeqa

\noindent
Substituting Eqs.~(\ref{eq:4.11}) into Eq.~(\ref{eq:4.9}) and
the result into Eq.~(\ref{current}), we 
integrate over $\epsilon$ and find  
the conductivity 

\begin{mathletters}
\beqa
\sigma = \sigma_D + \delta \sigma,
\label{eq:4.12}
\eeqa
\begin{eqnarray}
\frac{\delta\sigma}{\sigma_D}=
\int\limits_{-\infty}^\infty\frac{d\omega}{\pi}
\frac{\partial}{\partial\omega} &&
\left(\omega\coth \frac{\omega}{2T}\right)
\nonumber\\
&&
\nonumber\\
&&\times
\left[K_0(\omega) - K_1(\omega) -\frac{L_0(\omega)}{v_F\tau}\right],
\label{ccke}
\end{eqnarray}
\end{mathletters}

\noindent
where the Drude conductivity is $\sigma_D = e^2\nu v_F^2\tau/2$. 
Here we used the fact that in the absence of the magnetic field
all the kernels are diagonal, $K^{\alpha\beta}_{i} = 
\delta_{\alpha\beta}K_i$, 
$L^{\alpha\beta}_{0} = \delta_{\alpha\beta}L_0$.
We also used the identities

\begin{eqnarray*}
2\int\limits_{-\infty}^\infty d \epsilon f_F(\epsilon)
\frac{\partial f_F(\epsilon-\omega) }{\partial \epsilon}
=-1+\frac{\partial}{\partial\omega} 
\left(\omega\coth \frac{\omega}{2T}\right),
\end{eqnarray*}
\begin{eqnarray*}
\int\limits_{-\infty}^\infty 
d\omega K_{i}(\omega)= \int\limits_{-\infty}^\infty 
 d\omega L_{0}(\omega) = 0.
\end{eqnarray*}

In order to derive explicit expressions for the kernels $K_i$ and $L_0$ 
we have to solve Eq.~(\ref{Diffuson}) for the function $D$ in the absence 
of the magnetic field. 
The result can be written as
 
\beqa \label{eq:4.13}
D(\vec n,\vec n';\omega,\vec q) && = 
2\pi\delta(\widehat{\vec n  \vec n'})
D_0(\vec n,\omega,\vec q)
\\
&&
\nonumber\\
&& 
+D_0(\vec n,\omega,\vec q)D_0(\vec n',\omega,\vec q)
\frac{C}{C\tau-1}, 
\nonumber
\eeqa

\noindent
where $D_0$ denotes the solution of Eq.~(\ref{Diffuson}) without the 
angular averaged term (and in the absence of the magnetic field)

\beqa
D_0(\vec n,\omega,\vec q)=
\frac{1}{ -i\omega + i v_F \vec n \vec q+1/\tau}.
\nonumber
\eeqa

\noindent
Here we used the short-hand notation

\[
C=\sqrt{\left(-i\omega + 1/\tau\right)^2 + v_F^2 q^2},
\]

\noindent
which is similar to the notation $S$ used in Section~\ref{diagrams}
[in fact, $C=S^*$, see Eq.~(\ref{sqrt})].
Substituting Eq.~(\ref{eq:4.13}) into Eqs.~(\ref{K1}) -- (\ref{L0})
and performing the angular integration we arrive to

\begin{mathletters}
\beqa
&&K_1(\omega)=  -{\Im} \int \frac{q dq}{4\pi} {\cal D}^R(\omega, q) 
\\ 
&& 
\nonumber\\
&& \; \; 
\times\left\{ \frac{1}{v_F^2 q^2} 
\left( \frac{C - (-i\omega + 1/\tau)}{C-1/\tau} \right)^2 
+ \frac{C- (-i\omega +1/\tau)}{C (C-1/\tau)^2} \right\}, 
\nonumber\\ 
&& 
\nonumber\\
&&K_0(\omega)= {\Im} \int \frac{q dq}{4\pi}\ {\cal D}^R(\w, q)  
\\ 
&& 
\nonumber\\
&& \quad
\times\left\{ \frac{C-(-i\w + 1/\tau)}{C (C-1/\tau)^2} + 
\frac{(C-(-i\w + 1/\tau))^2}{C(C-1/\tau)} \frac{1}{v_F^2 q^2} \right\}, 
\nonumber\\ 
&& 
\nonumber\\
&&\frac{L_0(\omega)}{v_F\tau} = - {\Im} \int \frac{q dq}{4\pi} 
{\cal D}^R(\w, q) 
\\ 
&&
\nonumber\\
&& \quad\quad\quad\quad
\times\left\{ \frac{3}{2 \tau} \frac{v_F^2 q^2}{C^3 (C-1/\tau)^2} + 
\frac{v_F^2 q^2}{C^3} \frac{1/\tau^2}{(C-1/\tau)^3}
\right\}. 
\nonumber
\eeqa
\label{fullkern}
\end{mathletters}

\noindent
Together with the conductivity correction Eq.~(\ref{ccke}) the above 
expressions Eq.~(\ref{fullkern}) are identical to Eq.~(\ref{fock-int})
obtained in Section~\ref{diagrams} by means of the standard perturbation
theory. Thus the kinetic equation approach is completely equivalent to
such diagrammatic calculation. 

Integration over the wave vector $q$ requires the knowledge of the
interaction propagator. Substituting Eq.~(\ref{f-prop}) for the singlet 
channel and Eq.~(\ref{trip-prop}) for the triplet channel and performing 
the straightforward integration we arrive to the results in 
Section~\ref{results}.

\subsection{Derivation of the kinetic equation}
\label{derivation}

In this section we derive the kinetic equation discussed in 
Section~\ref{kinetic}. For simplicity we show the derivation for the
case of the singlet channel interaction Eq.~(\ref{Hrho}). The case
of the triplet channel can be treated in the same manner with minor
differences (introduction of extra spin indices) described in the end 
of this Section. To keep the discussion at the same level as in 
Section~\ref{diagrams}, we treat the Fermi liquid parameter $F^\rho$
in Eq.~(\ref{Hrho}) as a constant, similar to our treatment of the
triplet channel in Section ~\ref{triplet}.

\subsubsection{Keldysh formalism}

Here we summarize the results originally obtained by Keldysh
\cite{Keldysh} that enable us to calculate correlation functions for
any non-equilibrium distribution.

Let us first consider a Green's function of
electrons before disorder averaging.
The electron-electron interaction
is described by the Hamiltonian Eq.~(\ref{Hrho}). In the path-integral
formulation it can be decoupled from fermion operators using an 
auxiliary bosonic field $\phi$. Then the Green's function can be written
as

\beq 
\label{fullgreen}
\widehat G(x_1, x_2) = \int \left[ {\cal D} \phi \right] 
\widehat G(x_1, x_2 | \phi) e^{-i S_B[\phi]},
\eeq

\noindent 
with the action defined as 

\beq
S_B[\phi] = \int\limits_{-\infty}^{\infty} dt d^2r 
\left\{\frac{1}{2} \phi^T V_0^{-1} \sigma_3 \phi \right\}
+ i \log Z[\phi] ,
\label{sb}
\eeq

\noindent
where $-V_0$ is the (bare; following Eq.~(\ref{Hrho})
$V_0=V(q) +F_0^\rho/\nu$) electron-electron interaction propagator 
and $Z$ is the partition function, 

\beqa
Z[\phi] = \< {\bf T_C } e^{-i S_F[ \phi, \psi]} \> 
\label{z}
\eeqa
\beqa
S_{F}[ \phi, \psi] = 
\int\limits_{-\infty}^{\infty} dt d^2r \left\{  
 \psi^\dagger \phi_{\alpha} \hat\gamma^{\alpha} \psi
\right\},
\label{sfu}
\eeqa

\noindent
where $\hat\sigma_z={\rm diag}(-1,1)$ is the Pauli matrix in the Keldysh
space.

\begin{figure}
\vglue 0cm
\hspace{0.01\hsize}
\epsfxsize=0.9\hsize
\epsffile{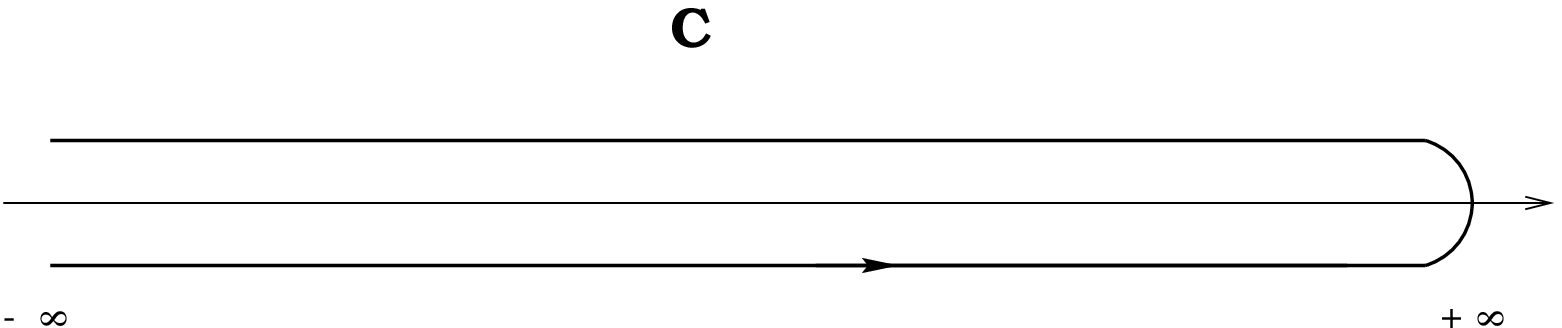}
\refstepcounter{figure} \label{contour}
{\center{\small Fig.\ \ref{contour}.
The Keldysh contour 
  \par}}
\end{figure}

\noindent
In the above expressions all the fields are defined on the Keldysh time
contour shown in Fig.~\ref{contour}. In particular, the fermionic fields
$\psi^\dagger$ and $\psi$ (as well as the bosonic field $\phi$) can be 
treated as doublets 

\beqa
\psi = \left( \matrix{\psi_+ \cr \psi_-} \right),
\eeqa

\noindent 
where we adopt the notation that fields
with a $-$ ($+$) subscript (also referred to by Greek letters in this 
Section) reside on the lower (upper) part of the
contour on Fig.~(\ref{contour}). 
The time dependent fermionic operators $\psi$ are taken in the
interaction represntation
\[
-i\partial_t \psi(t) = \left[\hat{H}_1(t); \psi (t)  \right],
\]
where $\hat{H}_1$ is the one-electron Hamiltonian which includes
the static disorder potential as well as external fields.

Consequently, the Green's function in
Eq.~(\ref{fullgreen}) is a $2\times2$ matrix. Time ordering along the 
contour is denoted in Eq.~(\ref{z}) by ${\bf T_C }$. 
Matrices $\hat\gamma^{\alpha}$ in Eq.~(\ref{sfu}) are defined as

\[
\hat\gamma^+ = \left(
\matrix{
-1 & 0 \cr
0 & 0
} 
\right)
\quad ;
\quad\quad
\hat\gamma^- = \left(
\matrix{
0 & 0 \cr
0 & 1
} 
\right)
\]

The Green's function $\widehat G(x_1, x_2 | \phi)$ in Eq.~(\ref{fullgreen})
is given by

\beq
\widehat G(x_1, x_2 | \phi) = \frac{1}{Z[\phi]} 
\< T_C \psi^\dagger_{\alpha}(x_1) \psi_{\beta}(x_2) 
e^{-i S_F[ \phi, \psi]} \>.
\label{fg2}
\eeq

\noindent
Here, as well as in Eq.~(\ref{z}) the angular brackets 
$\langle \dots \rangle$ denote quantum-mechanical averaging.  
In this section we will use the short hand notation

\[
x_i \equiv (t_i, \vec r_i).
\]

The bosonic action Eq.~(\ref{sb}) can be treated in the 
saddle point approximation:

\begin{mathletters}
\beqa
\< e^{-i S_B[\phi]}\> = e^{-i F[\phi]},
\eeqa
\beqa
F[\phi] = F[\phi = 0] + 
\frac{1}{2}\phi^T \widehat\Pi \phi + {\cal O}(\phi^3),
\label{fb}
\eeqa
\label{ba}
\end{mathletters}

\noindent
where $\Pi$ is the electronic polarization operator, defined as

\beq
\Pi_{\alpha \beta}(x_1, x_2) = \left.  
\frac{\delta^2 F}{\delta\phi_{\alpha}(x_1)\delta\phi_{\beta}(x_2)}
\right|_{\phi=0}.
\label{pder}
\eeq

\noindent
The quadratic expansion in Eq.~(\ref{fb}) is justified, provided that
the fields $\phi$ are slowly changing  on the scale much larger
than $\lambda_F$.

Let us now average the Green's function Eq.~(\ref{fullgreen}) over 
disorder: 

\beq 
\label{disorderedgreen}
\< \widehat G(x_1, x_2) \>_{dis} = 
\int \left[ {\cal D} \phi \right] 
\< \widehat G(x_1, x_2 | \phi) \>_{dis} e^{-i \< S_B[\phi]\>_{dis}},
\eeq
where $\<\dots  \>_{dis}$ hereafter denotes averaging over disorder.

Here we average the electronic Green's function Eq.~(\ref{fg2}) 
separately from the bosonic action Eq.~(\ref{ba}). This approximation
means that we neglect correlations between mesoscopic
fluctuations of the polarizability
 in Eq.~(\ref{fb})  and 
the fermionic operators in Eq.~(\ref{fg2}) (which describe the motion of 
conduction electrons).
This is the same approximation we used in Section~\ref{diagrams}.
It is justified by the well known fact that mesoscopic
fluctuations are smaller than average quantities by a factor of the 
order $1/(E_F\tau)^2$. 

It is convenient \cite{LO77} to rotate the Keldysh basis as follows 

\beqa
\widehat G \rightarrow \frac{1}{2}\hat\sigma_x 
\left(
\matrix{
1 & -1 \cr
1 & 1
} 
\right)
\widehat G 
\left(
\matrix{
1 & 1 \cr
-1 & 1
} 
\right).
\eeqa

\noindent
In the new basis the Green's function Eq.~(\ref{fg2}) has the form

\beq
\widehat G(x_1, x_2 | \phi) = 
\left(
\matrix{
G^R(x_1, x_2 | \phi) & G^K(x_1, x_2 | \phi) \cr
G^Z(x_1, x_2 | \phi) & G^A(x_1, x_2 | \phi)
} 
\right).
\label{nosense}
\eeq

\noindent
After the averaging over the bosonic field and over the disorder
according to Eq.~(\ref{disorderedgreen}) the entries in
Eq.~(\ref{nosense}) acquires the following meaning/ where after
integrating over the bosonic field $\phi$ the diagonal elements
$G^{R(A)}$ become the retarded (advanced) Green's functions of the
electron system

\begin{eqnarray*} 
\< G^R(t_1, t_2) \> = -i\eta(t_1-t_2)\langle \psi (t_1) \psi^\dagger(t_2)+
\psi^\dagger(t_2) \psi (t_1) \rangle,\\
\< G^A(t_1, t_2) \> = 
i\eta(t_2-t_1)\langle \psi (t_1) \psi^\dagger(t_2)+
\psi^\dagger(t_2) \psi (t_1) \rangle,
\end{eqnarray*} 

\noindent
where $\eta(t)$ is the Heaviside step function.  The lower diagonal
element vanishes due to the causality,

\[
\< G^Z(t_1, t_2) \>=0,
\]

\noindent 
even before the disorder averaging.  Finally, the upper off-diagonal
element (the so-called Keldysh Green's function) is related to the one
particle density matrix

\begin{eqnarray}
\< G^K(t_1, t_2) \> = -i \langle \psi (t_1) \psi^\dagger(t_2)-
\psi^\dagger(t_2) \psi (t_1) \rangle,
\label{GK}
\end{eqnarray}

\noindent
the quantum mechanical averaging is performed with an arbitrary
distribution function to be found from the solution of the kinetic
equation.

The bosonic field in the rotated basis has the two components

\beqa
\phi_{1(2)} = \frac{1}{2} \left( \phi_+ \pm \phi_- \right)
\eeqa

\noindent
which are described by the propagators
 
\begin{mathletters}
\label{Ds}
\beqa
&& 
\< \phi_1(t_1, \vec r_1) \phi_1(t_2, \vec r_2) \> = 
\frac{i}{2} {\cal D}^K(t_1,t_2; \vec r_1,\vec r_2), 
\\
&& 
\nonumber\\
&&
\< \phi_1(t_1, \vec r_1) \phi_2(t_2, \vec r_2) \> = 
\frac{i}{2} {\cal D}^R(t_1,t_2; \vec r_1,\vec r_2), 
\\
&& 
\nonumber\\
&&
\< \phi_2 (t_1, \vec r_1)\phi_1(t_2, \vec r_2) \> =
\frac{i}{2} {\cal D}^A(t_1,t_2; \vec r_1,\vec r_2), 
\\
&& 
\nonumber\\
&&
\< \phi_2 (t_1, \vec r_1)\phi_2(t_2, \vec r_2) \> = 0.
\label{do}
\eeqa
\label{prp}
\end{mathletters}

\noindent
The coupling Eq.~(\ref{sfu}) between the fermionic and bosonic fields 
in the rotated basis has the form:

\beq
\psi^\dagger \phi_{\alpha} \hat\gamma^{\alpha} \psi \rightarrow
\psi^\dagger
\left(
\matrix{
\phi_1 & \phi_2  \cr
\phi_2 & \phi_1
} 
\right) \psi.
\eeq

\noindent
The propagators Eq.~(\ref{prp}) are solutions of the Dyson equations

\beqa
\hat{\cal D}(1,2) = \hat{\cal D}_0(1,2) + \int d 3 d4 
\hat{\cal D}_0(1,3)\hat{\Pi}(3,4)\hat{\cal D}(4,2) 
\nonumber
\eeqa
\beqa
\hat{\cal D} =
\left(
\matrix{{\cal D}^R & {\cal D}^K \cr 0 & {\cal D}^A}
\right),
\quad
\hat{\Pi} =
\left(
\matrix{\Pi^R & \Pi^K \cr 0 & \Pi^A}
\right)
\label{Dysonb}
\eeqa

\noindent
and we introduced the short hand notation $(i) \equiv (t_i,\vec r_i)$.
The bare interaction propagators are

\beqa
{\cal D}_0^R = {\cal D}_0^A && = - \left[V(\r_1-\r_2) + 
F_0^\rho \delta(\r_1-\r_2)
\right] \delta(t_1-t_2), 
\nonumber \\
&&
\nonumber \\
{\cal D}_0^K && = 0.
\label{D00}
\eeqa

Any classical external field takes identical values on the two
branches of the contour and, hence, in the rotated basis has only a
diagonal component.

The matrix Green's function (\ref{nosense}) satisfies the equation

\beqa
&&\Bigg\{
i\partial_{t_1} + E_F
- \frac{\left[-i\vec\nabla_{r_1} + \vec A_{ext}(x_1)\right]^2}{2m}
- \hat{\phi}(x_1)
\label{eq1}
\\
&&
\nonumber\\
&&
\quad 
- U(\vec r_1) - \varphi_{ext}(x_1)
\Bigg\} \;
\widehat G(x_1, x_2 | \phi) = \hat{I}\delta(x_1-x_2)
\nonumber
\eeqa

\noindent
where  $U(\vec r)$ is the potential due to the static disorder,
$\vec A_{ext}(x_1)$ and $\varphi_{ext}(x_1)$ are the vector and
scalar potential due to the external electric and magnetic fields.

\beq
e{\vec E}= \partial_t {\vec A}_{ext} -
\vec\nabla
\varphi_{ext}, \quad
e{\vec B}=-\frac{1}{c}\vec\nabla \times {\vec A}_{ext} 
\label{EH}
\eeq

\noindent
The equation (\ref{eq1}) is the basis for the further consideration.
One can perform the disorder average in Eq.~(\ref{eq1}) in the leading
in $1/(E_F\tau)$ approximation, which amounts to summation over all
the non-intersecting impurity lines one obtains

\beqa
&&
\Bigg\{
i\partial_{t_1} + E_F 
- \frac{\left[-i\vec\nabla_{r_1} + \vec A_{ext}(x_1)\right]^2}{2m}
- \hat{\phi}(x_1)
\nonumber\\
&&
\nonumber\\
&&\quad\quad
- \varphi_{ext}(x_1)
\Bigg\} \;
\widehat G(x_1, x_2 | \phi) = 
\label{eq2}
\\
&&
\nonumber\\
&&\quad\quad\quad\quad
\hat{I}\delta(x_1-x_2) +
\int d x_3 \widehat\Sigma (x_1, x_3 | \phi) \widehat G(x_3, x_2 | \phi);
\nonumber
\eeqa
\beqa
\widehat\Sigma (x_1, x_2 | \phi) = \frac{\delta(r_1-r_2)}{2\pi\nu \tau}
\widehat G(x_1, x_2 | \phi).
\nonumber
\eeqa

\noindent
The equation (\ref{eq2}) allows for semi-classical treatment 
introduced  in Refs.~\onlinecite{Eilenberger,68}, and described in
great details in Ref.~\onlinecite{LO77}. 
Since we have already averaged
the equation of motion over disorder, the semi-classical approximation 
now amounts to averaging the Green's function $\widehat G(x_1, x_2 | \phi)$
over the distance from the 
Fermi surface. This is done in two steps:

\beqa
\widehat G(t_1, t_2;\vec p; \vec R)=
\int d^2r e^{i\vec P \cdot \vec r} \widehat G(x_1, x_2 | \phi),
\eeqa
\[
\vec r = \vec r_1 - \vec r_2 ; \quad 
\vec R = \frac{1}{2}(\vec r_1 + \vec r_2);
\]
\[
\vec P = \vec p - \frac{1}{2}
\left[\vec A_{ext}\left(t_1, \vec R\right)+
\vec A_{ext}\left(t_2, \vec R\right)\right];
\]
\beqa
\label{intg}
\hat g(t_1, t_2; \vec n, &&\vec r) = 
\\
&&
\nonumber\\
&&
\frac{i}{\pi} 
\int\limits_{-\infty}^{\infty} d\xi \widehat G\left(t_1,
t_2; \vec n \left[p_F + \frac{\xi}{v_F}\right]; \vec r\right),
\nonumber
\eeqa

\noindent
Since we follow the avenue of Ref.~\onlinecite{LO77}, we will skip
further intermediate steps, and use the semiclassical equation written
in the next subsection.

\subsubsection{Eilenberger equation}

The dynamics of the electron matrix Green's function is then described
by the Eilenberger equation\cite{Eilenberger}:

\beq 
\left[
{\tilde\partial_{t}} + v_F \vec n \tilde{\vec\nabla}  
+ \vec \omega_c
\left(\vec n \times 
\frac{\partial}{\partial \vec n} \right)
\right] \hat{g} = 
\frac{\hat{g}\< \hat{g} \>_n - \< \hat{g} \>_n\hat{g}}{2\tau},
\label{eil1} 
\eeq

\noindent
where angular averaging is defined as before 

\[
\langle \dots \rangle_n =
\int 
\frac{d\theta}{2\pi} \dots,
\quad \vec n = (\cos\theta, \sin\theta) ,
\]

\noindent
and the covariant derivatives in Eq.~(\ref{eil1}) are defined as

\begin{mathletters}
\beqa
\tilde{\partial_t} \hat{g} = 
\partial_{t_1} \hat{g}  + \partial_{t_2} \hat{g} + 
i\hat{\varphi}(t_1,r)\hat{g} - 
i\hat{g}\hat{\varphi}(t_2,r),
\eeqa
\beq
\tilde{\vec\nabla}
\hat{g} = 
\vec\nabla
\hat{g} + 
i\hat{\vec  A}(t_1, \vec r)\hat{g} -
i\hat{g} \hat{\vec A}(t_2,\vec r).
\eeq
\label{covariant}
\end{mathletters}

\noindent
Here $\hat{g}$ is a matrix in Keldysh space, 

\beq \hat{g}(t_1, t_2; \vec n, \vec r) = 
\left(
\matrix{
g^R & g^K \cr
g^Z & g^A
} 
\right),
\eeq

\noindent
and we will suppress the coordinate and the time arguments unless
otherwise is stated. A product of such matrices should be understood
as a matrix product in Keldysh space and a convolution in time:

\beqa
\label{convolution}
&&\big[\hat{g}(\vec n, \vec r)\hat{g}(\vec n_1, \vec r)\big]_{ij} 
\equiv 
\\
&&
\nonumber\\
&&
\quad 
\int\limits_{-\infty}^\infty 
d t_3 \sum_k\big[\hat{g}(t_1, t_3;\vec n, \vec r)\big]_{ik}
\big[\hat{g}(t_3, t_2; \vec n_1, \vec r)\big]_{kj},
\nonumber
\eeqa

\noindent
and solutions of the homogeneous equation ~(\ref{eil1}) are subject 
to the constraints

\beq
\label{constraint}
\hat{g}(\vec n, \vec r)\hat{g}(\vec n, \vec r)= 
{\hat I}^K, \quad
\int\limits_{-\infty}^\infty dt \; 
{\bf Tr} \; \hat{g}(t,t;\vec n, \vec r)=0,
\eeq 

\noindent
where

\[
[ I^K ]_{ij} = \delta_{ij} \delta(t_1-t_2).
\]

\noindent
The scalar and vector potentials in Eq.~(\ref{covariant}) have the
following structure in the Keldysh space

\beqa
&&\hat{\vec  A}(t,\vec r) =
\left(
\matrix{
{\vec  A}_{ext}(t,\vec r)  & 0 \cr
0  & {\vec  A}_{ext}(t,\vec r) 
} 
\right), 
\label{fields}
\\
&&
\nonumber\\
&&
\hat{\varphi}(t,\vec r) =
\left(
\matrix{
{\varphi}_{ext}(t,\vec r) +\phi_{1}(t,\vec r)  &  \phi_{2}(t,\vec r)\cr
\phi_{2}(t,\vec r)  &  {\varphi}_{ext}(t,\vec r) + \phi_{1}(t,\vec r) 
} 
\right),
\nonumber
\eeqa

\noindent
where $\varphi_{ext}$ and ${\vec A}_{ext}$ are the external (classical)
potentials due to the electric field ${\vec E}$,

\beq
e{\vec E}= \partial_t {\vec A}_{ext} 
-\vec\nabla
\varphi_{ext}
\label{E}
\eeq

\noindent
acting on the electron system, and $\phi_{1,2}(t,\vec r)$ are the
auxiliary fluctuating fields decoupling the interaction in the singlet
channel. Because the singlet channel describes processes with small
momentum transfers (smaller than $q*$, see Section~\ref{soft}), the
fields $\phi_{1,2}(t,\vec r)$ vary slowly on the scale of the $1/q*$.

The condition (\ref{do}) enforces causality of the physical response
functions. It is worth noticing that the decoupling of interaction can
be performed also using a fluctuating vector potential; our choice is
strictly a matter of taste.

In this formalism any observable quantity described by one electron
operator ${\cal O}(\hat{p},\hat{r})$ is given by [see Eqs.~(\ref{GK})
and (\ref{intg})]

\beqa
{\cal O}(t,{\vec r}) = -
&&
\nu \int \frac{d\theta}{2\pi}{\cal O}(p_F{\vec n},{\vec r}) 
\label{O}
\\
&&
\nonumber\\
&&
\times\lim_{t_1\to t} \left[ 
\frac{\pi }{2}
\langle g^K(t_1, t;  \vec n, \vec r)\rangle_{\phi} +
{\varphi}_{ext}(t,{\vec r})
\right],
\nonumber
\eeqa

\noindent
where $\langle\dots \rangle_{\phi}$ stands for averaging over both
auxiliary fields $\phi_{1,2}$ fluctuating according to
Eqs.~(\ref{Ds}).  The last term in brackets is a consequence of the
ultraviolet anomaly, and its form is enforced by the requirement of
the gauge invariance.

Finally, the electronic polarization operators are determined [see
Eqs.~(\ref{pder}) and (\ref{intg})] as variational derivatives of the
solutions to the Eilenberger equation (\ref{eil1}):

\beqa 
&&{\Pi}^R(1,2)= {\Pi}^A(2,1) 
\label{Pi}
\\
&&
\nonumber\\
&&
\quad\quad\quad\quad
=\nu \int \frac{d\theta}{2\pi}
\left(\delta_{12} + 
\frac{\pi \langle \delta g^K(t_1, t_1; \vec n, \vec r_1)\rangle_{\phi}}
{2\delta \phi_1(t_2,\r_2)}\right); 
\nonumber\\
&&
\nonumber\\
&&
{\Pi}^K(1,2)=\nu\pi 
\nonumber\\
&&
\nonumber\\
&&\quad\times
\int  \frac{d\theta}{2\pi}
\frac{ \langle \delta g^K(t_1, t_1; \vec n, \vec r_1)
+\delta g^Z(t_1, t_1;  \vec n, \vec r_1)
\rangle_{\phi}}
{2\delta \phi_2(t_2,\vec r_2)}.
\nonumber
\eeqa

\subsubsection{Derivation of the kinetic equation}

Our goal now is to obtain an equation for the Keldysh function
averaged over the fluctuating fields, $\langle g^K(t_1, t_1; \vec n,
\vec r_1)\rangle_{\phi}$. It is this quantity that determines physical
observables, see Eq.~(\ref{O}).  We will do this using the
non-crossing approximation for bosonic propagators (i.e. the first
loop approximation for the collision integral), see
Fig.~\ref{15}. This approximation is justified provided that the
resulting dynamics for the electrons (characterized by time
$\tau_\epsilon$) is slow in comparison with motion of relevant bosonic
mode, $T\tau_\epsilon \gg 1$.

To do so, we notice that only two components of the matrix $\hat{g}$
are independent, and the other two are fixed by the constraint
(\ref{constraint}). For our purposes, we choose to fix the diagonal
components

\beq
\label{normsolution}
g^R=\sqrt{1-g^K g^Z};
\quad g^A= -\sqrt{1-g^Z g^K},
\eeq

\noindent
where the square root should be understood in operator sense: as a sum
of its Taylor series, with all arising products hereafter being time
convolutions, similar to Eq.~(\ref{convolution}). The two remaining
independent components of the Eilenberger equation have the explicit
form

\begin{mathletters}
\label{exact}
\beqa 
&&\left[
\tilde\partial_t + v_F \vec n  
\tilde{\vec\nabla}
+\vec\omega_c
\left(\vec n \times 
\frac{\partial}{\partial \vec n} \right)
\right] g^Z = 
\label{exactZ}
\\
&&
\nonumber\\
&&\;
- i\big[\phi_1(t_1,\vec r)-\phi_1(t_2,\vec r) \big] g^Z
- i\phi_2(t_1,\vec r) g^R + i g^A \phi_2(t_2,\vec r)
\nonumber \\ 
&&
\nonumber \\ 
&&\quad\quad
+  
\frac{1}{2 \tau} 
\big[
g^Z \< g^R \>_n - \<g^Z\>_n g^R 
+ g^A \< g^Z \>_n - \< g^A \>_n g^Z 
\big], 
\nonumber
\eeqa
\beqa 
&&\left[
{\tilde\partial_{t}} + v_F \vec n 
\tilde{\vec\nabla}
+\vec\omega_c
\left(\vec n \times 
\frac{\partial}{\partial \vec n} \right)
\right]g^K
=
\label{exactK}
\\
&&
\nonumber \\ 
&&\;
- i\big[\phi_1(t_1,\vec r)-\phi_1(t_2,\vec r) \big] g^K
- i\phi_2(t_1,\vec r) g^A + i g^R \phi_2(t_2,\vec r)  
\nonumber\\
&& 
\nonumber \\ 
&&\quad\quad
+  
\frac{1}{2 \tau} 
\big[
g^K \< g^A \>_n - \<g^K\>_n g^A 
+ g^R \< g^K \>_n - \< g^R \>_n g^K 
\big],
\nonumber
\eeqa
\end{mathletters}

\noindent
and we redefine the covariant derivatives
Eq.~(\ref{covariant}) to include only the external scalar and vector
potentials:

\beqa 
&&\tilde{\partial_t} {g} = \partial_{t_1} {g} +
\partial_{t_2} {g} + i\left[{\varphi}_{ext}(t_1,r)
-{\varphi}_{ext}(t_2,r) \right]{g},
\label{covariant2}
\\
&&
\nonumber\\
&&
\tilde{\vec\nabla}
{g} = 
\vec\nabla
{g}
+ i\left[{\vec A}_{ext}(t_1,r) -{\vec A}_{ext}(t_2,r) \right] {g}.
\nonumber 
\eeqa 

Now we are prepared to derive the collision integral. We notice that
due to causality $\langle g_Z \rangle_{\phi} =0$ in all orders of the
perturbation theory. We separate slow and fast degrees of freedom as
follows

\beqa 
\label{approx}
g_K= \langle g_K
\rangle_{\phi} +\delta g_K; \quad g_Z = \delta g_Z, 
\eeqa

\noindent
where $\delta g$ is the contribution fluctuating with the auxiliary
fields and we calculate it to first order in $\phi$.  In the same
approximation Eq.~(\ref{normsolution}) becomes

\beqa
\label{anormsolution}
&&g^R= \delta(t_1-t_2)- \frac{1}{2} g^K \delta g^Z;
\nonumber\\
&&
\nonumber\\
&&
g^A= -\delta(t_1-t_2)+ \frac{1}{2} \delta g^Z g^K,
\eeqa

\noindent
[expansion up to the second order in $\delta g^Z$ is unnecessary
because terms of such kind vanish due to Eq.~(\ref{do})].

We now substitute Eqs.~(\ref{approx}) and (\ref{anormsolution}) into
Eqs.~(\ref{exact}) and obtain equations governing the behavior of the
fluctuating parts

\begin{mathletters}
\label{first}
\beqa 
\Bigg[
{\tilde\partial_{t}} + v_F \vec n  
\tilde{\vec\nabla}
+ &&
\vec\omega_c
\left(\vec n \times 
\frac{\partial}{\partial \vec n} \right)\Bigg] 
\delta g^Z - \frac{1}{ \tau} 
\left[ \delta g^Z - \<\delta g^Z\>_n \right]
\nonumber\\
&&
\nonumber\\
&& 
=- 2 i\phi_2(t_1,\r)\delta(t_1-t_2),
\label{firstZ}
\eeqa
\beqa 
&&\left[
{\tilde\partial_{t}} + v_F \vec n  
\tilde{\vec\nabla}
+
\vec\omega_c
\left(\vec n \times 
\frac{\partial}{\partial \vec n} \right) \right]\delta g^K  +  
\frac{1}{\tau} \left[\delta g^K - \<\delta g^K\>_n \right]
\nonumber\\
&&
\nonumber\\
&&
=
2 i\phi_2(t_1,\r)\delta(t_1-t_2)
- i\left[\phi_1(t_1,\r)-\phi_1(t_2,\r) \right] 
\langle g^K\rangle_{\phi}
\nonumber\\
&& 
\nonumber\\
&&
+ \frac{1}{4 \tau} 
\left[
\langle g^K \rangle_{\phi} 
\< \delta g^Z  \langle g^K \rangle_{\phi} \>_n 
- \langle \<g^K\>_n \rangle_{\phi} 
\delta g^Z \langle g^K \rangle_{\phi} 
\right]
\label{firstK}
\\
&& 
\nonumber\\
&&
-  \frac{1}{4 \tau} 
\left[
 \langle g^K \rangle_{\phi} \delta g^Z
\langle \< g^K \>_n \rangle_{\phi} 
- \< \langle g^K \rangle_{\phi}
\delta g^Z\>_n\langle g^K \rangle_{\phi}
\right],
\nonumber
\eeqa
\end{mathletters}

Solutions to the equations (\ref{first}) should be substituted into
the equation (\ref{exactK}) for the smooth part. Than the equation for
the smooth part should be averaged over the fluctuating fields
$\phi_{1,2}$ with the help of Eq.~(\ref{Ds}). As a result

\beqa 
&&\left[
{\tilde\partial_{t}} + v_F \vec n  
\tilde{\vec\nabla}
+
\vec\omega_c
\left(\vec n \times 
\frac{\partial}{\partial \vec n} \right)\right] 
\langle g^K \rangle_{\phi} 
\label{almostthere}
\\
&&
\nonumber\\
&&
\quad
= \St_{in}\left\{\langle g^K \rangle_{\phi}  \right\}
+  \St_{el}\left\{\langle g^K \rangle_{\phi}  \right\}.
\nonumber
\eeqa

\noindent
Here we separate the collision integrals into two contributions. The
physical meaning of such separation will be discussed shortly. The
first, inelastic part has the structure

\beqa
&&\St_{in}\left\{\langle g^K \rangle_{\phi}  \right\}
(t_1,t_2; {\vec n}, \vec r)
\label{inelastic1}
\\
&&
\nonumber\\
&&
\quad=- i\langle 
\left[\phi_1(t_1,\vec r)-\phi_1(t_2,\vec r) \right] 
\delta g^K(t_1,t_2; {\vec n}, \vec r)
 \rangle_{\phi}. 
\nonumber
\eeqa

\noindent
The second, elastic contribution has the form

\end{multicols}
\widetext

\beqa
&&\St_{el}\left\{\langle g^K \rangle_{\phi}\right\}
(t_1,t_2; \vec n; \vec r)
= \frac{1}{\tau}\Big[
\<\langle  g^K (t_1,t_2; \vec n; \vec r)\rangle_{\phi} \>_n
-\langle g^K(t_1,t_2; \vec n; \vec r) \rangle_{\phi} \Big]
\label{elastic1}
\\
&&
\nonumber\\
&&
\quad
+ 
\int d t_3 \int\frac{d\theta_1}{2\pi}
\left[
 \<\langle  g^K (t_1,t_3; \vec n_1, \vec r)\rangle_{\phi}
F^A(t_3,t_2; \vec n_1,  \vec n; \vec r) 
-  \<\langle  g^K (t_1,t_3; \vec n, \vec r)\rangle_{\phi}
F^A(t_3,t_2; \vec n,  \vec n_1; \vec r) 
\right]
\nonumber\\
&&
\nonumber\\
&&\quad+  
\int d t_3 \int\frac{d\theta_1}{2\pi}
\left[
F^R(t_1,t_3; \vec n,  \vec n_1;  \vec r) 
 \<\langle  g^K (t_3,t_2; \vec n_1, \vec r)\rangle_{\phi}
- F^R(t_1,t_3; \vec n_1,  \vec n; \vec r) 
 \<\langle  g^K (t_3,t_2; \vec n, \vec r)\rangle_{\phi}
\right],
\nonumber
\eeqa

\noindent
where the first term is just the ordinary impurity scattering and the
remaining terms characterize interaction effects. The kernels 
in Eq.~(\ref{elastic1}) are defined as

\beqa
F^R(t_1,t_2; \vec n, \vec n_1; \vec r)
= \frac{1}{4\tau}\int d t_3 
\langle \delta g^K (t_1,t_3; \vec n, \vec r) 
\left[\delta  g^Z (t_3,t_2; \vec n_1, \vec r)
-\delta  g^Z (t_3,t_2; \vec n, \vec r) \right]
\rangle_{\phi}
\nonumber\\
\nonumber\\
F^A(t_1,t_2; \vec n, \vec n_1; \vec r)
=\frac{1}{4\tau} \int d t_3 
\langle\left[\delta  g^Z (t_1,t_3; \vec n_1, \vec r)
-\delta  g^Z (t_1,t_3; \vec n, \vec r) \right]
\delta g^K (t_3,t_2; \vec n, \vec r) 
\rangle_{\phi}
\label{Fs}
\eeqa

\begin{multicols}{2}

The equations (\ref{almostthere}), (\ref{inelastic1}), (\ref{elastic1}),
(\ref{first}), (\ref{Ds}), and (\ref{Dysonb}) constitute a closed
system of kinetic equations. 
Although sufficient for description of interaction effects in 
disorder systems, these equations are inconvenient for analytical
calculations because the
expressions for the collision integral are nonlocal in space and time.
To simplify further calculations we will use the assumption that 
$\langle g^K (t_1,t_2;
\nn, \r)\rangle_{\phi}$ is a smooth function so that a gradient
expansion will be possible.

Before embarking on such calculation we pause to discuss the
physical distinction between the elastic (\ref{elastic1}) and
inelastic (\ref{inelastic1}) collision terms. One immediately notices
from Eq.~(\ref{elastic1}) that

\beq
\int d\theta\ \St_{el}(t_1,t_2; \vec n;  \vec r)=0,
\label{elastic2}
\eeq 

\noindent
for any $t_1$ and $t_2$.  This indicates that this part of the
collision integral preserves the number of particles on a given energy
shell [see below for explicit connections between time representation
and energy representation Eq.~(\ref{energy})].

The inelastic term (\ref{inelastic1}) does not vanish after angular
averaging. Therefore this part does promote electrons
between energy shells. However, we notice that

\beq
\St_{in}\left\{\langle g^K \rangle_{\phi}  \right\}
(t_1,t_1; {\vec n}, \r)=0,
\label{inelastic2}
\eeq

\noindent
for any direction ${\vec n}$.  Taking coinciding time arguments is
equivalent to integrating over the whole energy spectrum [see
Eq.~(\ref{energy})], so that not only the total number of particles is
conserved, but the total number of particles moving along a given
direction ${\vec n}$ is conserved (i.e. inelastic forward scattering).

Let us now perform the actual calculation of the collision integrals. We
solve Eq.~(\ref{firstZ}) and obtain

\beqa
&&\delta g^Z(t_1,t_2; \vec n,\vec r)=
2i\delta(t_1-t_2)
\label{gzsol}
\\
&&
\nonumber\\
&&
\quad\times 
\int d\r_1dt_3
\phi_2(\r_1,t_3)\int \frac{d\theta'}{2\pi}
D(t_3-t_1, \vec n', \vec n; \vec r_1, \vec r)
\nonumber\\
&&
\nonumber\\ 
&&
D(t; \vec n, \vec n'; \vec r_1, \vec r_2)=
\int \frac{d\omega d^2\q}{(2\pi)^3}
e^{i{\vec q}(\vec r_1-\vec r_2)-i\omega t}
D(\vec n, \vec n';  \omega, \vec q),
\nonumber
\eeqa

\noindent
where the diffuson propagator $D$ is defined in Eq.~(\ref{Diffuson}).

To simplify the analytic solution of Eq.~(\ref{firstK}), we assume
without loss of generality that $\langle g^K\rangle_{\phi}$
varies slowly on the spatial scale $L_T = v_F {
min}(1/T,\sqrt{\tau /T})$, and also a slow function of $t_1+t_2$ on the time
scale $\simeq 1/T$. These assumptions are consistent with the first
loop approximation we already invoked. 

In what follows we will keep only the zeroth and
first angular harmonics (which is consistent with assumption about
the spatial smoothness) in the direction dependence of the Keldysh
function:

\beqa
\langle g\left(t_1, t_2; \vec n, \vec r\right) \rangle_\phi \approx &&
\langle g\left(t_1, t_2; \vec n, \vec r\right) \rangle_n
\nonumber\\
&&
\nonumber\\
&&
+2 \vec n 
\langle \vec n' g\left(t_1, t_2; \vec n', \vec r\right) \rangle_{n'}
\label{angular}
\eeqa 

\noindent
This approximation does not affect results for any relevant
quantities. From now on we will suppress the explicit sign of averaging
over the fluctuating fields because we will not be dealing with
non-averaged quantities anymore.

We now substitute Eq.~(\ref{angular}) into the right-hand side of
Eq.~(\ref{firstK}) and obtain

\beqa
\delta g^K (t_1, t_2; \vec n, \vec r) = &&
 \delta g^K_1 (t_1, t_2; \vec n, \vec r) 
\nonumber\\
&&
\nonumber\\
&&
+
\delta g^K_2 (t_1, t_2; \vec n, \vec r) 
\label{gk12}
\eeqa

\noindent
The first term in Eq.~(\ref{gk12}) is proportional to the field
$\phi_1$ and gives contributions to both the elastic and the
inelastic parts of the collision integral. To obtain non-vanishing
contribution to the latter we have to do each one of the following:
(i) take into account the first angular harmonic; (ii) perform the 
first order gradient expansion; (iii) expand up to the first order 
in external fields, ${\vec A}_{ext}$. The result is

\end{multicols}
\widetext

\begin{mathletters}
\label{gk1}
\beqa
\delta g^K_1 (t_1, t_2; \vec n, \vec r) &=& 
-i\int dt
\left[\phi_1(\vec r_1,t_1-t) -\phi_1(\vec r_1,t_2-t)\right] 
\int \frac{d\vec n'}{2\pi} 
D(t, \vec n, \vec n'; \vec r, \vec r_1) 
\nonumber\\
&&
\nonumber\\
&&
\times \left.\Big\{
\<g^K (t_1-t, t_2-t; \vec n_1, \vec r)\>_{\vec n_1}
\right.
\label{gk1a}
\\
&&
\nonumber\\
&&\quad\quad+
2 \vec n' \<\vec n_1g^K (t_1-t, t_2-t; \vec n_1, \vec r)\>_{\vec n_1}
\label{gk1b}
\\
&&
\nonumber\\
&&\quad\quad+ \left(\vec r_1-\vec r\right)
\tilde{\vec\nabla}
\<g^K (t_1-t, t_2-t; \vec n_1, \vec r)\>_{\vec n_1}
\Big\}
\label{gk1c}
\eeqa
\end{mathletters}

\noindent
where the covariant derivative is defined in Eq.~(\ref{covariant2})
and we neglected higher order derivatives of the external fields.
Expansion in the time coordinate $t_1+t_2$ (using the covariant
derivative $\tilde{\partial}_t$) is not necessary because it produces a
negligible correction to the inelastic collision integral and does not
affect the elastic one.

The second term in the RHS of Eq.~(\ref{gk12}) is proportional to the
field $\phi_2$, and according to Eqs.~(\ref{Fs}) and (\ref{do}) it
does not contribute to the elastic collision integral. Therefore, it
is sufficient to keep only the zeroth angular component and neglect
gradient terms at all. This yields

\beqa
&&\delta g^K_2(t_1, t_2; \vec n, \vec r)
=\int \frac{d\theta'}{2\pi}\frac{d\theta''}{2\pi}
\int d\vec r_1 dt
D(t, \vec n, \vec n'; \vec r, \vec r_1)
\left. \Big\{
2 i\phi_2(\vec r_1,t_1-t) \delta(t_1-t_2)
\right. 
\label{gk2}
\\
&&
\nonumber\\
&&
\left.
+ \frac{i}{\tau} 
\langle g\left(t_1 - t, t_3; \vec n_1, \vec r\right) \rangle_{\vec n_1} 
 \langle g\left(t_3, t_2 - t; \vec n_1, \vec r\right) \rangle_{\vec n_1}
\left[
\<D(t_4-t_3, \vec n'', \vec n_1; \vec r_2, \vec r_1)\>_{\vec n_1}-
D(t_4-t_3, \vec n'', \vec n'; \vec r_2, \vec r_1)
\right].
\phi_2(\vec r_2,t_4)
\right\}
\nonumber
\eeqa

\begin{multicols}{2}

As we already mentioned, $g(t_1,t_2)$ has a much faster dependence on
the difference $t_1-t_2$ then on the sum $t_1+t_2$. Therefore it is more 
convenient to
use a temporal transformation of the Green's function

\beq
g^K(t_1, t_2; \vec n, \vec r) = \int \frac{d\epsilon}{2\pi} 
g\left(\frac{t_1+t_2}{2}, \epsilon; \vec n, \vec r\right) 
e^{i\epsilon(t_2-t_1)}, 
\label{energy}
\eeq  

\noindent
which defines the precise notion of energy $\epsilon$ in this
context.  We introduce the same transformation for the propagators of
auxiliary fields (\ref{Ds})

\begin{equation}
{\cal D}(t_1,t_2) = \int \frac{d\w}{2\pi} 
{\cal D}\left( \frac{t_1+t_2}{2},\omega\right)e^{i\omega (t_2-t_1)}
\label{dw}
\end{equation}

\noindent
The transformed functions have the symmetry property (hereafter we
omit the $K$ superscript for brevity since we are only dealing with
the Keldysh function) 

\beqa
g(t,\epsilon) &=& - g(t,-\epsilon); 
\label{symmetry}
\\
&&
\nonumber\\
{\cal D}^K\left(t,\omega;\vec r_1,\vec r_2\right)&=&
{\cal D}^K\left(t,-\omega;\vec r_2,\vec r_1\right);
\nonumber\\
&&
\nonumber\\
{\cal D}^R\left(t,\omega;\vec r_1,\vec r_2\right)&=&
{\cal D}^A\left(t,-\omega;\vec r_2,\vec r_1\right).
\nonumber
\eeqa

Now, we are ready to obtain the explicit form of the collision
integral. We start with the inelastic contribution and perform the
following three steps: (1) substitute Eq.~(\ref{gk1a}) and (\ref{gk2})
into Eq.~(\ref{inelastic}); (2) average over the fields $\phi_{1,2}$
with the help of Eq.~(\ref{Ds}); (3) perform the temporal
transformation (\ref{energy}) of the result. As a result we obtain
with the help of Eqs.~(\ref{dw}) and (\ref{symmetry}) the following
form of the collision integral:

\end{multicols}
\widetext

\beqa
&&\St_{in}\left\{g^K \right\}
(t,\epsilon; \vec r)=
- \frac{i}{2}
\int d^2 r_1 \int \frac{d\omega}{2\pi}
{\cal D}^K\left(t,\omega; \vec r, \vec r_1\right)
\left[\langle D(\omega;\vec r,\vec r_1)\rangle +
\langle D(-\omega;\vec r_1,\vec r)\rangle\right] 
\nonumber\\
&&
\nonumber\\
&&
\hspace*{5cm}
\times
\left[\langle g\left(t, \epsilon; \vec n, \vec r\right) \rangle_{\vec n}
-
\langle g\left(t, \epsilon -\omega ; \vec n, \vec r\right) \rangle_{\vec n}
\right]
\nonumber\\
&&
\nonumber\\
&&
\quad
+\frac{i}{2\tau}
\int d\vec r_1d\vec r_2 \int \frac{d\omega}{2\pi}
\left[{\cal D}^R\left(t,\omega; \vec r_1, \vec r_2\right)-
{\cal D}^A\left(t,\omega; \vec r_2, \vec r_1\right)\right]
\left[\langle D(\omega;\vec r,\vec r_1)\rangle
\langle D(-\omega;\vec r,\vec r_2)\rangle
- \langle D(\omega;\vec r,\vec r_1) D(-\omega;\vec r,\vec r_2)\rangle
\right]
\nonumber\\
&&
\nonumber\\
&&
\hspace*{5cm}
\times
\langle g\left(t, \epsilon +\omega ; \vec n, \vec r\right) \rangle_{\vec n}
\langle g\left(t, \epsilon; \vec n, \vec r\right) \rangle_{\vec n}
\label{inelastic3}
\eeqa

\noindent
where the angular averaging of the diffusons is defined after
Eqs.~(\ref{eqs4.6}). 

Now, we have to express the bosonic propagator in terms of the
fermionic polarization operators. The polarization operators are 
given by Eqs.~(\ref{Pi}), where we now substitute Eqs.~(\ref{gzsol}),
(\ref{gk1a}), and (\ref{gk2}). After the temporal 
transformation (\ref{energy}) we find

\begin{mathletters}
\beqa
\Pi^R(\omega;t, \vec r_1,\vec r_2)&=&\Pi^A(-\omega;t, \vec r_2,\vec r_1)
\nonumber\\
&&
\nonumber\\
&=&\nu 
\left[\delta(\vec r_1-\vec r_2) + \frac{i}{4}
\langle D(\omega; \vec r_1, \vec r_2 )\rangle
\int d \epsilon 
\left[\langle g\left(t, \epsilon; \vec n, \vec r\right) \rangle_{\vec n}
-
\langle g\left(t, \epsilon -\omega ; \vec n, \vec r\right) \rangle_{\vec n}
\right] \right]  \label{Pra}
\\
&&
\nonumber\\
&=&
\nu \left[\delta(\vec r_1-\vec r_2) 
+i\omega  \langle D(\omega; \vec r_1, \vec r_2 )\rangle\right] 
\label{Prb}
\eeqa
\end{mathletters}
\beqa
\Pi^K(t, \omega; \vec r_1,\vec r_2)&=&
\frac{i\nu}{4\tau} 
\int d \vec r
\Big[\langle D(\omega;\vec r,\vec r_1)\rangle
\langle D(-\omega;\vec r,\vec r_2)\rangle
- \langle D(\omega;\vec r,\vec r_1) D(-\omega;\vec r,\vec r_2)\rangle
\Big] 
\label{Pk}
\\
&&
\nonumber\\
&\times&\int {d\epsilon}  \left[
\langle g\left(t, \epsilon +\omega ; \vec n, \vec r\right) \rangle_{\vec n}
\langle g\left(t, \epsilon; \vec n, \vec r\right) \rangle_{\vec n}
-4 \right] 
\nonumber
\eeqa

\noindent
The last step in the calculation of the interaction propagators is to
solve Eq.~(\ref{Dysonb}) with the polarization operators
Eq.~(\ref{Pra}).  This gives ${\cal D}^{R,A}$ in the form given by
Eq.~(\ref{int-dyson}) and for the Keldysh component we obtain

\begin{mathletters}
\beq
{\cal D}^K={\cal D}^{R}\Pi^K{\cal D}^{A}.
\eeq

\noindent
Also we can relate the difference of the retarded and advanced 
propagators which enters the collision integral Eq.~(\ref{inelastic3})
to the polarization operators:
  
\beq
{\cal D}^{R}-{\cal D}^{A}= {\cal D}^{R}
\left[{\Pi}^{R}-{\Pi}^{A}\right]
{\cal D}^{A}.
\eeq
\label{dsolution}
\end{mathletters}

To obtain the final form of the inelastic part of the collision 
integral Eqs.~(\ref{eq:4.8}) we need to
substitute Eq.~(\ref{dsolution}) into Eq.~(\ref{inelastic3}), while
using Eq.~(\ref{Pra}) for $\Pi^R-\Pi^A$. In addition, we note that

\[
\frac{2}{\tau}
\Big[
\langle D(\omega; q) D(-\omega;-q)\rangle
- \langle D(\omega; q)\rangle\langle D(-\omega;-q)\rangle
\Big] = \langle D(\omega; q)\rangle + \langle D(-\omega;-q)\rangle.
\]

\noindent
Finally, we introduce the gauge invariant distribution function $f$ as

\begin{equation}
f(\epsilon,t; \vec n, \vec r)
= \frac{1}{2} - \frac{1}{4} 
g(\epsilon + \varphi_{ext}(\vec r),t; \vec n, \vec r)  
\label{fg}
\end{equation}

\noindent
and obtain Eqs.~(\ref{eq:4.8}).

The calculation of the elastic part of the collision integral is
completely analogous. We substitute Eqs.~(\ref{gk1a})--(\ref{gk1c})
and Eq.~(\ref{gzsol}) into Eqs.~(\ref{Fs}) and average over
fluctuating fields with the help of Eq.~(\ref{Ds}).  After the temporal
transformation (\ref{energy}) we find

\begin{mathletters}
\label{el4}
\label{elastic4}
\beqa
F^R(\epsilon,t; \vec n_1,\vec n_2,\vec r)&=&
F^A(\epsilon,t; \vec n_1,\vec n_2,\vec r)^*
= \frac{i}{4\tau}
\int \frac{d \omega}{2\pi}
\int d\vec r_1 d\vec r_2
{\cal D}^R(\omega, \vec r_1, \vec r_2) 
\nonumber\\
&&
\nonumber\\
&\times&
\int \frac{d\vec n_3}{2\pi}\frac{d\vec n_4}{2\pi}
\left[D(\omega;  \vec n_3, \vec n_2, \vec r_2, \vec r) - 
D(\omega;  \vec n_3, \vec n_1, \vec r_2, \vec r)\right]
D(\omega;  \vec n_1, \vec n_4, \vec r, \vec r_1)
\nonumber\\
&&
\nonumber\\
&&
\quad \times \Big\{
\<g (t, \epsilon-\w; \vec n', \vec r)\>_{\vec n'} 
\label{el4a}
\\
&&
\nonumber\\
&&
\quad +2 \vec n_4 \<\vec n' g (t, \epsilon-\w ; \vec n_1, \vec r)\>_{\vec n'}
\label{el4b}
\\
&&
\nonumber\\
&&
+\left(\vec r_1-\vec r\right)
\left(
\vec\nabla
+\frac{\partial {\vec A}_{ext}}{\partial t} 
\frac{\partial}{\partial \epsilon} 
\right) 
\<g (t, \epsilon -\w; \vec n', \vec r)\>_{\vec n'}
\Big\}
\label{el4c}
\eeqa
\end{mathletters}

\noindent
Deriving Eq.~(\ref{elastic4}) we use the fact that $\int d\omega {\cal
D}^R(\omega)D(\omega) = 0$.

We substitute Eqs.~(\ref{el4}) and (\ref{angular}) into
Eq.~(\ref{elastic2}). We expand the result into angular harmonics.
The zeroth angular harmonic vanishes because of the conservation law
(\ref{elastic2}), and the first harmonic gives (we write only
interaction correction to the collision integral)

\begin{mathletters}
\label{el5}
\beqa
\St_{el}(t; \epsilon;  \vec r)
=\frac{2}{\tau}\int \frac{d\omega}{2\pi}
&\Big[&
n_\alpha K_1^{\alpha\beta}(\w) 
\langle n_\beta g(t; \epsilon, \vec r,\vec n)\rangle_n
\langle g(t; \epsilon-\omega, \vec r,\vec n)\rangle_n
\label{el5a}
\\
&&
\nonumber\\
&+&n_\alpha K_0^{\alpha\beta}(\w) 
\langle n_\beta g(t; \epsilon-\omega, \vec r,\vec n)\rangle_n
\langle g(t; \epsilon, \vec r,\vec n)\rangle_n
\label{el5b}
\\
&&
\nonumber\\
&+&
\frac{n_\alpha L_0^{\alpha\beta}(\w)}{2}
\langle g(t; \epsilon, \vec r,\vec n)\rangle_n
\left(\nabla_\beta
+ \frac{\partial [A_{ext}]_\beta}{\partial t} 
\frac{\partial}{\partial \epsilon} 
\right) 
\langle g(t; \epsilon-\omega, \vec r,\vec n)\rangle_n
\Big].
\label{el5c}
\eeqa
\end{mathletters}

\noindent
Here the kernels $K$ and $L$ are given by Eq.~(\ref{eqs4.6}).  Each
labeled separately term in Eqs.~(\ref{el5}) corresponds to ones in
Eqs.~(\ref{elastic4}) and in Eqs.~(\ref{gk1}) so the origin of terms
can be easily traced.

\begin{multicols}{2}

Finally, we use the gauge invariant distribution function (\ref{fg})
instead of $g$ and we arrive to Eqs.~(\ref{eq:4.5}).

Closing this section, we remark that the above treatment can be easily
generalized to include other channels as well as the higher angular
harmonics of the Fermi liquid constant. For the latter task one has to
introduce angle dependent auxiliary fields $\phi_{1,2}(\vec r,\vec n,t)$ and
use $F_0^\rho \to F^\rho(\widehat{\vec n_1\vec n_2})$.

The triplet channel requires introduction of the coupling of the form
${\vec h}_{1,2}(\vec r,t) 
\hat{\vec\sigma}$,
where
$\hat{\sigma}^j,\ j=x,y,z$ are the Pauli matrices in the spin space,
and ${\vec h}_{1,2}(\vec r,t)$ are the auxiliary fields.  Accordingly, each
bosonic propagator from Eq.~(\ref{Ds}) becomes a $3 \times 3$ matrix. The
equation~(\ref{Dysonb}) retains the same form with the matrix multiplication
in Keldysh and spin spaces implied. The equation (\ref{D00}) becomes

\beq
[{\cal D}_0^R]_{ij} = [{\cal D}_0^A]_{ij} 
=- F_0^\sigma \delta_{ij}\delta(\vec r_1-\vec r_2)\delta(t_1-t_2), 
\eeq   

\noindent
and Eq.~(\ref{Pi}) is modified to

\beqa 
&&{\Pi}_{ij}^R(1,2)= {\Pi}_{ji}^A(2,1) 
\label{Pi2}
\\
&&
\nonumber\\
&&
\quad =\nu \int \frac{d\theta}{2\pi}
\left(\delta_{12}\delta_{ji} + 
\frac{\pi \langle  {\bf Tr}\sigma_i
\delta g^K(t_1, t_1;  \vec n, \vec r_1)\rangle_{\phi}}
{4\delta h_1^j(t_2,\vec r_2)}\right); 
\nonumber\\
&&
\nonumber\\
&&
{\Pi}^K(1,2)= \pi\nu
\nonumber\\
&&
\nonumber\\
&&
\int \frac{d\theta}{2\pi} 
\frac{ \langle  {\bf Tr}\sigma_i
 \delta g^K(t_1, t_1;  \vec n, \vec r_1)
+ {\bf Tr}\sigma_i\delta  g^Z(t_1, t_1;  \vec n, \vec r_1)
\rangle_{\phi}}
{4\delta h^j_2(t_2,\vec r_2)}
\nonumber
\eeqa

\noindent
where trace is performed in spin space.

Further derivation consists of a repetition of the steps described
in this section, and in the absence of the spin structure of the
distribution function, $f_{ij}= \delta_{ij}f$, results in
Eqs.~(\ref{r1}) and (\ref{r2}). The spin-orbit interaction or Zeeman
splitting by external magnetic field slightly changes the results, 
but we will
postpone the corresponding analysis until the future publication \cite{pre}.

Finally, the Cooper channel interaction
(\ref{106}) can be treated in the same manner by
introducing auxiliary fields in the Gorkov-Nambu space. We will not
discuss this question further in the present paper.

\section{Conclusions}
\label{projects}

This paper is an attempt to consistently describe the effect of 
electron-electron interaction on longitudinal conductivity of disordered
2D electron gas at $T\ll E_F$. Our results are valid for an arbitrary
relation between $T$ and $\hbar/\tau$ and are summarized in 
Section~\ref{results}. At low temperatures $T\tau\ll \hbar$ we reproduce the
logarithmically divergent Altshuler-Aronov correction. At higher
temperatures $T\tau > \hbar$, i.e. in the ballistic region, we found the
linear temperature dependence in accord with Refs.~\onlinecite{gdl,rei}.
However, even the sign of the slope of this dependence depends
on the strength of electron-electron interaction in contradiction to
the results of Refs.~\onlinecite{gdl,rei} (see Sections~\ref{qualitative}
and \ref{single} for discussions of this discrepancy).

We deliberately did not compare the theory with experimental data,
postponing this comparison until the publication of theoretical
results for Hall conductivity and magneto-resistance in the parallel 
field. For comparison with data obtained for Si-MOSFET samples the
valley degeneracy should be taken into account (the degenaracy may 
increase the numerical factor in Eq.~(\ref{tc}) by as much as a factor
of $5$ in the case of low intervalley scattering). 
We also relegate the
corresponding discussion to a separate publication.

Finally, we derived a kinetic equation to describe the effect of 
electron-electron interaction at arbitrary $T\tau$. The advantage of 
this approach is that it turns out to be more convenient for
practical calculations of transport properties in magnetic field
as well as thermal transport properties.

\acknowledgements

We are grateful to B.L. Altshuler, 
M.E. Gershenson, L.I. Glazman, A.I. Larkin, and M.Yu. Reizer
for numerous discussions. I.A. acknowledges support of
the Packard Foundation.

\appendix

\section*{}

{
\narrowtext
\begin{figure}[ht]
\vspace{0.5 cm}
\epsfxsize=6 cm
\centerline{\epsfbox{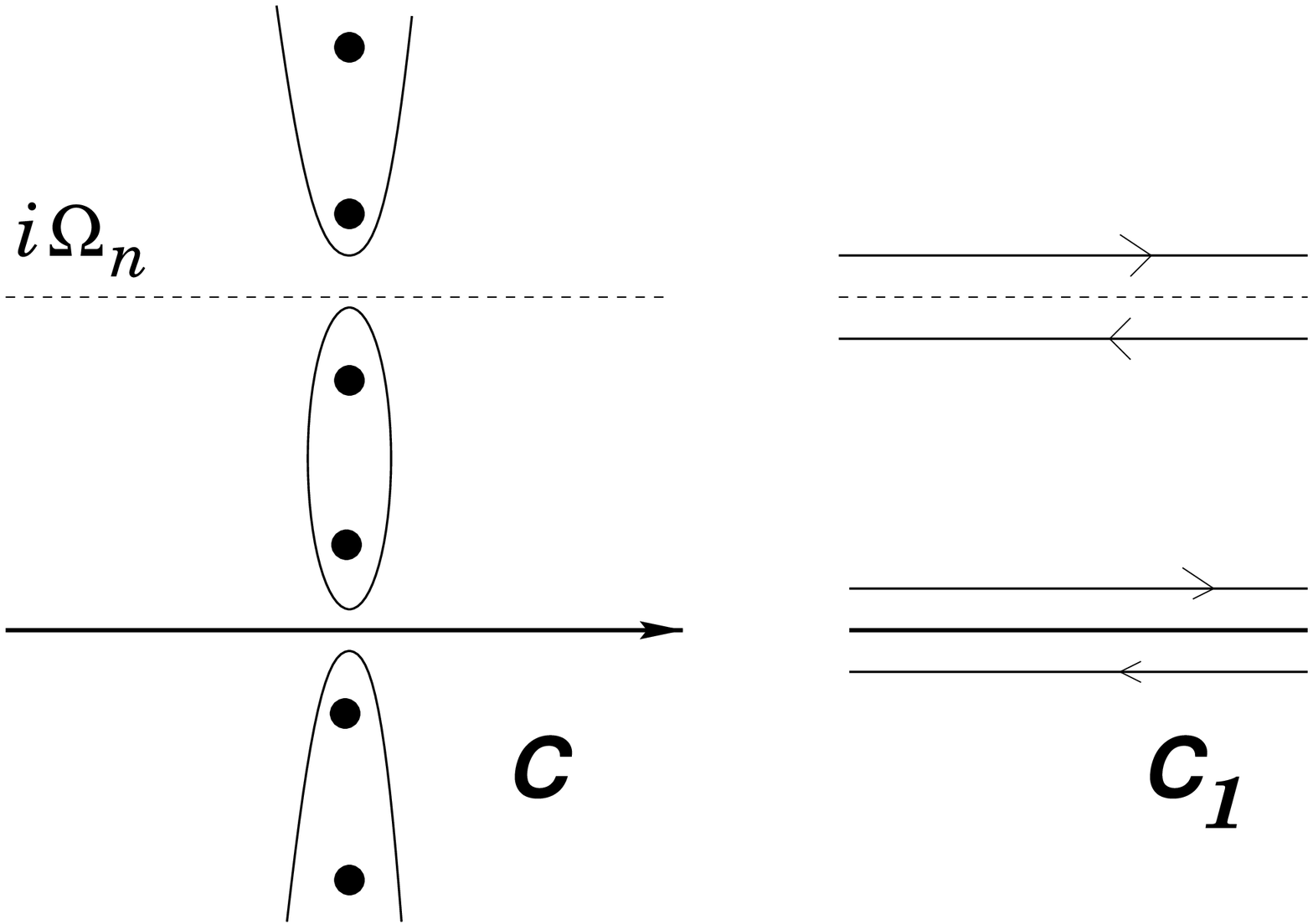}}
\vspace{0.5cm}
\caption{Integration contours for analytic continuation of
 Eqs.~(\protect\ref{aa1}).}
\label{figA1}
\end{figure}
}

In this Appendix we show in some detail the procedure of analytical 
continuation that leads to the expression for the interaction
correction Eq.~(\ref{cc}) to the conductivity in terms of exact Green's 
function of noninteracting disordered system.
The structure of the current correlator is  

\begin{mathletters}
\label{aa1}
\beqa
&&\int\limits_{0}^{1/T} d\tau \langle {\bf T_\tau} \hat j_\alpha(\tau)
\hat j_\beta(0) \rangle e^{i\Omega_n\tau} =
 \nonumber\\
 &&
\
- T\sum_{\epsilon_n}J_\alpha
G(i\epsilon_n+ i\Omega_n)
J_\beta G(i\epsilon_n)
\label{aa1a} \\
&&
\
- T\sum_{\epsilon_n}J_\alpha
G(i\epsilon_n+ i\Omega_n)
\Gamma_\beta (i\epsilon_n+ i\Omega_n, i\epsilon_n)
G(i\epsilon_n),
\label{aa1b}
\eeqa
\end{mathletters}

\noindent
where
$\epsilon_n= \pi T(2n+1)$ is the fermionic Matsubara frequency,
$G(i\epsilon_n)$ is the exact Green's function 
of the interacting system (diagrams $1$ and $2$ on Fig.~\ref{sr}
are the first order correction to the Green's function),
$\Gamma_\beta (i\epsilon_n+ i\Omega_n, i\epsilon_n)$ is the
vertex function (not to be confused with disorder averaged 
interaction vertex $\Gamma$ from Sec.~\ref{diagrams}).
Diagrams $3$ on Fig.~\ref{sr}
are the first order correction to the vertex function.
The current operator is defined in Eq.~(\ref{jop-def}).
Note, that 
we omit the spatial coordinates and the integration 
whenever it should cause no confusion.

We perform the analytic continuation in each term 
(\ref{aa1a}) and (\ref{aa1b}) separately. In Eq.~(\ref{aa1a})
we use the standard procedure
\beq
T\sum\limits_{\epsilon_n} (\dots ) = \frac{1}{4 \pi i}
\int\limits_{{\cal C}} d\epsilon \tanh\frac{\epsilon}{2T}(\dots ),
\label{trick}
\eeq
where integration contour is shown on Fig.~\ref{figA1}. We deform
this contour to form $C_1$, use the facts that
$\tanh (\epsilon + i\Omega_n)/2T=\tanh \epsilon/2T$,
and $G(\epsilon \pm i0)=G^{R(A)}(\epsilon)$ and we obtain
\beqa
&&M_1(i\Omega_n)=T\sum_{\epsilon_n}J_\alpha
G(i\epsilon_n+ i\Omega_n) J_\beta G(i\epsilon_n)
\nonumber\\
&&=\int\frac{d\epsilon}{4 \pi i}
\tanh\frac{\epsilon}{2T}
\Big\{
J_\alpha G(\epsilon + i\Omega_n) J_\beta
\left[G^R(\epsilon ) - G^A(\epsilon ) \right]
\nonumber\\
&&+ J_\alpha \left[G^R(\epsilon ) - G^A(\epsilon ) \right]
J_\beta G(\epsilon - i\Omega_n)
\Big\}
\label{aa2}
\eeqa
In the form (\ref{aa2}) frequency $\Omega$ is present only in
functions which may have singularities only on the real axis, so that
the required analytic continuation can be easily performed:

\beqa
&&M_1(\omega)=
M_1(i\Omega_n \to \omega + i0)
\nonumber\\
&&=\int\frac{d\epsilon}{4 \pi i}
\tanh\frac{\epsilon}{2T}
\Big\{
J_\alpha G^R(\epsilon + \omega) J_\beta
\left[G^R(\epsilon ) - G^A(\epsilon ) \right]
\nonumber\\
&&\quad + J_\alpha \left[G^R(\epsilon ) - G^A(\epsilon ) \right]
J_\beta G^A(\epsilon - \omega)
\Big\}
\label{aa3}
\eeqa
Thus one obtains for the quantity entering into conductivity
(\ref{kubo})
\beqa
&&N_1=- \lim_{\omega \to 0}
{\rm Im}
\left(
\frac{M_1(\omega)}{\omega}
\right) \nonumber\\
&&=
{\rm Re}
\int\frac{d\epsilon}{2 \pi}
\tanh\frac{\epsilon}{2T}
 J_\alpha G^A(\epsilon ) 
J_\beta \partial_\epsilon G^A(\epsilon)
\nonumber\\
&&
+ \int\frac{d\epsilon}{4 \pi}
\left(\frac{d}{d \epsilon}\tanh\frac{\epsilon}{2T}
\right)
J_\alpha  G^R(\epsilon ) J_\beta
G^A(\epsilon )
\label{aa4}
\eeqa
Equation (\ref{aa4}) can be further simplified for the calculation of
the symmetric part of the conductivity 
\beqa
&&N_1^{sym}=
{\rm Re}
 \int\frac{d\epsilon}{8\pi}
\left(\frac{d}{d \epsilon}\tanh\frac{\epsilon}{2T}
\right)\label{aa5}\\
&&\times
\Big[
- J_\alpha G^A(\epsilon ) 
J_\beta G^A(\epsilon)
+ J_\alpha  G^R(\epsilon ) J_\beta
G^A(\epsilon )
+(\alpha \leftrightarrow \beta)
\Big]
\nonumber
\eeqa

\end{multicols}
\widetext
Term (\ref{aa1b}) is considered analogously. 
We find similarly to Eq.~(\ref{aa2}) 
\beqa
&&M_2(i\Omega_n)=T\sum_{\epsilon_n}J_\alpha
G(i\epsilon_n+ i\Omega_n) 
\Gamma_\beta (i\epsilon_n+ i\Omega_n, i\epsilon_n) G(i\epsilon_n)
\nonumber\\
&&=\int\frac{d\epsilon}{4 \pi i}
\tanh\frac{\epsilon}{2T}
\Big\{
J_\alpha G(\epsilon + i\Omega_n)
\left[\Gamma_\beta (\epsilon + i\Omega_n, \epsilon + i0)
G^R(\epsilon ) - 
\Gamma_\beta (\epsilon + i\Omega_n, \epsilon - i0)
G^A(\epsilon ) \right]
\nonumber\\
&&+ J_\alpha \left[G^R(\epsilon ) 
\Gamma_\beta (\epsilon + i0, \epsilon - i\Omega_n)
- G^A(\epsilon ) 
\Gamma_\beta (\epsilon - i0, \epsilon - i\Omega_n)
\right]
G(\epsilon - i\Omega_n )
\Big\}
\label{aa6}
\eeqa
Using analytic properties of the Green's function and the vertex
function, we perform the analytic continuation and obtain
\beqa
&&M_2(\omega)= M_2(i\Omega_n \to \omega+i0)
\nonumber\\
&&
=
\lim_{\delta_1 \to 0^+}\lim_{\delta_2 \to 0^+}
\int\frac{d\epsilon}{4 \pi i}
\tanh\frac{\epsilon}{2T}
\Big\{
J_\alpha G^R(\epsilon + \omega)
\left[\Gamma_\beta (\epsilon + \omega + i\delta_1, \epsilon + i\delta_2)
G^R(\epsilon ) - 
\Gamma_\beta (\epsilon + \omega + i0, \epsilon - i0)
G^A(\epsilon ) \right]
\nonumber\\
&&+ J_\alpha \left[G^R(\epsilon ) 
\Gamma_\beta (\epsilon + i0, \epsilon - \omega - i0)
- G^A(\epsilon ) 
\Gamma_\beta (\epsilon - i\delta_2, \epsilon - \omega - i\delta_1)
\right]
 G^A(\epsilon - \omega )
\Big\}.
\label{aa7}
\eeqa
In the appropriate frequency limit, we find
\beqa
&&N_2= 
- \lim_{\omega \to 0}
{\rm Im}
\left(
\frac{M_2(\omega)}{\omega}
\right)
=
{\rm Re}
\int\frac{d\epsilon}{2\pi }
\tanh\frac{\epsilon}{2T}
J_\alpha 
G^A(\epsilon)
\left.
\frac{\partial}{\partial \epsilon_1}
\right|_{\epsilon_1=\epsilon }
\Big[
\lim_{\delta_1 \to 0^+}\lim_{\delta_2 \to 0^+}
\Gamma_\beta (\epsilon - i\delta_2, \epsilon_1 - i\delta_1)
G^A(\epsilon_1 )
\Big]
\nonumber\\
&&+
\int\frac{d\epsilon}{4 \pi}
\left(\frac{d}{d \epsilon}\tanh\frac{\epsilon}{2T}
\right)
J_\alpha  G^R(\epsilon ) 
\Gamma_\beta (\epsilon + i0, \epsilon - i0)
G^A(\epsilon )
\label{aa8}
\eeqa

\begin{multicols}{2}
Further calculation requires specification of the form of
the self-energy and the vertex function. We have to find both
in the first order in interaction propagator, and expand 
Eq.~(\ref{aa5}) up to the first order:
\beqa
&&\delta N_1^{sym}=
{\rm Re}
 \int\frac{d\epsilon}{4\pi}
\left(\frac{d}{d \epsilon}\tanh\frac{\epsilon}{2T}
\right)\label{aa50}\\
&&\times
\Big[-
 J_\alpha G^A(\epsilon ) \Sigma^A(\epsilon )  G^A(\epsilon ) 
J_\beta G^A(\epsilon)
\nonumber\\
&&
+ J_\alpha  G^R(\epsilon ) J_\beta
G^A(\epsilon ) 
\Sigma^A(\epsilon )  G^A(\epsilon ) 
+(\alpha \leftrightarrow \beta)
\Big].
\nonumber
\eeqa
For brevity we consider only the ``Fock'' contribution of
Fig.~\ref{sr} b:
\beq
\Sigma(i\epsilon_n )_{12} = T\sum_{\Omega_m}
{\cal D}_{12}(i\Omega_m)
G_{12}(i\epsilon_n-i\Omega_m ) 
\label{aa9}
\eeq
where ${\cal D}$ is the bosonic propagator, and we restored the
notation for spatial coordinates.
In the same order
\beqa
&&\left[\Gamma(i\epsilon_n, i\epsilon_m )_{\beta}\right]_{12} 
\label{aa10}\\
=
&&\quad 
T\sum_{\Omega_m}{\cal D}_{12}(\Omega_m)
\left[G(i\epsilon_n- i\Omega_m ) J_\beta 
G(i\epsilon_m- i\Omega_m ) \right]_{12}.
\nonumber
\eeqa

After analytic continuation similar to that in the derivation of
Eq.~(\ref{aa3}) we find
\beqa
&&\Sigma^A_{12}(\epsilon) =
- \int \frac{d \Omega}{2\pi} \coth \frac{\Omega}{2T}
\Big[{\rm Im} {\cal D}_{12}^A(\Omega)
\Big]G^A_{12}(\epsilon-\Omega)
\label{aa11}\\
&&+
i\int \frac{d \Omega}{4\pi} 
\tanh \frac{\epsilon-\Omega}{2T}
{\cal D}_{12}^A(\Omega)
\Big[G^A_{12}(\epsilon-\Omega)
- G^R_{12}(\epsilon-\Omega)
\Big]
\nonumber
\eeqa
and for the vertex function we have two cases:

\end{multicols}
\widetext
\begin{mathletters}
\label{aa12}
\beqa
\lim_{\delta_1 \to 0^+}\lim_{\delta_2 \to 0^+}
\Gamma_\beta (\epsilon - i\delta_2, \epsilon_1 - i\delta_1)&&
=
- \int \frac{d \Omega}{2\pi} \coth \frac{\Omega}{2T}
\Big[{\rm Im} {\cal D}_{12}^A(\Omega)
\Big]\Big[G^A(\epsilon-\Omega)J_\beta G^A(\epsilon_1-\Omega)\Big]_{12}
\nonumber\\
&&
+i\int \frac{d \Omega}{4\pi} \tanh \frac{\epsilon - \Omega}{2T}
{\cal D}_{12}^A(\Omega)
\Big[\Big(G^A(\epsilon -\Omega)
- G^R(\epsilon -\Omega)\Big)J_\beta G^A(\epsilon_1-\Omega)\Big]_{12}
\nonumber\\
&&
+i\int \frac{d \Omega}{4\pi} \tanh \frac{\epsilon_1 - \Omega}{2T}
{\cal D}_{12}^A(\Omega)
\Big[G^R(\epsilon -\Omega)J_\beta\Big(G^A(\epsilon_1-\Omega)
- G^R(\epsilon_1-\Omega)\Big) \Big]_{12}
\eeqa
\beqa
\Gamma_\beta (\epsilon + i0, \epsilon - i0) &&
=
- \int \frac{d \Omega}{2\pi} \coth \frac{\Omega}{2T}
\Big[{\rm Im} {\cal D}_{12}^A(\Omega)
\Big]\Big[G^R(\epsilon-\Omega)J_\beta G^A(\epsilon-\Omega)\Big]_{12}
\nonumber\\
&&
+i\int \frac{d \Omega}{4\pi} \tanh \frac{\epsilon - \Omega}{2T}
{\cal D}_{12}^R(\Omega)
\Big[\Big(G^A(\epsilon -\Omega)
- G^R(\epsilon -\Omega)\Big)J_\beta G^A(\epsilon-\Omega)\Big]_{12}
\nonumber\\
&&
+i\int \frac{d \Omega}{4\pi} \tanh \frac{\epsilon - \Omega}{2T}
{\cal D}_{12}^A(\Omega)
\Big[G^R(\epsilon -\Omega)J_\beta\Big(G^A(\epsilon -\Omega)
- G^R(\epsilon -\Omega)\Big) \Big]_{12}
\eeqa
\end{mathletters}

We now substitute Eq.~(\ref{aa11}) into Eq.~(\ref{aa50}).  We use the
fact that the combination containing only retarded or only advanced
Green's functions vanish upon the disorder averaging.  Moreover, the
average of the combinations like $G(\epsilon)G(\epsilon-\Omega_1)\dots
G(\epsilon-\Omega_N)$ does not depend on the energy $\epsilon$, which
enable us to perform the integration over $\epsilon$ using

\[
\int d\epsilon \tanh \frac{\epsilon-\Omega}{2T}
\frac{d}{d\epsilon}\tanh \frac{\epsilon}{2T}=
- 2 \frac{d}{d\Omega}\left(\Omega \coth \frac{\Omega}{2T}\right).
\] 
We find using $D^A(\Omega) = [D^A(-\Omega)]^*$
\begin{mathletters}
\label{aa500}
\beqa
&&\delta N_1^{sym}= {\rm Im}
 \int\frac{d\Omega}{8\pi^2}
 \left[\frac{d}{d\Omega}\left(\Omega \coth \frac{\Omega}{2T}\right)
\right]{\cal D}_{12}^A(\Omega)
\label{aa500a}\\
&&
\times
\Big[
 J_\alpha G^A(\epsilon ) 
G^R_{12} (\epsilon -\Omega) G^A(\epsilon ) 
J_\beta G^A(\epsilon) -
J_\alpha G^A(\epsilon ) 
G^R_{12} (\epsilon -\Omega) G^A(\epsilon ) 
J_\beta G^R(\epsilon)
- J_\alpha G^R(\epsilon ) 
G^R_{12} (\epsilon -\Omega) G^R(\epsilon ) 
J_\beta G^A(\epsilon)
\Big]
\nonumber\\
&&
-
\int\frac{d\Omega}{4\pi^2}
\Big[
\frac{\Omega}{2T\sinh^2\frac{\Omega}{2T}}
{\rm Im} {\cal D}_{12}^A(\Omega) 
\Big] {\rm Re}
\Big[ J_\alpha G^A(\epsilon ) 
G^A_{12} (\epsilon -\Omega) G^A(\epsilon ) 
J_\beta G^R(\epsilon)\Big] + (\alpha \leftrightarrow \beta).
\label{aa500b}
\eeqa
\end{mathletters}

The same manipulations are performed with substituition of
Eqs.~(\ref{aa12}) into Eq.~(\ref{aa8}). One finds for the symmetrized
part

\begin{mathletters}
\label{aa800}
\beqa
&&\delta N_2^{sym}= {\rm Im}
 \int\frac{d\Omega}{8\pi^2}
 \left[\frac{d}{d\Omega}\left(\Omega \coth \frac{\Omega}{2T}\right)
\right]{\cal D}_{12}^A(\Omega)
\label{aa800a}\\
&&
\times
\Bigg\{
\Big[ G^A(\epsilon )  J_\alpha G^A(\epsilon ) \Big]_{12}
 \Big[ G^R (\epsilon -\Omega)
J_\beta G^R(\epsilon- \Omega ) \Big]_{21}
- 2 \Big[ G^R(\epsilon )  J_\alpha G^A(\epsilon ) \Big]_{12}
 \Big[ G^R (\epsilon -\Omega)
J_\beta G^R(\epsilon- \Omega ) \Big]_{21}
\Bigg\}
\nonumber\\
&&
-
\int\frac{d\Omega}{8\pi^2}
\Big[
\frac{\Omega}{2T\sinh^2\frac{\Omega}{2T}}
{\rm Im} {\cal D}_{12}^A(\Omega) 
\Big] 
\Big\{ \Big[ G^R(\epsilon )  J_\alpha G^A(\epsilon ) \Big]_{12}
 \Big[ G^A (\epsilon -\Omega)
J_\beta G^R(\epsilon- \Omega ) \Big]_{21}\Big\} 
+ (\alpha \leftrightarrow \beta).
\label{aa800b}
\eeqa
\end{mathletters}

Total correction to the conductivity is just $N_1+N_2$.
In the Hartree-Fock approximation $D^A=-V(q)$ and we obtain
Eq.~(\ref{cc}). In the case for the stronger interaction
terms (\ref{aa500a}) and (\ref{aa800a}) are added to produce 
Eq.~(\ref{cci}) and terms (\ref{aa500b}), (\ref{aa800b})
give rise to the inelastic or so-called dephasing term:
\beqa
\delta\sigma^{deph}_{\alpha\beta}
&=& -
\int\frac{d\Omega}{8\pi^2}
\Big[
\frac{\Omega}{2T\sinh^2\frac{\Omega}{2T}}
{\rm Im} {\cal D}_{12}^A(\Omega) 
\Big] {\rm Re}
\Bigg\{
\Big[ 2 J_\alpha G^A(\epsilon ) 
G^A_{12} (\epsilon -\Omega) G^A(\epsilon ) 
J_\beta G^R(\epsilon)\Big] 
\nonumber\\
&+& 
\Big[ G^R(\epsilon )  J_\alpha G^A(\epsilon ) \Big]_{12}
 \Big[ G^A (\epsilon -\Omega)
J_\beta G^R(\epsilon- \Omega ) \Big]_{21}
\Bigg\}
+
(\alpha \leftrightarrow \beta).
\label{adeph}
\eeqa

\begin{multicols}{2}

In our leading approximation in $1/E_F\tau$ this term vanishes,
see Fig.~\ref{vanishing}. The role of this term in the temperature
dependence of weak localization correction is discussed in detail in 
Ref.~\onlinecite{aag}.   
{
\narrowtext
\begin{figure}[ht]
\vspace{0.2 cm}
\epsfxsize=7 cm
\centerline{\epsfbox{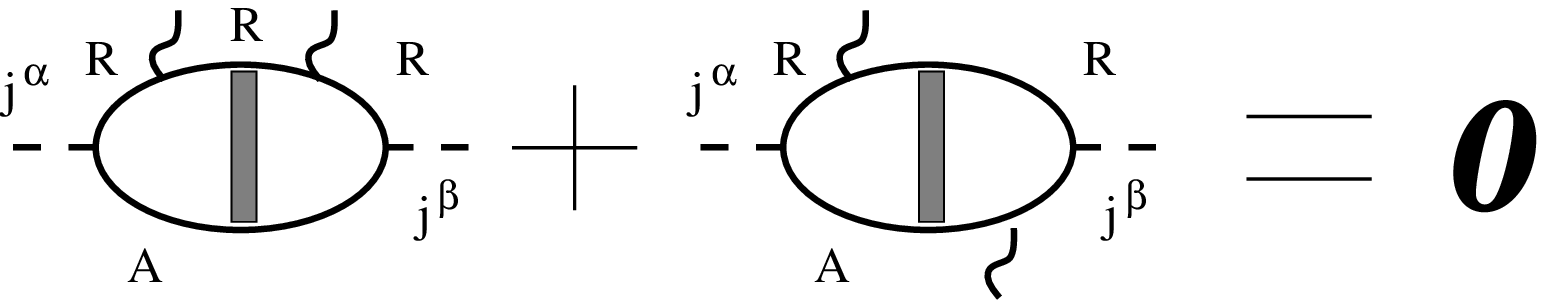}}
\vspace{0.2cm}
\caption{Cancellation of inelastic term (\protect\ref{adeph}) in
the leading ladder approximation.} 
\label{vanishing}
\end{figure}
}

\end{multicols}


\begin{references}
\bibitem{footnote1} Logarithmic renormalization of $F_0^\sigma$,
see Ref. \onlinecite{fin,Lee} will not be relevant for the further discussion.
\end{references}
\end{document}